\documentclass[a4paper,11pt]{report}
\usepackage[onehalfspacing]{setspace}
\usepackage{wrapfig}
\usepackage{cite}
\usepackage{caption}
\usepackage{subcaption}
\usepackage{changepage}
\usepackage{fancyhdr}
\usepackage{subfiles}
\usepackage[utf8]{inputenc}
\usepackage[left=2.5cm,right=2.5cm,top=2.5cm]{geometry}
\usepackage{amsmath}
\usepackage{amsfonts}
\usepackage{relsize}
\usepackage{multirow}
\usepackage{graphicx}
\usepackage{slashed}
\usepackage{hyperref}
\usepackage{color}
\usepackage{bbold} 
\usepackage{bigints}
\usepackage{float}
\usepackage{tikz} 
\usepackage[compat=1.1.0]{tikz-feynman}
\usetikzlibrary{decorations.pathmorphing}
\usetikzlibrary{decorations.markings}
\tikzset{
	graviton/.style={line width=.8pt, double, -latex,decorate, decoration={snake, segment length=7pt,amplitude=1.4pt, pre length=.1cm, post length=.1cm}},
	worldline/.style={gray, line width=1pt},
	worldlineBold/.style={black, line width=.6pt},
	zUndirected/.style={line width=1pt},
	zParticle/.style={line width=1pt,postaction={decorate},decoration={markings,mark=at position .6 with {\arrow[#1]{latex}}}},
	zParticleF/.style={line width=1pt,postaction={decorate}},
	cscalar/.style={line width=1pt,postaction={decorate},decoration={markings,mark=at position .6 with {\arrow[#1]{latex}}}},
	cscalar2/.style={line width=1pt,postaction={decorate},decoration={markings,mark=at position .8 with {\arrow[#1]{latex}}}},
	psiParticle/.style={line width=1pt,dashed,postaction={decorate}},
	psiParticleDirected/.style={line width=1pt,dashed,postaction={decorate},decoration={markings,mark=at position .56 with {\arrow[#1]{latex}}}},
	photon/.style={line width =.8pt, -latex, decorate, decoration={snake, segment length=7pt, amplitude=1.4pt,  pre length=.1cm, post length=.1cm}},
	photon2/.style={line width =.8pt, decorate, decoration={snake, segment length=7pt, amplitude=1.4pt,  pre length=.1cm, post length=0cm}},
	Scalar/.style={line width=1pt,dashed,postaction={decorate}},
}

\def\dd{\delta\!\!\!{}^-\!} 

\newcommand{\unchapter}[1]{\begingroup\chapter*{#1}\addcontentsline{toc}{chapter}{#1}\endgroup}

\begin{document}

\begin{titlepage}
\begin{center}
\vspace*{-3mm}
\huge{\textbf{MASTER THESIS}}\\ 
\vspace{.7cm}
\large{In fulfillment of the requirements \\ 
for the degree of Master of Science (M. Sc.) \\
in Physics}\\
\vspace{2cm}
\huge{\textbf{Relations between the Worldline Quantum Field Theory and scattering amplitudes for particles with spin}} \\ 
\vspace{2cm}
\large{submitted to\\
Mathematisch-Naturwissenschaftliche Fakultät \\
Institut für Physik 
Humboldt-Universität zu Berlin}
 \\ \vspace{2cm}
 
 by \\ \textbf{Raphael Kopp}  \\
574696
\\ \vspace{2cm}
Supervisors: \\
\textbf{Prof. Dr. Jan Plefka} \\
\textbf{Dr. Jan Steinhoff} \\
\vspace{2cm}
September 29, 2022
\end{center}
\end{titlepage}

\newgeometry{outer=25mm, inner=35mm, top=30mm, bottom=30mm, twoside}
\newpage
\setcounter{page}{1}
\pagestyle{empty}
\addtocontents{toc}{\protect\thispagestyle{empty}} 
\tableofcontents

\vspace{2cm}
\newpage
\paragraph{Conventions and Definitions}
\begin{itemize}
\item Natural units are used, where $c = \hbar = 1$.
\item We will work in the mostly minus metric convention, i.e. $\eta=\mathrm{diag}(+,-,-,-)$.
\item Index contractions will occasionally be abbreviated as $x^\mu x_\mu = x \cdot x$.
\item Clifford algebra: $\{\gamma^\mu,\gamma^\nu\}=2\eta^{\mu\nu}$

\item Fourier-Transformation convention (for in-going particles):
\begin{align*}
    F\left(x\right)=\int \frac{d^{D} p}{(2 \pi)^{D}} \tilde{F}(p) e^{-i x\cdot p} &&   \tilde{F}\left(p\right)=\int d^{D} x F(x) e^{i x\cdot p}
    \label{fourier-trafo convention}
\end{align*}
Note that for out-going particles in an scattering event the signs in the exponents have to be reversed.

\item At some points throughout the thesis the short-hand notation \begin{equation*}
    \begin{aligned}
      \int_\omega :=\int\frac{\mathrm{d}\omega}{2\pi},&\hspace{1.5cm}& \int_k :=\int\frac{\mathrm{d}^4k}{(2\pi)^4}
    \end{aligned}
\end{equation*} will be used.

\item Factors of $2\pi$ will sometimes be absorbed into delta functions:
\begin{equation*}
    \dd (k) :=(2\pi)^4\delta^{(4)} (k), \hspace{.5cm} \dd (\omega) :=(2\pi)\delta (\omega) .
\end{equation*}

\item The Christoffel symbol is given by:
\begin{equation*}
    \Gamma^\lambda_{\mu\nu} = \frac{1}{2}g^{\lambda\sigma}(\partial_\mu g_{\nu\sigma} + \partial_\nu g_{\sigma\mu}-\partial_\sigma g_{\mu\nu} ) .
\end{equation*}
\item The Riemann tensor is given by:
\begin{equation*}
{{R_{\mu \nu}}^{\rho}}_{\sigma}=\partial_{\mu} {\Gamma^{\rho}}_{\nu \sigma}-\partial_{\nu} {\Gamma^{\rho}}_{\mu \sigma}+{\Gamma^{\rho}}_{\mu \lambda} {\Gamma^{\lambda}}_{\nu \sigma}-{\Gamma^{\rho}}_{\nu \lambda} {\Gamma^{\lambda}}_{\mu \sigma} .
\label{riemann tensor}
\end{equation*}
\item The Ricci tensor is given by:
\begin{equation*}
   R_{\mu\nu}= {{R_{\rho\mu}}^{\rho}}_{\nu}.
   \label{ricci tensor}
\end{equation*}
\item The Ricci scalar is given by:
\begin{equation*}
   R=g^{\mu\nu}  R_{\mu\nu} .
   \label{ricci scalar}
\end{equation*}
\end{itemize}
\clearpage
\newpage
\pagestyle{plain} 
\unchapter{Introduction}
Since the first gravitational wave measurement in 2015 on September 14 by the LIGO gravitational wave detectors in Livingston, Louisiana, and Hanford, Washington and the increasing number of other detections that followed this event \cite{Gravitational_Wave_Detection1, Gravitational_Wave_Detection2,Gravitational_Wave_Detection3,Gravitational_Wave_Detection4,Gravitational_Wave_Detection5}, there has been a growing need for precise calculations of waveforms and other observables related to the measurement of gravitational waves. The signal of the first detected gravitational wave GW150914 originated from a black hole merger occurring about 1.3 billion light-years away \cite{Gravitational_Wave_Detection1}. The distinctive gravitational wave signature created by this or other events allows us to draw conclusions about the physics happening at the source and can be used to confirm existing theories or may open up hints on new physics in the future. Therefore, theoretical predictions of observables deducible from gravitational wave measurements become more and more important as the measurements become more accurate. 

Until now, all detected gravitational waves come from the two-body merger of two black holes, two neutron stars, or a black hole and a neutron star, but conceivable sources for future high-accuracy detectors could also be supernovae, pulsars in our galaxy, or even the stochastical background from the big bang. The merger of two massive objects is typically categorized into three phases: The early inspiral, the actual merger, and a ringdown phase.
In the early phase of a binary inspiral, the velocity of the coalescing objects is usually relatively small, and the gravitational field is relatively weak since the objects are still relatively far away from each other. Here, it makes sense to use the Post-Newtonian (PN) expansion or the Post-Minkowskian (PM) expansion. In the PN expansion, one expands in terms of the velocity $v/c$ and the gravitational coupling simultaneously, whereas in the PM expansion, one only expands in the gravitational coupling. For the later phases, this expansion breaks down since the velocities, as well as the strength of the gravitational field, increase. Hence, the computation of gravitational wave signals emitted during the merger or the ringdown call for other methods typically given by numerical relativity or descriptions in terms of quasi normal modes, respectively.
A very effective theory that aims to describe all different phases of the two-body dynamics is the effective one-body formalism (EOB) \cite{Buonanno_1999, Effective_One_Body_2,DAMOUR_2008}.
The EOB uses the best information available in the PN theory but resums the PN expansion in a suitable way. It maps the real description to an effective one-body description in a deformed black hole spacetime. Therefore it gives much better results than the PN expansion even when the binary approaches the merger, and the velocities are increasing.

Apart from bound problems like the binary inspiral, it is also interesting to consider scattering processes of massive astrophysical objects such as black holes or neutron stars. Although scattering processes have not been the source of any detected gravitational wave up to now, it has been pointed out that bound orbits like in a binary inspiral can be related to scattering observables \cite{K_lin_2020, K_lin_2020II, Cho_2022, Saketh_2021}. In addition to that, events during a gravitational scattering process may represent interesting targets for future gravitational wave searches \cite{Jakobsen_2021_gravitational_bremstrahlung}.
Gravitational scattering processes are naturally described in a PM weak field expansion, where the magnitude of the velocities is not limited, but one can still expand in the velocity later on, e.g. when relating it to a bound system.

In general, scattering processes are very naturally described in quantum field theory (QFT). Since the tools and methods available in QFT have been proven to be very efficient and well-developed, it is promising to tackle the scattering of classical objects like black holes or neutron stars with QFT methods. In fact, there have been many approaches to computing classical scattering from QFT amplitudes in the recent years.
For black holes, the motivation for this is very evident: Because black holes have an event horizon, it is impossible to access any internal information from inside the black hole. According to the \textit{no-hair theorem}, the properties of a black hole are entirely described by its mass, charge, and spin \cite{carroll_spacetime_2016}. When additionally looking from very far away, also their spatial size does not matter. This reminds very strongly of elementary particle physics, where we consider point particles described by their mass, charge, and spin. Consequently, treating black holes as point particles gives a good approximation as long as they do not get too close to each other and as long as we are only looking from far away. 

The usual approach of the QFT-based methods is to calculate amplitudes via ordinary QFT-computations and then take the classical limit ($\hbar\rightarrow 0$) in the end. A formalism for computing classically measurable quantities directly from on-shell quantum scattering amplitudes is, for example, given in \cite{Kosower_2019} and for spinning particles in \cite{Maybee_2019}. The disadvantage of these approaches is that the classical limit is very subtle. Moreover, the computation overshoots the needed result since it keeps all quantum contributions until taking the classical limit in the end.

In the effective field theories (EFT) approach \cite{Goldberger_2006} (or for reviews \cite{Les_Houches_Lectures_on_Effective_Field_Theories, Porto_2016}), one integrates out the graviton in the path integral but keeps the wordline trajectory $x^\mu(\tau)$ of black holes as classical background sources. In the recently established worldline quantum field theory (WQFT) \cite{Jan_2021} one goes a step further and quantizes both the graviton field $h_{\mu\nu}$ and the fluctuations about the bodies worldline trajectories $z^\mu(\tau)$. Observables are then calculated via expectation values of operators in a path integral like formalism from a worldline action. 
The advantage of the WQFT is that the $\hbar$-counting is very apparent since it provides a direct expansion in $\hbar$ through the order of internal loops in WQFT diagrams. Accordingly, in the WQFT, the leading order of the $\hbar$-expansion is simply given by tree-level diagrams. In contrast to ordinary QFT, quantum contributions can be neglected right from the start. Hence, the WQFT gives an efficient tool for calculating classical scattering but still gives the possibility to build on well-developed QFT methods. 
By now, the original WQFT approach for non-spinning objects coupled to a weak gravitational field has been extended to spinning bodies up to quadratic order in spin and has proven to be efficient for explicit computations \cite{jakobsen_2021gravitational_bremstrahlung_spinning,Jakobsen_2021_Susy,Jakobsen_22_Conservative_and_Radiative, Shi_2021_Classical_double_copy}.

In \cite{Jan_2021}, where the WQFT was first introduced, the authors presented an interesting link between QFT amplitudes and the WQFT for scalars in a gravitational background that bridges the gap between the QFT amplitudes-based methods and the WQFT. This work aims to find a similar link between the WQFT and QFT amplitudes for spin-1/2 particles. Although it might seem counterintuitive to consider spin-1/2 since quantized spin does not incorporate classical rotations, it has been shown that the WQFT description of a spin-1/2 particle corresponds to the linear order in spin of a classical rotation. In the same manner, spin-1 corresponds to the quadratic order in spin, etc. \cite{Jakobsen_2021_Susy}.


To start with, we will study a much simpler case of a scalar in an abelian background and relate it to a WQFT describing a non-spinning particle coupled to an abelian background in chapter one. To begin with, the WQFT for non-spinning particles coupled to an electromagnetic field will be introduced. As we will show, S-matrices in QFT can be expressed via dressed propagators in the classical limit, which we will then use to relate scalar QED to the WQFT for non-spinning particles coupled to an electromagnetic field. In order to do that, we will derive a worldline path integral representation of the dressed scalar propagator, which is akin to the WQFT expression and provides a formal link of scalar QED dressed propagators to its WQFT analogue. This formal link will then be used to finally relate S-matrices in scalar QED to the non-spinning WQFT.
In chapter two, this approach will be extended to spin-1/2 particles, generally repeating the same strategy. In the last chapter, we will finally couple the spin-1/2 particle to gravity.

\chapter{Massive point particle coupled to an electromagnetic field}
The first chapter is dedicated to gaining a general understanding and establishing the tools and methods used in this thesis in the most simple instance. The easiest to study for our purpose is a massive point particle coupled to an electromagnetic field. In the next chapters, we will then expand the approach by adding spin and moving on to gravity. This way, we can concentrate on the features exclusively related to the introduction of spin or curved space, respectively, later on in the thesis and now focus on the features already present in the simple case.

\section{Worldline action and the WQFT}
To start with, let us look at the appropriate action describing a massive point particle coupled to an electromagnetic field:
\begin{equation}
    S=S_{\mathrm{pm}} + S_{\mathrm{A}} + S_{\mathrm{int}} + S_\mathrm{gf}.
\end{equation}
The worldline action of a classical relativistic point-mass particle is given by:
\begin{equation}\label{Spm}
    S_{\mathrm{pm}}=-m \int \mathrm{d}s=-m\int\mathrm{d}\tau\sqrt{\dot{x}^\mu\dot{x}_\mu},
\end{equation}
where the dot denotes a derivative with respect to $\tau$.
Introducing an additional field, the einbein $e(\tau)$, it can be brought to the equivalent Polyakov form:
\begin{equation}
  S_{\mathrm{pm}}=-\frac{m}{2}\int_{-\infty}^\infty  \mathrm{d}\tau \left(e^{-1}(\tau)\dot{x}^\mu\dot{x}_\mu+e(\tau)\right).
\end{equation}
Since the equation of motion for the einbein is $e(\tau)=\sqrt{\dot{x}^\mu\dot{x}_\mu}$, one recovers \eqref{Spm} if it is inserted into the action again. Gauge fixing $e(\tau)=1$, the Polyakov action yields
\begin{equation}
  S_{\mathrm{pm}}=-\frac{m}{2}\int_{-\infty}^\infty  \mathrm{d}\tau \left(\dot{x}^\mu\dot{x}_\mu+1\right).
\end{equation}
In the following, the plus one term will be omitted since it does not play a role in any case. Coupling the point-mass particle to an electromagnetic field, the interaction term is given by:
\begin{equation}
    S_{\mathrm{int}}=-e\int \mathrm{d}x^\mu A_\mu(x)=-e\int\mathrm{d}\tau\dot{x}^\mu A_\mu(x).
\end{equation}
The kinetic term for the electromagnetic field is the usual Maxwell action:
\begin{equation}
    S_\mathrm{A}=-\frac{1}{4}\int\mathrm{d}^D x F_{\mu\nu}F^{\mu\nu},
\end{equation}
where $F_{\mu\nu} = \partial_\mu A_\nu -\partial_\nu A_\mu$. In addition to that, $A_\mu$ can be gauge-fixed by adding an appropriate gauge fixing term $S_\mathrm{gf}$.

\noindent In the recently established worldline quantum field theory \cite{Jan_2021} we expand $x^\mu$ around a straight line 
\begin{equation}\label{eq:straight_line_expansion}
    x(\tau )=b^\mu+v^\mu \tau + z^\mu (\tau)
    \end{equation}
and calculate observables as path integral like expectation values:
\begin{equation}
\begin{array}{l}
\displaystyle
\left\langle\mathcal{O}\left(A,\left\{x_{i}\right\}\right)\right\rangle_{\text {WQFT }}:=\mathcal{Z}_{\text {WQFT }}^{-1} \int D\left[A_{\mu}\right] \int \prod_{i=1}^{n} D\left[z_{i}\right] \mathcal{O}\left(A,\left\{x_{i}\right\}\right) e^{iS},
\end{array}
\end{equation}
where $n$ is the number of the involved particles. 
The partition function $\mathcal{Z}_{\mathrm{WQFT}}$ is given by
\begin{equation}\label{eq:WQFT_partition_function}
\begin{array}{r}\displaystyle
\mathcal{Z}_{\mathrm{WQFT}}:=\text { const } \times \int D\left[A_{\mu}\right] \int \prod_{i=1}^{n} D\left[z_{i}\right] e^{iS}
\end{array}
\end{equation}
where the overall constant ensures that $\mathcal{Z}_{\mathrm{WQFT}}=1$ in the non-interacting case ($e=0$).

\section{WQFT Feynman rules}\label{sec:WQFT_feynman_rules}

For the evaluation of general expectation values in the WQFT, it is useful to derive Feynman rules, which can be read off the action of our WQFT theory:
\begin{equation}
    S=-m\int\mathrm{d}\tau\ \left(\frac{1}{2}\dot{x}^2+e \ \dot{x}\cdot A\right) + S_\mathrm{A}+ S_\mathrm{gf} .
\end{equation}
Redefining the proper time $\tau = 2m\tilde{\tau}$ will make the comparison with quantum field theory easier later on and will give a uniform mass counting independent of the loop order in the WQFT diagrams. Note that this redefinition also causes a redefinition of the velocity in \eqref{eq:straight_line_expansion} $v^\mu\rightarrow v^\mu/(2m)$. This has to be kept in mind in order to get the right dimensions and have the correct physical interpretation of the background parameter $v^\mu$. With this redefinition, the worldline action yields
\begin{equation}\label{eq:worldline_action_redifined_propertime}
    S=-\int\mathrm{d}\tilde{\tau}\ \left(\frac{1}{4}\dot{x}^2+e \ \dot{x}\cdot A\right) + S_\mathrm{A}+ S_\mathrm{gf},
\end{equation}
where the dots now denote derivatives with respect to the redefined ``proper time" parameter (actually $\tilde{\tau}$ does not have the dimension of a proper time anymore).
From now on, the tilde will be omitted, and we will proceed with the redefined parameter $\tau$.

Expanding $x^\mu$ as a deviation of a straight line trajectory $x^\mu=b^\mu+v^\mu \tau+z^\mu(\tau)$, we obtain:
\begin{equation}
    S=-\int \mathrm{d}\tau \left(\frac{v^2}{4}+\frac{v\cdot \dot{z}(\tau)}{2}+\frac{\dot{z}(\tau)^2}{4}+e\left(v+\dot{z}(\tau)\right)\cdot A\right)+ \dots
\end{equation}
For now, we will ignore the first term (a constant) and the second term (a boundary term). Although the boundary term does not vanish, we will neglect it nevertheless. A more careful treatment of the boundary term can be done with the so-called in-in formalism \cite{Jakobsen_2022_All_things_retarded}.
Here, two copies of the action are taken, and the boundary term cancels as it is the same in both copies. Alternatively, we can argue that the term is a constant since we keep the boundary fixed in the path integral or when using the principle of least action, and therefore it would not contribute to the equations of motion. As long as we are only interested in classical observables, the equations of motion should still be our guidance. 

Since Feynman rules are most naturally described in momentum space and energy space, we Fourier-transform the fields $z^\mu(\tau)$ and $A^\mu(x)$:
\begin{equation}\label{eq:fourierfields}
    \begin{aligned}
      z^\mu (\tau)=\int_\omega e^{i\omega \tau}z^\mu(-\omega), &\hspace{1cm}& \hspace{1cm}& A^\mu(x)=\int_k A^\mu(-k) e^{ik\cdot x},
      \end{aligned}
\end{equation}
where the shorthand notation
\begin{equation}
    \begin{aligned}
      \int_\omega :=\int\frac{\mathrm{d}\omega}{2\pi},&\hspace{1.5cm}& \int_k :=\int\frac{\mathrm{d}^4k}{(2\pi)^4}
    \end{aligned}
\end{equation}
was introduced. The minus signs in the arguments of the Fourier-fields are introduced to have the worldline field $z^\mu$ and the electromagnetic field $A^\mu$ as out-going.

The kinetic term in the worldline action gives us the worldline propagator:
\begin{align}\label{eq:propagatorz}
    \begin{tikzpicture}[baseline={(current bounding box.center)}]
    \coordinate (in) at (-2,0);
    \coordinate (out) at (2,0);
    \coordinate (x) at (1,0);
    \coordinate (y) at (-1,0);
    \draw [dotted] (in) -- (x);
    \draw [dotted] (y) -- (out);
    \draw [zUndirected] (x) -- (y) node [midway, below] {$\omega$};
    \draw [fill] (x) circle (.08) node [above]{$z^\nu(\tau_2)$};
    \draw [fill] (y) circle (.08) node [above]{$z^\mu(\tau_1)$};
    \end{tikzpicture}=-2i\eta^{\mu\nu}\int_\omega\frac{e^{-i\omega(\tau_1-\tau_2)}}{(\omega\pm i\epsilon)^2} = i\eta^{\mu\nu} \left( | \tau_1 - \tau_2 | \pm ( \tau_1 - \tau_2 )\right).
\end{align}
The dotted line indicates that $z$ is propagating on the worldline. This notation is meant to distinguish the fields that live only on the worldline from fields defined everywhere in spacetime, like the electromagnetic field $A^\mu$.

Depending on the sign of the $i\epsilon$ prescription the propagator given above is the retarded or advanced version of the propagator, which is nonzero if $\tau_1 > \tau_2$ or  $\tau_1 < \tau_2$ respectively. As explained in \cite{Jan_2021}, the choice of $i\epsilon$ prescription determines the interpretation of the background parameters $b$ and $v$. For the retarded prescription $b^\mu$ and $v^\mu$ describe the initial worldline at $\tau \rightarrow - \infty$ and if advanced propagators are used $b^\mu$ and $v^\mu$ depict the final worldline at $\tau \rightarrow + \infty$. Hence, the preferred choice of the $i\epsilon$ prescription depends on the physical problem, e.g. for radiating systems the retarded one would be the one of choice \cite{Jan_2021}.
For scattering processes, one usually uses time symmetric propagators as the Feynman propagator in quantum field theory. In the WQFT the time-symmetric propagator is given by the average of the retarded and the advanced propagator. In this case, the background parameters $b^\mu$ and $v^\mu$ describe an in-between state at $\tau=0$. To leading order in the coupling, they are given by the average of the retarded and advanced background parameters. As we will explicitly confirm later, the time-symmetric propagator directly corresponds to the Feynman propagator in quantum field theory.

The photon propagator is just the usual one known from quantum electrodynamics and can be found in any textbook on quantum field theory (e.g. \cite{Peskin:1995ev} or \cite{SchwartzMatthewD2014Qfta}). Since it will not be used explicitly in this thesis, it will not be listed here.

To read off the vertices, we can look at the interaction part of the action:
\begin{equation}\label{eq:Sint}
    S_\text{int}=-\int \mathrm{d}\tau e\left(v+\dot{z}(\tau)\right)\cdot A(x).
\end{equation}
When we look at these $A^\mu$ dependent terms, we have to take into account its $x$ dependence. If we apply the straight line expansion as before and move on to energy space, the $A^\mu$-field inherits a $\tau$ dependence:
\begin{equation}\label{eq:Atau}
  A^\mu(x(\tau))=\int_k A^\mu(-k) e^{ik\cdot x}=  \sum_{n=0}^\infty\frac{i^n}{n!}\int_{k,\omega_1,\dots,\omega_n}\hspace{-2mm}e^{ik\cdot b}e^{i\left(k\cdot v+\sum_{i=1}^n\omega_i\right)\tau}\left(\prod_{i=1}^{n}k\cdot z(-\omega_i)\right) A^\mu(-k).
\end{equation}
Using \eqref{eq:Atau} and \eqref{eq:fourierfields} the interaction part of the action \eqref{eq:Sint} can be written as:
\begin{equation}\label{eq:Sint_A_expanded}
\begin{aligned}
    S_\text{int}=&-e\sum_{n=0}^\infty\frac{i^n}{n!}\int_{k,\omega_1,\dots,\omega_n }\dd(k\cdot v+\textstyle\sum_{i=1}^{n}\omega_i)\displaystyle e^{ik\cdot b}\left(\prod_{i=1}^{n} z^{\rho_i}(-\omega_i)\right) \\
    &\times A_\mu(k) \left(\left(\prod_{i=1}^{n}k_{\rho_i}\right)v^\mu+\sum_{i=1}^n\omega_i\left(\prod_{j\neq i}^nk_{\rho_j}\right)\delta^\mu_{\rho_i}\right),
\end{aligned}
\end{equation}
where factors of $2\pi$ were absorbed into the $\delta$-functions:
\begin{equation}
    \dd (\omega) := (2\pi)\delta (\omega).
\end{equation}
From the interaction part of the action in \eqref{eq:Sint_A_expanded} we can read off the vertices of our WQFT theory. There is only one vertex of zeroth order in $z^\mu$:
\begin{align}\label{eq:vertexA}
	  \begin{tikzpicture}[baseline={(current bounding box.center)}]
	  \coordinate (in) at (-1,0);
	  \coordinate (out) at (1,0);
	  \coordinate (x) at (0,0);
	  \node (k) at (0,-1.3) {$A_{\mu}(k)$};
	  \draw [dotted] (in) -- (x);
	  \draw [dotted] (x) -- (out);
	  \draw [photon] (x) -- (k);
	  \draw [fill] (x) circle (.08);
	  \end{tikzpicture}=-ie \ e^{ik\cdot b}\dd(k\cdot v)v^\mu
\end{align}
and one vertex of linear order in $z^\mu$:
\begin{align}\label{eq:vertexAZ}
\begin{aligned}
  \begin{tikzpicture}[baseline={(current bounding box.center)}]
  \coordinate (in) at (-1,0);
  \coordinate (out) at (1,0);
  \coordinate (x) at (0,0);
  \node (k) at (0,-1.3) {$A_{\mu}(k)$};
  \draw (out) node [right] {$z^\rho(\omega)$};
  \draw [dotted] (in) -- (x);
  \draw [zUndirected] (x) -- (out) node [midway, above] {$\underrightarrow{\omega}$};
  \draw [photon] (x) -- (k);
  \draw [fill] (x) circle (.08);
  \end{tikzpicture}=&e\ e^{ik\cdot b}\dd(k\cdot v+\omega)\left(k_\rho v^\mu+\delta^{\mu}_\rho\omega\right),
\end{aligned}
\end{align}
where as a matter of course, the sign of $\omega$ has to be reversed for diagrams with the inverse flux of the parameter $\omega$.
All other vertices are of higher order in $z^\mu$ and will not contribute if we only look at tree-level diagrams up to a two photon background. If we are only interested in the classical limit ($\hbar \rightarrow 0$) this is sufficient since the propagator scales with $\hbar$ and open $z$-legs need to be connected to other open worldline legs by propagators. Nevertheless, we can also give an expression for a general vertex to arbitrary order in $z$:
\begin{align}\label{eq:vertexAZ_allOrder}
\begin{aligned}
  \begin{tikzpicture}[baseline={(current bounding box.center)}]
  \coordinate (in) at (-1,0);
  \coordinate (out1) at (1,0);
  \coordinate (out2) at (1,0.5);
  \coordinate (outn) at (1,1.3);
  \coordinate (x) at (0,0);
  \node (k) at (0,-1.3) {$A_{\mu}(k)$};
  \node (dots) at (0.7,1) {$\vdots$};
  \draw (out1) node [right] {$z^{\rho_1}(\omega_{\rho_1})$};
  \draw (out2) node [right] {$z^{\rho_2}(\omega_{\rho_2})$};
  \draw (outn) node [right] {$z^{\rho_n}(\omega_{\rho_n})$};
  \draw [dotted] (in) -- (x);
  \draw [zUndirected] (x) -- (out1);
  \draw [zUndirected] (x) to [out=45, in=180] (out2);
  \draw [zUndirected] (x) to [out=90, in=180] (outn);
  \draw [photon] (x) -- (k);
  \draw [fill] (x) circle (.08);
  \end{tikzpicture}=&e\ i^{n-1} e^{ik\cdot b} \ \dd\left(k\cdot v+\textstyle\sum^n_{i=1}\omega_i\right) \left(\prod^n_{i=1} k_{\rho_i} v^\mu + \sum_{i=1}^n\omega_i \left(\prod_{i\neq j}^n k_{\rho_j}\right)\delta^\mu_{\ \rho_i}\right) .
\end{aligned}
\end{align}
Here again, all open lines are taken as out-going, but the arrows that indicate the flux of the "energies" $\omega_i$ are omitted to keep the notation clean and readable. Moreover, the following convention is used for products:
\begin{equation}
    \prod_{i=1}^l a_i = \begin{cases}
    0 \text{ if } l <0 \\
    1 \text{ if } l=0 \\
     \prod_{i=1}^l a_i \text{ otherwise} .
    \end{cases}
\end{equation}
With this set of propagators and vertices, one can now start to build Feynman diagrams in our WQFT as we will explicitly do for the dressed propagator in section \ref{Dressed propagator in Scalar QED}. As in any quantum field theory, one has to integrate over the $\omega_i$ parameter of any internal $z$-line and the four momenta $k$ of any internal photon line.
In contrast to ordinary quantum field theory, where the classical limit is very difficile, the WQFT provides a direct expansion in $\hbar$.

\section{S-matrices in quantum field theory and the classical limit}\label{sec:S-matrices}
Before proceeding with explicit calculations in the WQFT we should take a step back and reflect on how we can relate the WQFT to ordinary quantum field theory. In quantum field theory, we usually compute S-matrices, which are obtained from time-ordered correlators via the LSZ reduction and give the probability of scattering processes between asymptotically free states. The QFT avatar of a massive point particle with no internal degrees of freedom (i.e. particle without spin) is a scalar field $\phi$. Coupled to an electromagnetic field $A^\mu(x)$, it is scalar quantum electrodynamics (scalar QED).
The action of scalar QED for $n$ scalar particles is given by
\begin{equation}\label{eq:scalar_QED_action}
    S_\mathrm{ScalarQED}=\sum_{i}^n\underbrace{ -\int\mathrm{d}^Dx \ \phi_i^*( D^2 +m^2) \phi_i}_{\mathrm{S_i}} \underbrace{-\int\mathrm{d}^Dx\frac{1}{4}F_{\mu\nu}F^{\mu\nu}}_{\mathrm{S_\mathrm{A}}}+S_{\mathrm{gf}},
\end{equation}
where $D_\mu=\partial_\mu + ieA_\mu$.
Considering a scattering process of two scalars with or without a final state photon (therefore written in parentheses), we begin with the time-ordered correlator
\begin{equation}
\begin{aligned}\label{full_5pt_correlator}
\langle\Omega| T &\ \{\left(A_{\mu}(x)\right) \phi_{1}\left(x_{1}\right) \phi^\dagger_{1}\left(x_{1}^{\prime}\right) \phi_{2}\left(x_{2}\right) \phi^\dagger_{2}\left(x_{2}^{\prime}\right)\}|\Omega\rangle \\
&=\tilde{\mathcal{Z}}^{-1} \int \mathcal{D}\left[A_{\mu},\phi_{1}^\dagger, \phi_{1}, \phi_{2}^\dagger, \phi_{2}\right] \left(A_{\mu}(x)\right) \phi_{1}\left(x_{1}\right) \phi^\dagger_{1}\left(x_{1}^{\prime}\right) \phi_{2}\left(x_{2}\right) \phi^\dagger_{2}\left(x_{2}^{\prime}\right) e^{i S_\mathrm{ScalarQED}} .
\end{aligned}
\end{equation}
The expression above is the full quantum correlator, but if we are only interested in the classical limit ($\hbar\rightarrow 0$), many diagrams will not contribute. In order to see which diagrams contribute and which do not, one could expand in a power series of $\hbar$. But this expansion is very subtle in QFT, 
because it depends on the choice of $\hbar$ independent quantities. For our purpose, a detailed discussion of the classical limit is not necessary, but it can be found, for instance, in \cite{Kosower_2019}. Even without a thorough analysis of the classical limit some contributions are clearly quantum and can therefore be neglected, e.g. pure scalar loops mediated between photon lines. If we ignore diagrams that include such loops, we can integrate out the scalars treating the gauge field as a constant background and express the correlator \eqref{full_5pt_correlator} via dressed scalar propagators (two-point functions of two scalars in a fixed Abelian background). The dressed scalar propagators are defined as:
\begin{equation}\label{dressed_scalar_propagator_definition}
G_{A,i}\left(x, x^{\prime}\right)=\langle A|T\{\phi_i(x)\phi^\dagger_i(x^\prime)\}|A\rangle=\mathcal{Z}_{i}^{-1} \int \mathcal{D}\left[\phi^\dagger_i,\phi_{i}\right] \phi_{i}(x) \phi^\dagger_{i}\left(x^{\prime}\right) e^{i S_{i}},
\end{equation}
where we write $\langle A| \dots | A\rangle$ instead of $\langle \Omega | \dots | \Omega\rangle$ to indicate that the matrix elements are taken in the presence of an external field rather than the vacuum.
As argued above, we can now use the dressed propagators to rewrite \eqref{full_5pt_correlator} in the classical limit:
\begin{equation}
\begin{aligned} \label{5pt_correlator_via_dressed_propagator}
\langle\Omega| T &\ \{\left(A_{\mu}(x)\right) \phi_{1}\left(x_{1}\right) \phi^\dagger_{1}\left(x_{1}^{\prime}\right) \phi_{2}\left(x_{2}\right) \phi^\dagger_{2}\left(x_{2}^{\prime}\right)\}|\Omega\rangle \\
&=\mathcal{Z}^{-1} \int \mathcal{D}\left[A_{\mu}\right] \left(A_{\mu}(x)\right) G_{A,1}\left(x_{1}, x_{1}^{\prime}\right) G_{A,2}\left(x_{2}, x_{2}^{\prime}\right) e^{i\left(S_{\mathrm{A}}+S_{\mathrm{gf}}\right)}.
\end{aligned}
\end{equation}
Note that the expression above can still contain some quantum contributions. The statement here is only that it equals \eqref{full_5pt_correlator} in the classical limit. 

From \eqref{5pt_correlator_via_dressed_propagator} we can move on to the S-matrix via LSZ reduction, i.e. set the external legs on-shell and replace the external propagators by polarization vectors and Fourier transform to momentum space:
\begin{equation}
\begin{array}{l}\label{Smatrix_via_dressed_propagator}
\displaystyle
\left\langle\phi_{1} \phi_{2}(A)|S| \phi_{1} \phi_{2}\right\rangle=\mathcal{Z}^{-1} \int \mathrm{d}^{D}\left[x_{i}, x_{i}^{\prime}, x\right] e^{i p_{i} \cdot x_{i}-i p_{i}^{\prime} \cdot x^{\prime}_{i} (-i k \cdot x)} \\ \displaystyle
\qquad\int \mathcal{D}\left[A_{\mu}\right]\left(\varepsilon^{\mu}(k) A_{\mu}(x)\right) \widehat{G}_{A,1}\left(x_{1}, x_{1}^{\prime}\right)\widehat{G}_{A,2}\left(x_{2}, x_{2}^{\prime}\right) e^{i\left(S_{\mathrm{A}}+S_{\mathrm{gf}}\right)}.
\end{array}
\end{equation}
The $\widehat{G}_{A,i}\left(x_{1}, x_{1}^{\prime}\right)$ are the amputated on-shell dressed propagators (with the external scalar legs on-shell, but the photon legs are still off-shell).

With equation \eqref{Smatrix_via_dressed_propagator} we have argued that S-matrices can be obtained from dressed propagators in the classical limit. If we want to find out how the WQFT is related to scattering amplitudes in quantum field theory, it will be valuable to first of all establish an understanding of the relation between dressed propagators in quantum field theory and the analogous expressions in the WQFT. 
In the next sections, we will therefore derive a worldline representation of the dressed QFT propagator and then compare it to the analogous expression in the WQFT.



\section{Dressed propagators}
\begin{figure}[b]
 \centering
\begin{equation*}
    G_A(x,x^\prime) \ = \ \begin{tikzpicture}[baseline]
       \begin{feynman}
       \coordinate (in) at (-1.1,0);
       \coordinate (out) at (1.1,0);
       \diagram{(in) -- [fermion] (out)};
       \end{feynman}
    \end{tikzpicture} \ + \
    \begin{tikzpicture}[baseline]
       \begin{feynman}
       \coordinate (in) at (-1.1,0);
       \coordinate (out) at (1.1,0);
       \coordinate (x) at (0,0);
       \coordinate (k) at (0, 1.2);
       \diagram{(in) -- [fermion] (x) -- [fermion] (out)};
       \draw [photon] (x) -- (k);
       \end{feynman}
    \end{tikzpicture} \ + \ 
    \begin{tikzpicture}[baseline]
       \begin{feynman}
       \coordinate (in) at (-1.1,0);
       \coordinate (out) at (1.1,0);
       \coordinate (x) at (-.4,0);
       \coordinate (y) at (.4,0);
       \coordinate (k1) at (-.4cm, 1.2);
        \coordinate (k2) at (.4cm, 1.2);
       \diagram{(in) -- [fermion] (x) -- [fermion] (y) -- [fermion] (out)};
       \draw [photon] (x) -- (k1);
       \draw [photon] (y) -- (k2);
       \end{feynman}
    \end{tikzpicture} \ + \ \dots
\end{equation*}
    \caption{Dressed propagator schematically pictured in terms of Feynman diagrams}
    \label{fig:Dressed_Propagator_pictured}
\end{figure}
In order to finally find a relation between S-matrices in quantum field theory and expectation values in the WQFT later on, we will now begin with deriving expressions for the dressed propagator in scalar QED and its analogue in the WQFT. The dressed propagator is the propagator that arises if we integrate out the scalar field or the worldline field $x^\mu(\tau)$, respectively, and treat the gauge field $A^\mu$ as a constant background. Consequently, the dressed propagator is a function of the field that was not integrated out, in our case, the electromagnetic field $A^\mu$. The dressed propagator in terms of Feynman diagrams is graphically depicted in figure \ref{fig:Dressed_Propagator_pictured}. As a warm-up and to get a better understanding of the theories we seek to compare in the end, we will initially compare the first diagrams depicted in figure \ref{fig:Dressed_Propagator_pictured} in both theories and then move on to deriving general expressions that include an arbitrary number of photon legs.


\subsection{Dressed propagator in the WQFT}
In the classical limit, the WQFT analogue to the dressed propagator is built up from tree-level diagrams, which can easily be written down using the Feynman rules derived in section \ref{sec:WQFT_feynman_rules}. For higher orders in the $\hbar$ expansion, the procedure becomes a bit more involved since non-trivial worldline integrals have to be solved, which we will explicitly do towards the end of this section.

\subsubsection{Tree-level contributions}
The one photon contribution is trivial and just given by the zeroth order vertex \eqref{eq:vertexA}:
\begin{equation}\label{eq:1photon_WQFT_diagram}
    \begin{aligned}
      \begin{tikzpicture}[baseline={(current bounding box.center)}]
        \coordinate (in) at (-1,0);
        \coordinate (out) at (1,0);
        \coordinate (x) at (0,0);
        \draw [fill] (x) circle (.08);
        \node (k) at (0,-1.3) {$\mu$};
        \draw [dotted] (in) -- (out);
        \draw [photon] (x) -- (k) node [midway, right]{$k$};
      \end{tikzpicture} =-ie \ e^{ik\cdot b}\dd(k\cdot v)v_\mu.
    \end{aligned}
\end{equation}
For the two photon contribution, we need the vertex \eqref{eq:vertexAZ} and a worldline propagator. As we will see in a bit and later on in a more general manner in section \ref{sec:Comparison_to_WQFT_propagator}, the worldline propagator that corresponds to the Feynman propagator in QFT is the time-symmetric worldline propagator, i.e. the average of the retarded and the advanced propagator:
\begin{align}\label{eq:propagatorz_averaged}
    \begin{tikzpicture}[baseline={(current bounding box.center)}]
    \coordinate (in) at (-2,0);
    \coordinate (out) at (2,0);
    \coordinate (x) at (1,0);
    \coordinate (y) at (-1,0);
    \draw [dotted] (in) -- (x);
    \draw [dotted] (y) -- (out);
    \draw [zUndirected] (x) -- (y) node [midway, below] {$\omega$};
    \draw [fill] (x) circle (.08) node [above]{$\nu$};
    \draw [fill] (y) circle (.08) node [above]{$\mu$};
    \end{tikzpicture}=-i\eta^{\mu\nu}\left(\frac{1}{(\omega+i\epsilon)^2}+\frac{1}{(\omega-i\epsilon)^2}\right).
\end{align}
With this time symmetric propagator the two photon contribution is given by
\begin{figure}[H]\vspace{-1.5cm}
\begin{equation}\label{eq:two_photon_WQFT_diagram}  \def\arraystretch{2} 
    \begin{aligned}
      \begin{tikzpicture}
        \coordinate (in) at (-1.5,0);
        \coordinate (out) at (1.5,0);
        \coordinate (x) at (-0.6,0);
        \coordinate (y) at (0.6,0);
        \node (k1) at (-0.6,-1.3) {$\mu_1$};
        \node (k2) at (0.6,-1.3) {$\mu_2$};
        \draw [fill] (x) circle (.08);
        \draw [fill] (y) circle (.08);
        \draw [dotted] (in) -- (out);
        \draw [photon] (x) -- (k1) node [midway, right]{$k_1$};
        \draw [photon] (y) -- (k2) node [midway, right]{$k_2$};
        \draw [zUndirected] (x) -- (y) node [midway, above] {$\underrightarrow{\omega}$};
      \end{tikzpicture}  \begin{array}{ll}  & \\
      =&\displaystyle -i e^2 \ e^{i(k_1+k_2)\cdot b} \bigintsss_\omega \dd(k_1\cdot v + \omega) \dd(k_2\cdot v -\omega) \\
      &\displaystyle \hspace{.8cm}\times\left( k_{1\rho}v_{\mu_1}+\omega \eta_{\mu_1\rho}\right)\left(k_2^\rho v_{\mu_2} -\omega \delta_{\mu_2}^\rho\right)\left(\textstyle\frac{1}{(\omega+i\epsilon)^2}+\frac{1}{(\omega-i\epsilon)^2}\right)\\
      =&\displaystyle-i e^2 \ e^{i(k_1+k_2)\cdot b} \ \textstyle \dd\left( (k_1+k_2)\cdot v)\right) \left[\left( \frac{k_1 \cdot k_2}{(v\cdot k_2 + i\epsilon)^2} + \frac{k_1 \cdot k_2}{(v\cdot k_2 - i\epsilon)^2}\right) v_{\mu_1} v_{\mu_2} \right. \\
      &\displaystyle \left. + \left(\textstyle\frac{1}{(v\cdot k_1 +i\epsilon)}+\frac{1}{(v\cdot k_1 -i\epsilon)}\right)\displaystyle\left(k_{2\mu_1}v_{\mu_2} - k_{1\mu_2}v_{\mu_1}\right) \ - \ 2 \ \eta_{\mu_1\mu_2}\right] .
      \end{array} 
    \end{aligned}
\end{equation}\end{figure}
\noindent Since the vertex \eqref{eq:vertexAZ} contains a linear power of $\omega$, it reduces the second order propagator to a first order propagator or cancels entirely against it in some of the terms so that we end up with integrands of the order $\omega^0$, $1/\omega$ and $1/\omega^2$. For this cancellation, the $i\epsilon$ can be ignored because it gives contributions that are zero in the limit $\epsilon\rightarrow 0$. From the first to the second line in equation \eqref{eq:two_photon_WQFT_diagram} we evaluated the integral, which is trivial in this case since it just exploits one of the delta functions.

\subsubsection{n-loop worldline propagator}
If we include higher orders in the $\hbar$ expansion, i.e. include worldline loops, solving the integrals is not trivial anymore, because the number of integrals exceeds the number of available delta functions. Before evaluating any integrals, the two photon contribution to the dressed propagator with an arbitrary number of internal worldline propagators is given by
\begin{figure}[H]\vspace{-1cm}\begin{equation}\def\arraystretch{2.5}
\begin{aligned}\label{WQFT_4pt_diagramm1}
  \begin{tikzpicture}
  \coordinate (in) at (-2,0);
  \coordinate (out) at (2,0);
  \coordinate (x) at (-1,0);
  \coordinate (y) at (1,0);
  \coordinate (top) at (0,1.5);
  \node (second) at (0,0.5) {$\underrightarrow{\omega_2}$};
  \node (topw) at (0,1.8) {$\underrightarrow{\omega_n}$};
  \node (k1) at (-1,-1.5) {$\mu_1$};
  \node (k2) at (1,-1.5) {$\mu_2$};
  \node (dots) at (0,1.2) {$\vdots$};
  \draw (out) node [right] {};
  \draw [dotted] (in) -- (x);
  \draw [zUndirected] (x) -- (y) node [midway, below] {$\overrightarrow{\omega_1}$};
  \draw [zUndirected] (x) to [out=40, in=200] (second) to [out=-20, in=140] (y);
  \draw [zUndirected] (x) to [out=90, in=180] (top)  to [out=0, in=90]  (y);
  \draw [dotted] (y) -- (out);
  \draw [photon] (x) -- (k1) node [midway, right]{$k_1$};
  \draw [photon] (y) -- (k2) node [midway, right]{$k_2$};
  \draw [fill] (x) circle (.08);
  \draw [fill] (y) circle (.08);
  \end{tikzpicture}& \begin{array}{ll}& \\ =&\displaystyle\frac{1}{n!} e^2 (-1)^n i^{3n-2} \ e^{i(k_1+k_2)\cdot b} \ \dd((k_1+k_2)\cdot v) \\
  & \times\displaystyle\bigintss_{\omega_1, \dots, \omega_n} \hspace{-0.8cm}\dd\left(\textstyle \sum^n_{i=1}\omega_i - k_2\cdot v\right) \displaystyle\left[ \ \prod_{i=1}^n D^2(\omega_i) \ (k_1\cdot k_2)^n v_{\mu_1}v_{\mu_2} \right. \\
  & - \ n (n-1) \displaystyle\prod_{i=1}^{n-2} D^2(\omega_i) \ D^1(\omega_{n-1}) \ D^1(\omega_n) \ (k_1\cdot k_2)^{n-2} \ k_{1\mu_2}k_{2\mu_1}\\ 
  &\displaystyle + \ n \prod_{i=1}^{n-1} D^2(\omega_i) \  D^1(\omega_n) \ (k_1\cdot k_2)^{n-1} \left( k_{2\mu_1}v_{\mu_2}-k_{1\mu_2}v_{\mu_1}\right) \\
  &\displaystyle\left. - \ n \prod_{i=1}^{n-1} D^2(\omega_i) (k_1\cdot k_2)^{n-1} \ 2 \eta_{\mu_1\mu_2} \ \right] ,
  \end{array}
\end{aligned}
\end{equation}\end{figure}
\noindent where we defined
\begin{equation}
    D^n(\omega):=\left(\frac{1}{(\omega + i \epsilon )^n} + \frac{1}{(\omega - i \epsilon )^n}\right)
\end{equation}
to shorten the notation. Here again, the different powers in the propagators $D^n(\omega)$ arise from the cancelation of $\omega$'s in the vertices and the propagators. The integrals appearing in \eqref{WQFT_4pt_diagramm1} can be solved even for arbitrary numbers of internal lines as shown in appendix \ref{appendix:worldline_integrals}. Using the master formulas given in appendix \ref{appendix:worldline_integrals} the two photon dressed WQFT propagator with an arbitrary number of internal lines yields
\begingroup
\addtolength{\jot}{1em}
\begin{equation}\label{WQFT_4pt_diagram_result}
  \begin{aligned}
\eqref{WQFT_4pt_diagramm1} =& (-ie)^2 i \ e^{i(k_1+k_2)\cdot b} \ \dd((k_1+k_2)\cdot v) \left[\left(\frac{(-1)^{n-1}(k_1\cdot k_2)^n}{(k_2\cdot v -i\epsilon)^{n+1}} + \frac{(k_1\cdot k_2)^n}{(k_2 \cdot v + i\epsilon)^{n+1}}\right)v_{\mu_1}v_{\mu_2} \right. \\ 
&- \left(\frac{(-1)^{n-1}(k_1\cdot k_2)^{n-2}}{(k_2\cdot v -i\epsilon)^{n-1}} + \frac{(k_1\cdot k_2)^{n-2}}{(k_2 \cdot v + i\epsilon)^{n-1}}\right) k_{1\mu_2}k_{2\mu_1}\ \delta_{n\geq2} \\ 
&  + \left. \left(\frac{(-1)^{n-1}(k_1\cdot k_2)^{n-1}}{(k_2\cdot v -i\epsilon)^{n}} + \frac{(k_1\cdot k_2)^{n-1}}{(k_2 \cdot v + i\epsilon)^{n}}\right) \left(k_{2\mu_1}v_{\mu_2}- k_{1\mu_2}v_{\mu_1}\right) \ \delta_{n \geq 1} \ -  2 \eta_{\mu_1\mu_2} \ \delta_{1n}\right] .
  \end{aligned}
\end{equation}
\endgroup
where the $\delta_{n\geq2}$ and $\delta_{n \geq 1}$ indicate that the corresponding terms only contribute for $n\geq 2$ and $n\geq1$, respectively.
Note that the formula also contains the previously derived tree-level version \eqref{eq:two_photon_WQFT_diagram} as well as a disconnected diagram with no internal $z$ line. In equation \eqref{WQFT_4pt_diagram_result} every second order of the terms in round brackets are just disguised derivatives of delta functions due to
\begin{equation}
    \frac{\dd^{(n)}(\omega)}{(-1)^n n!} = \frac{i}{(\omega+ i0)^{n+1}}- \frac{i}{(w-i0)^{n+1}} .
\end{equation}
Hence, we either have an $n$-th order time-symmetric (averaged of retarded and advanced) propagator $D^n(k_2\cdot v)$, or derivatives of delta functions $ \frac{\dd^{(n)}(k_2\cdot v)}{(-1)^n n!} $. To have a uniform expression for all orders we stick to the notation in \eqref{WQFT_4pt_diagram_result}.

Now that we have derived a general expression, we can sum up all the diagrams described by
\eqref{WQFT_4pt_diagram_result}:
\begin{adjustwidth}{-3.5mm}{3.5mm}
\hspace{-1cm}\begin{equation}\label{WQFT_4pt_diagramm_sum}
\boxed{ 
\begin{aligned}\ \\ 
  \begin{tikzpicture}[baseline]
  \coordinate (in) at (-1,0);
  \coordinate (out) at (1,0);
  \coordinate (x) at (-0.4,0);
  \coordinate (y) at (0.4,0);
  \node (k1) at (-0.4,-1) {};
  \node (k2) at (0.4,-1) {};
  \draw [dotted] (in) -- (out);
  \draw [photon] (x) -- (k1);
  \draw [photon] (y) -- (k2);
  \draw [fill] (x) circle (.06);
  \draw [fill] (y) circle (.06);
  \end{tikzpicture} +  \begin{tikzpicture}[baseline]
  \coordinate (in) at (-1,0);
  \coordinate (out) at (1,0);
  \coordinate (x) at (-0.4,0);
  \coordinate (y) at (0.4,0);
  \node (k1) at (-0.4,-1) {};
  \node (k2) at (0.4,-1) {};
  \draw [dotted] (in) -- (out);
  \draw [photon] (x) -- (k1);
  \draw [photon] (y) -- (k2);
  \draw [fill] (x) circle (.06);
  \draw [fill] (y) circle (.06);
  \draw [zUndirected] (x) -- (y);
  \end{tikzpicture}  + 
  \begin{tikzpicture}[baseline]
  \coordinate (in) at (-1,0);
  \coordinate (out) at (1,0);
  \coordinate (x) at (-0.4,0);
  \coordinate (y) at (0.4,0);
  \node (k1) at (-0.4,-1) {};
  \node (k2) at (0.4,-1) {};
  \draw [dotted] (in) -- (out);
  \draw [photon] (x) -- (k1);
  \draw [photon] (y) -- (k2);
  \draw [fill] (x) circle (.06);
  \draw [fill] (y) circle (.06);
  \draw [zUndirected] (x) -- (y);
  \draw [zUndirected] (x) to [out=50, in=130] (y);
  \end{tikzpicture}  + 
  \begin{tikzpicture}[baseline]
  \coordinate (in) at (-1,0);
  \coordinate (out) at (1,0);
  \coordinate (x) at (-0.4,0);
  \coordinate (y) at (0.4,0);
  \coordinate (third) at (0,0.4);
  \node (k1) at (-0.4,-1) {};
  \node (k2) at (0.4,-1) {};
  \draw [dotted] (in) -- (out);
  \draw [photon] (x) -- (k1);
  \draw [photon] (y) -- (k2);
  \draw [fill] (x) circle (.06);
  \draw [fill] (y) circle (.06);
  \draw [zUndirected] (x) -- (y);
  \draw [zUndirected] (x) to [out=50, in=130] (y);
  \draw [zUndirected] (x) to [out=70, in=180] (third) to [out=0, in=110] (y);
  \end{tikzpicture}  +  
   \begin{tikzpicture}[baseline]
  \coordinate (in) at (-1,0);
  \coordinate (out) at (1,0);
  \coordinate (x) at (-0.4,0);
  \coordinate (y) at (0.4,0);
  \coordinate (third) at (0,0.4);
  \coordinate (fourth) at (0,0.6);
  \node (k1) at (-0.4,-1) {};
  \node (k2) at (0.4,-1) {};
  \draw [dotted] (in) -- (out);
  \draw [photon] (x) -- (k1);
  \draw [photon] (y) -- (k2);
  \draw [fill] (x) circle (.06);
  \draw [fill] (y) circle (.06);
  \draw [zUndirected] (x) -- (y);
  \draw [zUndirected] (x) to [out=50, in=130] (y);
  \draw [zUndirected] (x) to [out=70, in=180] (third) to [out=0, in=110] (y);
  \draw [zUndirected] (x) to [out=80, in=180] (fourth) to [out=0, in=100] (y);
  \end{tikzpicture}  + 
   \begin{tikzpicture}[baseline]
  \coordinate (in) at (-1,0);
  \coordinate (out) at (1,0);
  \coordinate (x) at (-0.4,0);
  \coordinate (y) at (0.4,0);
  \coordinate (third) at (0,0.4);
  \coordinate (fourth) at (0,0.6);
  \coordinate (fifth) at (0,0.8);
  \node (k1) at (-0.4,-1) {};
  \node (k2) at (0.4,-1) {};
  \draw [dotted] (in) -- (out);
  \draw [photon] (x) -- (k1);
  \draw [photon] (y) -- (k2);
  \draw [fill] (x) circle (.06);
  \draw [fill] (y) circle (.06);
  \draw [zUndirected] (x) -- (y);
  \draw [zUndirected] (x) to [out=50, in=130] (y);
  \draw [zUndirected] (x) to [out=70, in=180] (third) to [out=0, in=110] (y);
  \draw [zUndirected] (x) to [out=80, in=180] (fourth) to [out=0, in=100] (y);
  \draw [zUndirected] (x) to [out=90, in=180] (fifth) to [out=0, in=90] (y);
  \end{tikzpicture}  +  \dots \  \\ \ \\ \begin{aligned}
=& (-ie)^2 i \ e^{i(k_1+k_2)\cdot b} \ \dd((k_1+k_2)\cdot v) &&\\ & \ \\ &\left[\sum_{n=0}^\infty\left(\frac{(-1)^{n-1}(k_1\cdot k_2)^n}{(k_2\cdot v -i\epsilon)^{n+1}} + \frac{(k_1\cdot k_2)^n}{(k_2 \cdot v + i\epsilon)^{n+1}}\right)\left(v_{\mu_1}v_{\mu_2}-k_{1\mu_2}k_{2\mu_1}\right) \right. && \\ & \ \\
 & + \left. \sum_{n=0}^\infty\left(\frac{(-1)^{n-1}(k_1\cdot k_2)^n}{(k_2\cdot v -i\epsilon)^{n+1}} - \frac{(k_1\cdot k_2)^n}{(k_2 \cdot v + i\epsilon)^{n+1}}\right) \left( k_{1\mu_2}v_{\mu_1}-k_{2\mu_1}v_{\mu_2}\right) \ - 2 \eta_{\mu_1\mu_2}\right] , & \hspace{8mm}& \\ & \
     \end{aligned} \end{aligned}}
\end{equation}\end{adjustwidth}
where some of the sums were shifted to have all the sums starting at the same index. As the diagrammatic picture indicates, also the disconnected diagram is considered in the sum because we may want to connect the dressed propagator to a second worldline when using the argument presented in section \ref{sec:S-matrices}. 

\subsection{Dressed propagator in Scalar QED}\label{Dressed propagator in Scalar QED}
Now let us compare the one and two photon WQFT diagram to the corresponding expressions in scalar QED. The Feynman rules for scalar QED are listed in appendix \ref{apendix:FeynmanRules}. The amputated one photon contribution to the dressed propagator (the second diagram in figure \ref{fig:Dressed_Propagator_pictured}) is just given by the trivalent vertex:
\begin{equation}\label{eq:1photon_scalar_QED_dressed_propagator}
\begin{aligned}
    \begin{tikzpicture}[baseline]
    \begin{feynman}
    \coordinate (x) at (0,0);
    \coordinate (in) at (-1.2,0);
    \coordinate (out) at (1.2,0);
    \coordinate (k) at (0,1.2);
    \diagram{ (in) -- [charged scalar,  momentum={[arrow shorten=0.12mm] \(p\)}] (x) --[charged scalar,  momentum={[arrow shorten=0.12mm] \(p^\prime\)}] (out)};
    \draw [photon] (x) -- (k) node [above] {$\mu$};
    \draw [fill] (x) circle (.04);
    \end{feynman}
    \end{tikzpicture}  = -ie \ (p+p^\prime)_\mu.
\end{aligned}
\end{equation}
Apart from a factor of $e^{ik\cdot b}\dd(k\cdot v)$ this looks similar to the WQFT vertex \eqref{eq:1photon_WQFT_diagram} if we identify $v^\mu=(p+p^\prime)^\mu$. According to the Feynman rules given in appendix \ref{apendix:FeynmanRules}, the amputated scalar propagator dressed with two photons is given by
\begin{equation}\label{eq:2photon_scalar_QED_dressed_propagator}\def\arraystretch{2.5}
\begin{array}{l}
\begin{tikzpicture}[baseline]
   \begin{feynman}
   \coordinate (in) at (-1.8,0);
   \coordinate (x) at (-0.6,0);
   \coordinate (y) at (0.6,0);
   \coordinate (out) at (1.8,0);
   \coordinate (k1) at (-0.6,1.2);
   \coordinate (k2) at (0.6,1.2);
   \diagram{(in) -- [charged scalar,  momentum={[arrow shorten=0.12mm] \(p\)}] (x) -- [charged scalar,  momentum'={[arrow shorten=0.12mm] \({p^\prime+k_2}\)}] (y) -- [charged scalar,  momentum={[arrow shorten=0.12mm] \(p^\prime\)}] (out)};
   \end{feynman}
   \draw [photon] (x) -- (k1) node [above] {$\varepsilon_1 k_1$};
   \draw [photon] (y) -- (k2) node [above] {$\varepsilon_2 k_2$};
   \draw [fill] (x) circle (.04);
   \draw [fill] (y) circle (.04);
\end{tikzpicture} \ + \
\begin{tikzpicture}[baseline]
   \begin{feynman}
   \coordinate (in) at (-1.8,0);
   \coordinate (x) at (-0.6,0);
   \coordinate (y) at (0.6,0);
   \coordinate (out) at (1.8,0);
   \coordinate (k1) at (-0.6,1.2);
   \coordinate (k2) at (0.6,1.2);
   \diagram{(in) -- [charged scalar,  momentum={[arrow shorten=0.12mm] \(p\)}] (x) -- [charged scalar,  momentum'={[arrow shorten=0.12mm] \({p^\prime+k_1}\)}] (y) -- [charged scalar,  momentum={[arrow shorten=0.12mm] \(p^\prime\)}] (out)};
   \end{feynman}
   \draw [photon] (x) -- (k2) node [above] {$\varepsilon_2 k_2$};
   \draw [photon] (y) -- (k1) node [above] {$\varepsilon_1 k_1$};
    \draw [fill] (x) circle (.04);
   \draw [fill] (y) circle (.04);
\end{tikzpicture} \ + \
\begin{tikzpicture}[baseline]
\begin{feynman}
   \coordinate (in) at (-1.2,0);
   \coordinate (x) at (0,0);
   \coordinate (out) at (1.2,0);
   \coordinate (k1) at (-0.6,1.2);
   \coordinate (k2) at (0.6,1.2);
   \diagram{(in) -- [charged scalar,  momentum={[arrow shorten=0.12mm] \(p\)}] (x) -- [charged scalar, momentum={[arrow shorten=0.12mm] \(p^\prime \)}] (out)};
   \draw [photon] (x) -- (k1) node [above] {$\varepsilon_1 k_1$};
   \draw [photon] (x) -- (k2) node [above] {$\varepsilon_2 k_2$}; 
    \draw [fill] (x) circle (.04);
\end{feynman}
\end{tikzpicture}  \\
=  (-ie)^2\displaystyle  \left(\left[\frac { i } { ( p ^ { \prime } + k _ { 1 } ) ^ { 2 }-m^2 + i\epsilon } \biggl(\left(p^{\prime}+p-k_{2}\right)_{\mu_1} \left(p^{\prime}+p+k_{1}\right)_{\mu_2}\biggr) +(1 \leftrightarrow 2) \right]-2 i\eta_{\mu_1\mu_2}\right) .
\end{array}
\end{equation}
Arranging the contribution coming from the trivalent vertices (written in square brackets) into symmetric and anti-symmetric parts under the exchange of photon 1 and 2 it can be written as
\begin{equation}\label{eq:2photon_scalar_QED_dressed_propagator_sym_sorted}
\begin{aligned}
  \eqref{eq:2photon_scalar_QED_dressed_propagator}  =& (-ie)^2 \left[ \left( \frac{i}{(p^\prime+k_1)^2-m+i\epsilon} + \frac{i}{(p^\prime+k_2)^2-m+i\epsilon}\right)\left(v_{\mu_1}v_{\mu_2} - k_{2_{\mu_1}}k_{1_{\mu_2}}\right) \right. \\ 
  & \left. \left( \frac{i}{(p^\prime+k_1)^2-m+i\epsilon} - \frac{i}{(p^\prime+k_2)^2-m+i\epsilon}\right)\left(v_{\mu_1}k_{1_{\mu_2}} - k_{2_{\mu_1}}v_{\mu_2}\right) \ - \ 2i  \eta_{\mu_1\mu_2}\right]
  \end{aligned}
\end{equation}
where $v^\mu=(p+p^\prime)^\mu$ was anticipated to shorten the notation. Written in this way the amputated scalar propagator dressed with two photons exhibits quite some similarities with the summed-up analogue in the WQFT in equation \eqref{WQFT_4pt_diagramm_sum}. The parts coming from the vertices match up perfectly and again it differs by a factor of $e^{i(k_1+k_2)\cdot b} \ \dd((k_1+k_2)\cdot v)$. But the parts originating from the internal propagator do not look the same at first glance. However, \eqref{WQFT_4pt_diagramm_sum} is effectively written as an expansion in $\hbar$ even though it is not written explicitly, whereas \eqref{eq:2photon_scalar_QED_dressed_propagator_sym_sorted} is not. Hence, we have to consider the classical limit of the internal scalar propagators in \eqref{eq:2photon_scalar_QED_dressed_propagator_sym_sorted} for a proper comparison. For this, we write the photon momenta in terms of wave numbers $k_i \rightarrow \hbar \tilde{k}_i$ to make the $\hbar$ explicit. Thus, taking the classical limit in this context is achieved by taking the soft limit $k_i \rightarrow \hbar \tilde{k}_i \rightarrow 0$. Applying the soft limit on the internal scalar propagators yields the following Taylor expansion:
\begin{equation}
    \frac{i}{(p^\prime+k_1)^2-m^2+i\epsilon} = \frac{i}{v\cdot k_1-k_1\cdot k_2 +i\epsilon}= 
    \frac{i}{v\cdot k_1+i\epsilon}\left(\frac{1}{1-\frac{k_1\cdot k_2}{v\cdot k_1+i\epsilon}}\right)=
    i\sum_{n=0}^\infty \frac{(k_1\cdot k_2 )^n}{(v\cdot k_1+i\epsilon)^{n+1}}
\end{equation}
where the external momenta of the scalars were assumed to be on-shell, and momentum conservation was used. Again using momentum conservation: $v\cdot k_1 = -v \cdot k_2$ the Taylor expansion can be written in terms of $v \cdot k_2$ instead of $v\cdot k_1$:
\begin{equation}\label{eq:taylor_expansion_internal_propagator1}
   \frac{i}{(p^\prime+k_1)^2-m^2+i\epsilon} =  i\sum_{n=0}^\infty (-1)^{n+1}\frac{(k_1\cdot k_2 )^n}{(v\cdot k_2-i\epsilon)^{n+1}} .
\end{equation}
Taylor expanding the second internal propagator yields
\begin{equation} \label{eq:taylor_expansion_internal_propagator2}
    \frac{i}{(p^\prime+k_2)^2-m^2+i\epsilon}=  i\sum_{n=0}^\infty \frac{(k_1\cdot k_2 )^n}{(v\cdot k_2+i\epsilon)^{n+1}}.
\end{equation}
Note that both internal propagators yield the same expression when expanded in the soft limit, except that they have the opposite sign in the $i\epsilon$ prescription and have an alternating relative sign in every second order.
As a result, we can write the two photon dressed propagator in terms of the soft limit expansion of the internal scalar propagators:
\begin{equation}
\begin{aligned}
    =(-ie)^2 i \ 
    \left[\sum_{n=0}^\infty\left(\frac{(-1)^{n+1}(k_1\cdot k_2)^n}{(k_2\cdot v -i\epsilon)^{n+1}} + \frac{(k_1\cdot k_2)^n}{(k_2 \cdot v + i\epsilon)^{n+1}}\right)\left(v_{\mu_1}v_{\mu_2}-k_{1\mu_2}k_{2\mu_1}\right) \right. \\ \ \\
  + \left. \sum_{n=0}^\infty\left(\frac{(-1)^{n+1}(k_1\cdot k_2)^n}{(k_2\cdot v -i\epsilon)^{n+1}} - \frac{(k_1\cdot k_2)^n}{(k_2 \cdot v + i\epsilon)^{n+1}}\right) \left( k_{1\mu_2}v_{\mu_1}-k_{2\mu_1}v_{\mu_2}\right) \ - 2 \eta_{\mu_1\mu_2}\right].
  \end{aligned}
\end{equation}
In fact, written this way the relation to the WQFT expression \eqref{WQFT_4pt_diagramm_sum} is fairly obvious: If we identify $v^\mu = (p+p^\prime)^\mu$, both expressions match up apart from a factor of $e^{i(k_1+k_2)\cdot b} \ \dd((k_1+k_2)\cdot v)$.

By comparing the explicit one and two photon dressed propagators, we found that the dressed propagator in scalar QED seems to be quite closely related to the WQFT analogue and just differs by an overall factor. Most importantly, this relation does not only hold in the classical limit, it even holds to all orders in the $\hbar$-expansion if we sum up all worldline loop orders in the WQFT.

\subsection{Worldline representation of the dressed scalar propagator}
After having compared the one and two photon propagators in the previous section, we now turn to comparing the $N$-photon propagators. For this, we will write the dressed scalar propagator as a first quantized quantum mechanical path integral, similar to the WQFT approach. To that end, we will derive a Feynman-Schwinger representation of the scalar propagator in an abelian background field. The approach used here was first proposed by Richard Feynman in the birth phase of QED \cite{Feynman_1950, Feynman_1951}, but we will follow a derivation similar to the one in \cite{SchwartzMatthewD2014Qfta} in this subsection.

The Schwinger proper time formalism is based on the mathematical identity 
\begin{equation}
    \frac{i}{a+i\epsilon} = \int_0^\infty \mathrm{d}T e^{iT(a+i\epsilon)},
\end{equation}
which holds for $a\in \mathbb{R}$ and $\epsilon > 0$. It allows to rewrite the Feynman propagator for scalars as
\begin{equation}
    \frac{i}{p^2-m^2+i\epsilon} = \int_0^\infty \mathrm{d}T e^{iT (p^2-m^2+i\epsilon)}
\end{equation}
and can be used in a similar way for the dressed propagator.

In analogy to non relativistic quantum mechanics, we can further introduce a one-particle Hilbert space spanned by $|x\rangle$ and momentum operators $\hat{p}^\mu$ with $\hat{p}^\mu |p\rangle = p ^\mu |p\rangle$. Then, we can write the position space representation of the scalar propagator as
\begin{equation}
\begin{aligned}
    G(x,x^\prime) & = \int \frac{\mathrm{d}^4p}{(2\pi)^4}e^{ip\cdot (x-x^\prime)}\frac{i}{p^2-m^2+i\epsilon} \\
    & = \int \frac{\mathrm{d}^4p}{(2\pi)^4} \langle x^\prime | p\rangle\langle p | \frac{i}{\hat{p}^2-m^2+i\epsilon}|x\rangle = \langle x^\prime | \frac{i}{\hat{p}^2-m^2+i\epsilon}|x\rangle ,
    \end{aligned}
\end{equation}
where we used the completeness relation $\int\frac{\mathrm{d}^4p}{(2\pi)^4} | p\rangle\langle p | = \mathbb{1}$. Further introducing Schwinger proper time we can write
\begin{equation}
     \langle x^\prime | \frac{i}{\hat{p}^2-m^2+i\epsilon}|x\rangle = \int_0^\infty \mathrm{d}T e^{-T\epsilon} e^{-iTm^2}\langle x^\prime |e^{iT \hat{p}^2} | x \rangle ,
\end{equation}
which already looks very much like a quantum mechanical transition amplitude for a state $| x\rangle$ to propagate to a state $| x^\prime \rangle$. Such transition amplitudes can be written as a quantum mechanical path integral (a derivation can be found in \cite{SchwartzMatthewD2014Qfta} for instance):
\begin{equation}\label{eq:qm_path_intgeral}
    \langle x^\prime|e^{-i\hat{H}T}|x\rangle = \int_{x(0)=x}^{x(T)=x^\prime} Dx \ \exp\left(i\int\mathrm{d}\tau\mathcal{L}(x,\dot{x})\right) ,
\end{equation}
where $\mathcal{L}$ is the Legendre transform of $\hat{H}$ and with $\hat{H} = -\hat{p}^2$ in our case. 

Let us now use the same procedure for the dressed propagator. The dressed propagator contains the interaction of the propagating scalar field with a constant abelian background and arises from the kinetic term in the action. If we treat the photon field as a constant background, the kinematic part of the scalar QED-lagrangian \eqref{eq:scalar_QED_action} is $\phi_i^*( D^2 +m^2) \phi_i$. Consequently, the Green's function we are looking for has to satisfy a variation of the Klein-Gordon equation:
\begin{equation}
   \left(D^2+m^2\right) G_A(x,x^\prime) = -i\delta^{(D)}(x-x^\prime) ,
\end{equation}
where $D_\mu=\partial_\mu+ieA_\mu$. In other words, the position space expression of our Green's function is
\begin{equation}
G_A(x,x^\prime) = \langle x^\prime| -i \left[ D^2+m^2 \right]^{-1} | x \rangle ,
\end{equation}
where we again introduced a one-particle Hilbert space. Using operator notation $\partial_\mu\rightarrow -i\hat{p}_\mu$, the Green's function becomes
\begin{equation}
    \hat{G}_A = \frac{i}{\left(\hat{p}-eA(\hat{x})\right)^2-m^2+i\epsilon} .
\end{equation}
Introducing a Schwinger-proper-time parameter, the position space expression for the dressed scalar propagator can be rewritten as we did with the regular propagator:
\begin{equation}
   G_A(x,x^\prime) = \langle x^\prime |  \hat{G}_A |x\rangle =\int_0^\infty \mathrm{d}T e^{-T\epsilon} e^{-iTm^2} \langle x^\prime| e^{-iT\hat{H}}|x\rangle
\end{equation}
but now with
\begin{equation}
    \hat{H}=-(\hat{p}-eA(\hat{x})^2.
\end{equation}
In order to further rewrite it as a quantum mechanical path integral as in \eqref{eq:qm_path_intgeral} we need the Legendre transform $\mathcal{L}$ of $\hat{H}$:
\begin{equation}
\label{lagranian_lagrange_transf}
   \mathcal{L} = \hat{p}^\mu\frac{\partial\hat{H}}{p^\mu}-\hat{H} = -(\hat{p}-eA\left(\hat{x})\right)^2-2eA(\hat{x})\cdot(\hat{p}-eA(\hat{x})).
\end{equation} 
The Heisenberg equations of motion for translations in $\tau$:
\begin{equation}
    \dot{\hat{x}}^\mu=\frac{\mathrm{d}\hat{x}}{d\tau}=i\left[\hat{H},\hat{x}^\mu\right]=i\left[-(\hat{p}-eA(\hat{x}))^2,\hat{x}^\mu\right]=2\left(\hat{p}-eA(\hat{x})\right)^\mu,
\end{equation}
where $\left[\hat{p}^\mu,\hat{x}^\nu\right]=i\eta^{\mu\nu}$ and $\left[\hat{x}^\mu,\hat{x}^\nu\right]=0$, allow us to rewrite the Lagrangian \eqref{lagranian_lagrange_transf} as
\begin{equation}
  \mathcal{L} =-\left(\frac{\dot{x}}{2}\right)^2-e\dot{x}\cdot A,
\end{equation}
which looks exactly like the worldline Lagrangian \eqref{eq:worldline_action_redifined_propertime}.
Finally, we found a path integral expression for the dressed propagator, which is very similar to the one in the WQFT:
\begin{equation}\label{eq:master_formula_dressed_propagator}
      G_A(x,x^\prime)=\int_0^\infty \mathrm{d}T e^{-T\epsilon} e^{-iTm^2} \int_{x(0)=x}^{x(T)=x^\prime} Dx \  e^{-i\int_0^T\mathrm{d}\tau\frac{1}{4}\dot{x}^2+e\dot{x}\cdot A}.
\end{equation}
The only difference to the WQFT is the different range of the $\tau$ integration in the exponent (from $0$ to $T$ instead of $-\infty$ to $\infty$) and the proper time $T$ dependent prefactor and integration.

\subsection{N-photon dressed propagator}
To get the N-photon dressed propagator, we write the gauge field $A^\mu$ as a sum of $N$ plane waves $A^\mu(x)=\sum_{i=1}^N\varepsilon_i^\mu e^{ik_i\cdot x}$ and only keep terms containing each polarisation vector linearly. Further defining the vertex operator 
\begin{equation}\label{eq:vertex_operator}
V^{x^{\prime} x}[k, \varepsilon]=\int_{0}^{T} \mathrm{d} \tau \ \varepsilon \cdot\dot{x} \  e^{i k \cdot x(\tau)} 
\end{equation}
we can write the scalar propagator dressed with $N$ photons as
\begin{equation}
\begin{aligned}
G_{N}^{x^{\prime} x}\left(k_{1}, \varepsilon_{1} ; \ldots ; k_{N}, \varepsilon_{N}\right)=&(-i e)^{N} \int_{0}^{\infty} \mathrm{d} T \ e^{-im^{2} T}  \int_{x(0)=0}^{x(T)=0} D x \ e^{-i \int_{0}^{T} \mathrm{d} \tau \frac{\dot{x}^{2}}{4}} \\
& \times  V^{x^{\prime} x}\left[k_{1}, \varepsilon_{1}\right] \cdots V^{x^{\prime}}\left[k_{N}, \varepsilon_{N}\right] .
\end{aligned}
\end{equation}
Introducing an auxiliary polarization vector $\alpha^\mu$ allows to write the vertex operator as
\begin{equation}
V^{x^{\prime} x}[k, \varepsilon]
=\left. \varepsilon \cdot\partial_\alpha \ \int_{0}^{T} \mathrm{d} \tau \  e^{\alpha\cdot\dot{x} + i k \cdot x(\tau)}\right|_{\alpha=0},
\end{equation}
so that all $x$ dependent terms are moved to the exponent of the exponential function.
Additionally, expanding the path integration variable $x$ around a straight-line trajectory 
\begin{equation}
x(\tau)=x+(x^\prime -x)\frac{\tau}{T}+q(\tau)
\end{equation}
(note that this implies Dirichlet boundary conditions $q(0)=q(T)=0$) we get
\begin{equation}
\begin{aligned}
G_{N}^{x^{\prime} x}\left(k_{1}, \varepsilon_{1} ; \ldots ; k_{N}, \varepsilon_{N}\right)=&(-i e)^{N} \int_{0}^{\infty} \mathrm{d} T \ e^{-im^{2} T} e^{-\frac{1}{4} i \frac{\left(x-x^{\prime}\right)^{2}}{T}} \int_{q(0)=0}^{q(T)=0} D q \ e^{-i \int_{0}^{T} \mathrm{d} \tau \frac{\dot{q}^{2}}{4}} \\
&  \times V^{x^{\prime} x}\left[k_{1}, \varepsilon_{1}\right] \cdots V^{x^{\prime}}\left[k_{N}, \varepsilon_{N}\right]
\end{aligned}
\end{equation}
where the photon vertex operator is now
\begin{equation}
V^{x^{\prime} x}[k, \varepsilon]=\left.\varepsilon \cdot\partial_\alpha \ \int_{0}^{T} \mathrm{d} \tau \  e^{\alpha\cdot\left(\frac{x^{\prime}-x}{T}+\dot{q}\right)} e^{i k \cdot\left(x+\left(x^{\prime}-x\right) \frac{\tau}{T}+q(\tau)\right)}\right|_{\alpha=0}.
\end{equation}
Eventually, the path integral over $q$ has a simple Gaussian form and can be easily evaluated. The appropriate worldline Green's function for the $q$-integral has to satisfy $\frac{\partial^2}{\partial\tau^2}\Delta(\tau,\tau^\prime)=\delta(\tau-\tau^\prime)$ and Dirichlet boundary conditions $q(0)=q(T)=0$ and is thus given by:
\begin{equation}\label{eq:greens_function_for_scalar_integral}
\begin{aligned}
\left\langle q^{\mu}(\tau) q^{\nu}\left(\tau^{\prime}\right)\right\rangle &=2i \Delta\left(\tau, \tau^{\prime}\right) \eta^{\mu \nu} \\
\Delta\left(\tau, \tau^{\prime}\right) &=\frac{\left|\tau-\tau^{\prime}\right|}{2}-\frac{\tau+\tau^{\prime}}{2}+\frac{\tau \tau^{\prime}}{T},
\end{aligned}
\end{equation}
where the expectation value is defined as
\begin{equation}
    \left\langle \mathcal{O}(q)\right\rangle = \int_{q(0)=0}^{q(T)=0} D q \ \mathcal{O}(q) \  e^{-i \int_{0}^{T} \mathrm{d} \tau \frac{\dot{q}^{2}}{4}} .
\end{equation}
Since the boundary conditions break translation invariance and since this property is inherited to the Green's function, one has to distinguish between derivatives with respect to the first argument and the second argument. In the following, derivatives with respect to the first argument will be denoted with a dot on the left and derivatives with respect to the second argument with a dot on the right, respectively:
\begin{equation}
\begin{aligned}
{ }^{\bullet} \Delta\left(\tau, \tau^{\prime}\right) &=\frac{\tau^{\prime}}{T}+\frac{1}{2} \operatorname{sign}\left(\tau-\tau^{\prime}\right)-\frac{1}{2} \\
\Delta^{\bullet}\left(\tau, \tau^{\prime}\right) &=\frac{\tau}{T}-\frac{1}{2} \operatorname{sign}\left(\tau-\tau^{\prime}\right)-\frac{1}{2} \\
{}^\bullet\Delta^{\bullet}\left(\tau, \tau^{\prime}\right) &=\frac{1}{T}-\delta\left(\tau-\tau^{\prime}\right).
\end{aligned}
\end{equation}
Using this notation, the $q$-integral yields in line with appendix \ref{appendix:gaussian_integrals}:
\begin{multline}
\int_{q(0)=0}^{q(T)=0} D q \ e^{\int_{0}^{T} \mathrm{d} \tau\left[- i\frac{\dot{q}^{2}}{4}+ q(\tau)\cdot\sum\limits_{i=1}^N\left(ik_i-\alpha_i\cdot\partial_{\tau_i}\right)\delta(\tau_i-\tau_j)\right]}=\\Z_0[q] \ e^{-i\sum\limits_{i,j=1}^N\left(k_i\cdot k_j\Delta(\tau_i,\tau_j)-\alpha_i\cdot\alpha_j{}^\bullet\Delta^\bullet(\tau_i,\tau_j)-ik_i\cdot\alpha_j\Delta^\bullet(\tau_i,\tau_j)-ik_j\cdot\alpha_i{}^\bullet\Delta(\tau_i,\tau_j)\right)}
\end{multline}
with
\begin{equation}
   Z_0[q]=\int_{q(0)=0}^{q(T)=0} D q \ e^{-i\int_{0}^{T} \mathrm{d} \tau\frac{\dot{q}^{2}}{4}}. 
\end{equation}
In the next step, we transform to momentum space treating $p$ as in-going and $p^\prime$ as out-going
\begin{equation}
G_{N}^{p^{\prime} p}\left(k_{1}, \varepsilon_{1} ; \ldots ; k_{N}, \varepsilon_{N}\right)=\int d^{D} x \int d^{D} x^{\prime} \mathrm{e}^{-i p \cdot x+i p^{\prime} \cdot x^{\prime}} G_{N}^{x^{\prime} x}\left(k_{1}, \varepsilon_{1} ; \ldots ; k_{N}, \varepsilon_{N}\right)
\end{equation}
and change the spacetime variables to
\begin{align}
    x_+=\frac{1}{2}\left(x+x^\prime\right), && x_-=x^\prime-x.
\end{align}
The $x_-$-integral is again simply a Gauss integral, and the $x_+$-integral just produces a delta-function imposing momentum conservation. The steps described above lead finally to the following result:
\begin{equation}
\begin{aligned}
G_{N}^{p^{\prime} p}\left(k_{1}, \varepsilon_{1} ; \ldots ; k_{N}, \varepsilon_{N}\right)=&(-i e)^N (2\pi)^D\delta^D\left(p-p^\prime-\sum\limits_{i=1}^{N}k_i\right)\int_{0}^{\infty} \mathrm{d} T \ e^{iT\left({p^\prime}^2-m^{2}+i\epsilon\right)}\\&\left. \left(\prod\limits_{i=1}^{N}\int_0^T\mathrm{d}\tau_i \ \varepsilon_i\cdot\partial_{\alpha_i}\right) 
 e^\mathrm{Exp}\right|_{\alpha_i=0}
\end{aligned}
\end{equation}
with
\begin{equation}
\begin{aligned}
    \mathrm{Exp}=&\sum\limits_{i=1}^N\left(ik_i\tau_i+\alpha_i\right)\cdot\left(p^\prime+p\right)-i\frac{1}{2}\sum\limits_{i,j=1}^N k_i\cdot k_j |\tau_i-\tau_j|
    -i\sum\limits_{i,j=1}^N \alpha_i\cdot\alpha_j\delta(\tau_i-\tau_j)\\&+\sum\limits_{i,j=1}^N k_i\cdot\alpha_j\mathrm{sign}(\tau_i-\tau_j)
\end{aligned}
\end{equation}

\subsection{Cutting of external legs}\label{sec:Cutting of external scalar legs}
The external scalar legs can be stripped off using a simple trick observed in \cite{Jan_2021}, which will be repeated here.
First, we can multiply the dressed propagator by a factor of $-i( p^{\prime 2} -m^2 +i\epsilon)$ to strip off the $p^\prime$-propagator:
\begin{equation}
    -i( p^{\prime 2} -m^2 +i\epsilon)G_{N}^{p^{\prime} p}\left(k_{1}, \varepsilon_{1} ; \ldots ; k_{N}, \varepsilon_{N}\right) .
\end{equation}
Defining
\begin{equation}
\begin{aligned}
\Omega(T)=&(-i e)^N (2\pi)^D\delta^D\left(p-p^\prime-\sum\limits_{i=1}^{N}k_i\right) \left. \left(\prod\limits_{i=1}^{N}\int_0^T\mathrm{d}\tau_i \ \varepsilon_i\cdot\partial_{\alpha_i}\right) 
 e^\mathrm{Exp}\right|_{\alpha_i=0}
\end{aligned}
\end{equation}
this can be rewritten as
\begin{equation}\begin{aligned}
    & -i( p^{\prime 2} -m^2 +i\epsilon)G_{N}^{p^{\prime} p}\left(k_{1}, \varepsilon_{1} ; \ldots ; k_{N}, \varepsilon_{N} \right) \\
   &  \hspace{3cm} = -i( p^{\prime 2} -m^2 +i\epsilon)\int_{0}^{\infty} \mathrm{d} T \ e^{iT\left({p^\prime}^2-m^{2}+i\epsilon\right)} \ \Omega(T) \\
   &  \hspace{3cm} = - \int_{0}^{\infty} \mathrm{d} T \ \frac{\partial}{\partial T}e^{iT\left({p^\prime}^2-m^{2}+i\epsilon\right)} \ \Omega(T) \\
   & \hspace{3cm} =  - \int_{0}^{\infty} \mathrm{d} T \ \frac{\partial}{\partial T}\left( e^{iT\left({p^\prime}^2-m^{2}+i\epsilon\right)} \ \Omega(T)\right) + \int_{0}^{\infty} \mathrm{d} T e^{iT\left({p^\prime}^2-m^{2}+i\epsilon\right)} \ \Omega^\prime (T).
\end{aligned}
\end{equation}
The first term is a boundary term that goes to zero since $\lim_{T\rightarrow\infty} e^{-T\epsilon} \rightarrow 0$ and $\Omega (0) = 0$. Setting the $p^\prime$-leg on shell, i.e. imposing $p^{\prime 2}-m^2=0$, and taking the limit $\epsilon \rightarrow 0$, we can write further:
\begin{equation}
    -i( p^{\prime 2} -m^2 +i\epsilon)G_{N}^{p^{\prime} p}\left(k_{1}, \varepsilon_{1} ; \ldots ; k_{N}, \varepsilon_{N} \right)\Bigg|_{\small{p^\prime}^2=m^2}   = \int_0^\infty \mathrm{d} T \ \Omega^\prime (T) = \Omega(\infty ), 
\end{equation}
where we again used that $\Omega(0)=0$. 

In the next step, we now want to also strip of the $p$-propagator. For this purpose, we first introduce new $\tau$ coordinates:
\begin{equation}\begin{aligned}\label{eq:new_centerOfMass_proper_time_coordinates}
    \Tilde{\tau_i}:=\tau_i -\tau_+&&\mathrm{with}&& \tau_+=\frac{1}{N}\sum_{i=1}^N\tau_i 
\end{aligned}\end{equation}
and rewrite the $\tau_i$-integrations as
\begin{equation}\label{eq:integral_with_centerOfMass_proper_time_coordinates}
    \prod\limits_{i=1}^N\int_{-\infty}^\infty\mathrm{d}\tau_i=\prod\limits_{i=1}^N\int_{-\infty}^\infty\mathrm{d}\Tilde{\tau}_i\int_{-\infty}^\infty\mathrm{d}\tau_+ \delta\left(\frac{1}{N}\sum\limits_{i=1}^N\Tilde{\tau}_i\right) .
\end{equation}
Note that with introducing the new ``center of mass" proper time coordinates we pick up the constraint $\sum_{i=1}^N\Tilde{\tau}_i=0$. Since the introduction of the new proper time coordinates does not change translation invariant quantities, particularly $\tau_i - \tau_j= \tilde{\tau}_i - \tilde{\tau}_j$, the only term contributing to the $\tau_+$-integral is $\exp{\left(\sum_{i=1}^Nik_i\cdot (p+p^\prime)\ \tau_i\right)}$. Introducing a regulator, using momentum conservation, and setting $p^\prime$ on shell, the $\tau_+$ integral yields:
\begin{equation}\begin{aligned}
   \int_0^\infty \mathrm{d}\tau_+ \ e^{i(p+p^\prime)\cdot\sum_{i=1}^Nk_i  \tau_i-\epsilon\tau_+}\Bigg|_{\small{p^\prime}^2=m^2}  &= e^{i(p+p^\prime)\cdot\sum_{i=1}^N k_i  \tilde{\tau}_i} \int_0^\infty \mathrm{d}\tau_+e^{i\tau_+\left((p+p^\prime)\cdot(p-p^\prime)+i\epsilon\right)}\Bigg|_{\small{p^\prime}^2=m^2} \\
   & = \frac{ie^{i(p+p^\prime)\cdot\sum_{i=1}^N k_i  \tilde{\tau}_i}}{p^2-m^2+i\epsilon}.
\end{aligned}\end{equation}
Consequently, omitting the $\tau_+$ integral and setting $\tau_+=0$ strips off the external $p$-propagator.

As a result, we can summarize all the previous steps into a simple recipe:
If we are only interested in an on-shell expression we can cut of external scalar legs by extending the $\tau$-integrals to $\mathbb{R}$, dropping the $T$-integral and inserting a delta function factor $N\delta\left(\sum\limits_{i=1}^N\tau_i\right)$. The amputated photon dressed propagator then yields:
\begin{equation}
\begin{aligned}
 \widehat{G}_{N}^{p^{\prime} p}\left(k_{1}, \varepsilon_{1} ; \ldots ; k_{N}, \varepsilon_{N}\right)&=({p^\prime}^2-m^2) G_{N}^{p^{\prime} p} (p^2-m^2)\\
 &=(-i e)^N (2\pi)^D\delta^D\left(-p+p^\prime+\sum\limits_{i=1}^{N}k_i\right)\\& \left(\prod\limits_{i=1}^{N}\int_{-\infty}^\infty\mathrm{d}\tau_i \ \varepsilon_i\cdot\partial_{\alpha_i}\right) N\delta\left(\sum\limits_{i=1}^N\tau_i\right)
 e^\mathrm{Exp}  \Bigg|_{\small\begin{aligned}\alpha_i=&0 \\ p^2=&m^2={p^\prime}^2\end{aligned}}    .
\end{aligned}
\label{form factor}
\end{equation}

\subsection{From the master formula to explicit diagrams}
Before moving on with the comparison to the WQFT let us test if the master formula for the amputated scalar propagator dressed with $N$ photons \eqref{form factor} indeed reproduces the scalar QED dressed propagator diagrams discussed in section \ref{Dressed propagator in Scalar QED}. For that, we will have to solve the $\tau_i$ integrals in the master formula, one for each photon leg.\\

\subsubsection{Three-point function}Solving for the amputated one photon dressed propagator is trivial and results in:
\begin{equation}
\begin{aligned}
\widehat{G}_{1}^{p^{\prime} p}\left(k, \varepsilon\right)&=-i e (2\pi)^D\delta^D\left(p-p^\prime-k\right) \int_{-\infty}^\infty\mathrm{d}\tau \ \delta\left(\tau\right) \varepsilon\cdot\partial_{\alpha} 
 e^{\alpha\cdot (p^\prime+p)} \Bigg|_{\small\alpha=0}  \\
 &= -i e (2\pi)^D\delta^D\left(p-p^\prime-k\right) \varepsilon\cdot (p^\prime+p) ,
\end{aligned}
\end{equation}
which indeed matches \eqref{eq:1photon_scalar_QED_dressed_propagator}.

\subsubsection{Four-point function}The expression for the amputated 2 photon dressed propagator is
\begin{equation}
\begin{aligned}
 \widehat{G}_{2}^{p^{\prime} p}\left(k_{1}, \varepsilon_{1} ; k_{2}, \varepsilon_{2}\right)
 =&(-i e)^2 (2\pi)^D\delta^D\left(p-p^\prime-k_1-k_2\right)\\&\times \int_{-\infty}^\infty\mathrm{d}\tau_1 \int_{-\infty}^\infty\mathrm{d}\tau_2 \ 2\delta\left(\tau_1+\tau_2\right) \varepsilon_1\cdot\partial_{\alpha_1}\varepsilon_2\cdot\partial_{\alpha_2} 
 e^\mathrm{Exp}  \Bigg|_{\small\begin{aligned}\alpha_i=&0 \\ p^2=&m^2={p^\prime}^2\end{aligned}} 
\end{aligned}
\end{equation}
with
\begin{equation}
\begin{aligned}
    \mathrm{Exp}=&\sum\limits_{i=1}^2\left(ik_i\tau_i+\alpha_i\right)\cdot\left(p^\prime+p\right)-i k_1\cdot k_2 |\tau_1-\tau_2|
    -2i \alpha_1\cdot\alpha_2\ \delta(\tau_1-\tau_2)\\&+ (k_1\cdot\alpha_2-k_2\cdot\alpha_1)\ \mathrm{sign}(\tau_1-\tau_2).
\end{aligned}
\end{equation}
Performing the partial derivatives yields
\begin{equation}
    \begin{aligned}
        \widehat{G}^{p^\prime p}_2 = &(-ie)^2 (2\pi)^D \delta^D(p-p^\prime -k_1-k_2) \ \mathrm{symb}^{-1}\int_{-\infty}^\infty \mathrm{d}\tau_1\int_{-\infty}^\infty \mathrm{d}\tau_2 \ 2\ \delta(\tau_1+\tau_2) \\
        &\times \ \biggl[ -i 2\varepsilon_1\cdot\varepsilon_2\delta(\tau_1-\tau_2)-k_1\cdot \varepsilon_2 k_2\cdot \varepsilon_2 +\varepsilon_1\cdot (p^\prime +p)\epsilon_2\cdot (p^\prime+p) \biggr.\\
        &\qquad+\biggl( \biggr. \varepsilon_1\cdot(p^\prime+p)\varepsilon_2\cdot k_1-\varepsilon_1\cdot k_2\varepsilon_2\cdot(p^\prime+p) \biggl. \biggr) \ \mathrm{sign}(\tau_1-\tau_2) \biggl. \biggr]\\
        &\times
        \exp\left({\sum\limits_{i=1}^2i\tau_i k_i\cdot\left(p^\prime+p\right)-i k_1\cdot k_2 |\tau_1-\tau_2|}\right).
    \end{aligned}
\end{equation}
In the expression above, there are only three integral types that have to be solved. The first one is trivial:
\begin{equation} \label{eq:integral1}
   \int_{-\infty}^\infty \mathrm{d}\tau_1\int_{-\infty}^\infty \mathrm{d}\tau_2 \ 2\ \delta(\tau_1+\tau_2) \delta(\tau_1-\tau_2)e^\mathrm{Exp(\alpha=0)} = 1,
\end{equation}
but the other two need to be solved. The first non-trivial integral yields
\begin{equation}
    \begin{aligned} 
     \int_{-\infty}^\infty \mathrm{d}\tau_1\int_{-\infty}^\infty \mathrm{d}\tau_2 \ 2\ \delta(\tau_1+\tau_2) \  \mathrm{exp}\left({\sum_{i=1}^2(ik_i\tau_i)\cdot (p^\prime+p)-ik_1\cdot k_2 |\tau_1-\tau_2|}\right) \\
      = \int_{-\infty}^\infty \mathrm{d}\tau_1 \ 2 \ \mathrm{exp}\left({i\tau_1(k_2-k_1)\cdot (p^\prime+p)-i2k_1\cdot k_2 |\tau_1|}\right).
    \end{aligned}
\end{equation}
Since it contains absolute values, it has to be broken up into two parts:
\begin{equation}
    \begin{aligned}
     \int_{-\infty}^0 \mathrm{d}\tau_1 \ 2 \ e^{i\tau_1\left((k_2-k_1)\cdot (p^\prime+p)+2k_1\cdot k_2\right)}\ + \int_{0}^\infty \mathrm{d}\tau_1 \ 2 \ e^{i\tau_1\left((k_2-k_1)\cdot (p^\prime+p)-2k_1\cdot k_2\right)}\\
     =\int_{0}^\infty \mathrm{d}\tau_1 \ 2 \ e^{i2\tau_1\left((p^\prime+k_1)^2 -m^2\right)}\ + \int_{0}^\infty \mathrm{d}\tau_1 \ 2 \ e^{i2\tau_1\left((p^\prime+k_2)^2 -m^2\right)},
    \end{aligned}
\end{equation}
where we used momentum conservation for the exponents in the last line. Introducing a regulator\footnote{Actually, the regulator was always there and stems from the $i\epsilon$ prescription in the original Feynman propagator, but we just did not drag it along all the way here.}, these two integrals can now be solved:
\begin{equation}\label{eq:integral2}
    \begin{aligned}
     \int_{0}^\infty \mathrm{d}\tau_1 \ 2 \ e^{i2\tau_1\left((p^\prime+k_1)^2 -m^2+i\epsilon\right)}\ + \int_{0}^\infty \mathrm{d}\tau_1 \ 2 \ e^{i2\tau_1\left((p^\prime+k_2)^2 -m^2+i\epsilon\right)}   \\
     = \frac{i}{(p^\prime+k_1)^2-m^2+i\epsilon} + \frac{i}{(p^\prime+k_2)^2-m^2+i\epsilon}.
    \end{aligned}
\end{equation}
The last integral is of a very similar type, and thus, we will skip the detailed steps and just give the result:
\begin{equation}\label{eq:integral3}
    \begin{aligned}
       \int_{-\infty}^\infty \mathrm{d}\tau_1\int_{-\infty}^\infty \mathrm{d}\tau_2 \ 2\ \delta(\tau_1+\tau_2)\ \mathrm{sign}(\tau_1-\tau_2) \  \mathrm{exp}\left({\sum_{i=1}^2(ik_i\tau_i)\cdot (p^\prime+p)-ik_1\cdot k_2 |\tau_1-\tau_2|}\right) 
       \\
     = -\frac{i}{(p^\prime+k_1)^2-m^2+i\epsilon} + \frac{i}{(p^\prime+k_2)^2-m^2+i\epsilon}.
    \end{aligned}
\end{equation}
Putting everything together, we finally find:
\begin{equation} \label{eq: 4pt scalar propagator}
\begin{aligned}
\widehat{G}_{2}^{p^{\prime} p}= &(-ie)^2 (2\pi)^D \delta^D(p-p^\prime -k_1-k_2)\\&\times\left\{-2 i\varepsilon_{1} \cdot \varepsilon_{2}+\left[\frac { i } { ( p ^ { \prime } + k _ { 1 } ) ^ { 2 }-m^2 } \biggl(\varepsilon_{1} \cdot\left(p^{\prime}+p-k_{2}\right) \varepsilon_{2} \cdot\left(p^{\prime}+p+k_{1}\right) \biggr)+(1 \leftrightarrow 2) \right]\right\},
\end{aligned}
\end{equation}
which again reproduces our former result in \eqref{eq:2photon_scalar_QED_dressed_propagator}.

\subsection{Comparison to the WQFT dressed propagator}\label{sec:Comparison_to_WQFT_propagator}
Now that we have found a master formula of a worldline representation of the dressed scalar propagator, which looks very similar to the WQFT expression, let us investigate how it is precisely related to the analogue of a dressed propagator in the WQFT. The WQFT analogue to a dressed propagator is given by 
\begin{equation}
     G_A^{\mathrm{WQFT}}\left(b,v\right) = \int D \left[ z \right] e^{iS_{\text{pm}}+iS_{\text{int}}},
\end{equation}
where we integrate out the worldline field $x$ but do not integrate over the gauge field $A^\mu$ yet. Hence, from the WQFT perspective, the starting point is very similar to \eqref{eq:master_formula_dressed_propagator}, but here we integrate over infinitely extended proper times and do not have a Schwinger-proper-time parameter anymore:
\begin{equation}\label{eq:WQFT_dressed_propagator_startingpoint}
    G_A^{\mathrm{WQFT}}\left(b,v\right)=\int Dx \ \exp{\left[-i\int_{-\infty}^{\infty}\mathrm{d}\tau\left( \frac{1}{4}\dot{x}^2+e\dot{x}\cdot A\right)\right]} .
\end{equation}
Although the WQFT expression looks very similar to the worldline representation of the dressed scalar propagator in \eqref{eq:master_formula_dressed_propagator} it is conceptually different. In the WQFT approach, we do not consider a scattering process of ingoing and outgoing free states with fixed positions or momenta anymore. Instead, we think of infinite trajectories. As stated in the introduction, physical quantities are calculated as expectation values of all deviations from straight-line trajectories (no scattering or no interaction). Hence, the in- and outgoing momenta or positions are not the natural parameters of our description anymore. In the WQFT, we rather think in terms of the impact parameter $b$ and the velocity of the object traveling along the trajectory $v$ as the parameters of our results.

Starting from \eqref{eq:WQFT_dressed_propagator_startingpoint} we proceed exactly like in the derivation of our worldline representation of the $N$ photon dressed scalar propagator, keeping in mind that there are some subtle differences because of the different boundary conditions.\\
Again writing the gauge field $A^\mu$ as a collection of plane waves $A^\mu(x)=\sum_{i=1}^N\varepsilon_i^\mu e^{ik_i x}$ and considering the background field expansion 
\begin{equation}
    x(\tau )=b^\mu+v^\mu \tau + z^\mu (\tau),
    \end{equation}the $N$-photon dressed WQFT-propagator yields
\begin{equation}
\begin{aligned}
  G^{\mathrm{WQFT}}_N\left(b,v,\{k_1,\varepsilon_1,...,k_N,\varepsilon_N\}\right)=&\left(-ie\right)^N\int Dz \ e^{-\int_{-\infty}^{\infty}\mathrm{d}\tau\frac{i}{4}(v+\dot{z})^2} \\&
  \hspace{.5cm}\times V^{b,v}\left[k_1,\varepsilon_1\right]...V^{b,v}\left[k_N,\varepsilon_N\right] ,
\end{aligned}
\end{equation}
where the vertex operator $V^{b,v}\left[k,\varepsilon\right]$ is just the vertex operator $V^{x,x^\prime}\left[k,\varepsilon\right]$ from before (defined in \eqref{eq:vertex_operator}) with $\frac{x^\prime-x}{T}$ replaced by $v$, $q$ replaced by $z$ and $x$ replaced by $b$.
Since we do not have vanishing boundary conditions anymore the kinetic term of the spacetime path integral produces a non-vanishing boundary term $e^{-\frac{1}{2}\int_{-\infty}^\infty iv\cdot \dot{z}(\tau)}$. As in section \ref{sec:WQFT_feynman_rules}, we will ignore this term and assume that all other boundary terms can also be ignored. This allows us to integrate by parts without producing any boundary terms. The path integral over $z$ can be performed as before in the finite worldline case, but this time we do not specify the Green's functions, i.e. we just impose a generic translation invariant Green's function:
\begin{equation}
\begin{aligned}
    \left\langle z^\mu(\tau)z^\nu(\tau^\prime) \right\rangle=& \ 2i\Delta(\tau -\tau^\prime)\eta^{\mu\nu} \hspace{0.5cm}&\mathrm{with}&&\hspace{0.5cm} \frac{\partial^2}{\partial\tau^2}\Delta(\tau-\tau^\prime)=&\delta(\tau-\tau^\prime) . \\
    \end{aligned}
\end{equation}
With this generic Green's function, the dressed WQFT-propagator becomes
\begin{equation}
\begin{aligned}
G^{\mathrm{WQFT}}_N\left(b,v,\{k_1,\varepsilon_1,...,k_N,\varepsilon_N\}\right)=& \ G^\mathrm{WQFT}_0[z] \  e^{
i\sum_{i=1}^N k_i \cdot b} (-ie)^N \\& \
\hspace{.5cm} \times \left. \left(\prod\limits_{i=1}^{N}\int_{-\infty}^\infty\mathrm{d}\tau_i \ \varepsilon_i\cdot\partial_{\alpha_i}\right) 
 e^\mathrm{Exp}\right|_{\alpha_i=0}
\end{aligned}
\label{wqft-propagator}
\end{equation}
with
\begin{equation}
\begin{aligned}
    \mathrm{Exp}=&\sum\limits_{i=1}^N\left(ik_i\tau_i+\alpha_i\right)\cdot v-i\sum\limits_{i,j=1}^N k_i\cdot k_j \Delta(\tau_i-\tau_j)
    +i\sum\limits_{i,j=1}^N \alpha_i\cdot\alpha_j{}^\bullet\Delta^\bullet(\tau_i-\tau_j)\\&-2\sum\limits_{i,j=1}^N k_i\cdot\alpha_j{}^\bullet\Delta(\tau_i-\tau_j) .
\end{aligned}
\end{equation}
and
\begin{equation}
    G^\mathrm{WQFT}_0[z]:=\lim\limits_{T\rightarrow\infty} e^{-i\frac{v^2}{2}T}Z^0[z],
\end{equation}
where $Z^0[z]$ is the free path integral normalization factor.
The WQFT-propagator \eqref{wqft-propagator} already looks very similar to the expression we had for the amputated dressed scalar propagator on a finite worldline $\widehat{G}^{p^\prime p}_N$ in \eqref{form factor}. In order to bring it into a even more similar form we introduce ``center of mass" proper time coordinates as in \eqref{eq:new_centerOfMass_proper_time_coordinates} and rewrite the $\tau_i$-integrations as in \eqref{eq:integral_with_centerOfMass_proper_time_coordinates}.
Since the Green's function $\Delta (\tau-\tau^\prime )$ is translation invariant, the only term contributing to the $\tau_+$-integral is $\exp{\left(\sum_{i=1}^Nik_i\cdot v \ \tau_i\right)}=\exp{\left(\sum_{i=1}^Nik_i\cdot v(\Tilde{\tau}_i+\tau_+)\right)}$. Consequently, the $\tau_+$-integral just produces a $\delta\left(\sum_{i=1}^Nk_i\cdot v\right)$. With this time coordinate change, the final expression for the dressed WQFT-propagator is:
\begin{equation}
\begin{aligned}
G^{\mathrm{WQFT}}_N\left(b,v,\{k_1,\varepsilon_1,...,k_N,\varepsilon_N\}\right)=&G^\mathrm{WQFT}_0[z] \ \delta\left(\sum\limits_{i=1}^N k_i\cdot v\right) e^{
i\sum_{i=1}^N k_i \cdot b} (-ie)^N  \\&
\left. \left(\prod\limits_{i=1}^{N}\int_{-\infty}^\infty\mathrm{d}\tau_i \ \varepsilon_i\cdot\partial_{\alpha_i}\right) N\delta\left(\sum\limits_{i=1}^N \tau_i\right)
 e^\mathrm{Exp}\right|_{\alpha_i=0}.
\end{aligned}
\label{wqft-propagator final}
\end{equation}
By comparing \eqref{wqft-propagator final} and \eqref{form factor} and ignoring the delta function imposing momentum conservation in \eqref{form factor} we find:
\begin{equation}
\boxed{
   \frac{G^\mathrm{WQFT}_N(b,v;\{\varepsilon_i,k_i\})}{G^\mathrm{WQFT}_0}=\delta\left(\sum\limits_{i=1}^N k_i\cdot v\right) e^{i\sum\limits_{i=1}^N k_i\cdot b} \widehat{G}^{p^\prime p}_N(k_i,\varepsilon_i;\dots;k_N,\varepsilon_N)}
   \label{link WQFT and scalar QED propagator}
\end{equation}
if we identify $v=(p^\prime +p)$ and choose the following Green's functions:
\begin{equation}
\Delta(\tau)=\frac{|\tau|}{2} .
\end{equation}
With this Green's function, we have
\begin{equation}\label{eq:greens_prop}
    \left\langle z^\mu(\tau)z^\nu(\tau^\prime) \right\rangle=  2i\Delta(\tau -\tau^\prime)\eta^{\mu\nu} = i \eta^{\mu\nu} \ | \tau -\tau^\prime | .
\end{equation}
If we recall the definition of the WQFT worldline propagator in \eqref{eq:propagatorz} we indeed confirm that \eqref{eq:greens_prop} is given by the average of the retarded and the advanced propagator as anticipated in the previous sections.
With equation \eqref{link WQFT and scalar QED propagator} we have now found a general link between the dressed propagators in scalar QED and the WQFT that holds to an arbitrary number of photon legs.
\section{S-matrices in scalar QED vs. expectation values in the WQFT}\label{sec:S-matrices in scalar QED vs. expectation values in the WQFT}
To find out what the previous observations and particularly equation \eqref{link WQFT and scalar QED propagator} mean for general S-matrices and their relation to the WQFT, let us revisit the discussion in section \ref{sec:S-matrices}. First of all, we can write \eqref{Smatrix_via_dressed_propagator} in terms of momentum space dressed propagators
\begin{equation}
\begin{aligned}
\displaystyle
\left\langle\phi_{1} \phi_{2}|S| \phi_{1} \phi_{2}\right\rangle=&\mathcal{Z}^{-1} \int \mathrm{d}^{D}\left[x_{i}, x_{i}^{\prime}\right] e^{i p_{i} \cdot x_{i}-i p_{i}^{\prime} \cdot x^{\prime}_{i}} \int\mathcal{D}\left[A_{\mu}\right]\widehat{G}_{A,1}\left(x_{1}, x_{1}^{\prime}\right)\widehat{G}_{A,2}\left(x_{2}, x_{2}^{\prime}\right) e^{i\left(S_{\mathrm{A}}+S_{\mathrm{gf}}\right)}\\ 
= & \mathcal{Z}^{-1}\int\mathcal{D}\left[A_{\mu}\right]\widehat{G}_{A,1}\left(p_{1}, p_{1}^{\prime}\right)\widehat{G}_{A,2}\left(p_{2}, p_{2}^{\prime}\right) e^{i\left(S_{\mathrm{A}}+S_{\mathrm{gf}}\right)} .
\end{aligned}\label{eq:S-matrix_via_momentum_dressed_prop}
\end{equation}
In principle, we can now use \eqref{link WQFT and scalar QED propagator} to replace the dressed scalar propagators by the WQFT analogue, but we need to be careful with how the photon legs are ``glued together" in the two theories by the path integral over the gauge field $A_\mu$. 
\begin{figure}[h]
\centering
\begin{subfigure}[t]{0.5\textwidth}
   \begin{tikzpicture}[baseline]
   \begin{feynman}
   \coordinate (in1) at (-3,1);
   \coordinate (k1) at (-1,1);
   \coordinate (k2) at (1,1);
   \coordinate (out1) at (3,1);
   \coordinate (in2) at (-3,-1);
   \coordinate (k1') at (-1,-1);
   \coordinate (k2') at (1,-1);
   \coordinate (out2) at (3,-1);
     \draw [fill] (k1) circle (.04);
   \draw [fill] (k2) circle (.04);
   \draw [fill] (k1') circle (.04);
   \draw [fill] (k2') circle (.04);
   \diagram{ (in1) -- [charged scalar, momentum={[arrow shorten=0.12mm] \(p_1\)}] (k1) -- [charged scalar, momentum={[arrow shorten=0.12mm] \(p_1-k_1\)}] (k2) -- [charged scalar, momentum={[arrow shorten=0.12mm] \(p_1^\prime\)}] (out1)};
   \diagram{ (in2) -- [charged scalar, momentum'={[arrow shorten=0.12mm] \(p_2\)}] (k1') -- [charged scalar, momentum'={[arrow shorten=0.12mm] \(p_2+k_1\)}] (k2') -- [charged scalar, momentum'={[arrow shorten=0.12mm] \(p_2^\prime\)}] (out2)};
   \draw [photon2] (k1) -- (k1') node [midway, right] {$\downarrow{k_1}$} ;
   \draw [photon2] (k2) -- (k2') node [midway, right] {$\downarrow{k_2=p_1-k_1-p_1^\prime}$} ;
   \end{feynman}
   \end{tikzpicture}
    \caption{4-pt function in scalar QED: Only $k_1$ is not fixed by the external momenta and momentum conservation. Hence, only $k_1$ is integrated over.}
    \label{fig:dressed_scalar_prop_glued_together}
    \end{subfigure}
    \hfill
    \begin{subfigure}[t]{0.4\textwidth}
      \begin{tikzpicture}[baseline]
   \begin{feynman}
   \coordinate (in1) at (-2.3,1);
   \coordinate (k1) at (-1,1);
   \coordinate (k2) at (1,1);
   \coordinate (out1) at (2.3,1);
   \coordinate (in2) at (-2.3,-1);
   \coordinate (k1') at (-1,-1);
   \coordinate (k2') at (1,-1);
   \coordinate (out2) at (2.3,-1);
   \draw [dotted] (in1) node [left]{$v_1, b_1$} -- (out1);
   \draw [zUndirected] (k1) -- (k2)  node [midway, above] {$\underrightarrow{w_1}$};
   \draw [dotted] (in2) node [left]{$v_2,b_2$} -- (out2);
   \draw [photon2] (k1) -- (k1') node [midway, right] {$\downarrow{k_1}$} ;
   \draw [photon2] (k2) -- (k2') node [midway, right] {$\downarrow{k_2}$} ;
   \draw [fill] (k1) circle (.08);
   \draw [fill] (k2) circle (.08);
   \draw [fill] (k1') circle (.08);
   \draw [fill] (k2') circle (.08);
   \end{feynman}
   \end{tikzpicture}
   \\ \ \\ \vspace{1.2mm}
    \caption{Tree-level 4-pt function in the WQFT: Both, $k_1$ and $k_2$ are not fixed, therefore both are integrated over.}
    \label{fig:WQFT-prop_glued_together} 
\end{subfigure}
\caption{Two photon exchange between two massive point particles in scalar QED (a) and the WQFT (b). When gluing dressed propagators together, i.e. connecting the photon legs of two dressed propagators, the momentum of one photon leg is fixed by the external momenta in QFT amplitudes, whereas all photon momenta are integrated over in the WQFT.}
\label{fig:prop_glued_together}\vspace{.5cm}
\end{figure}
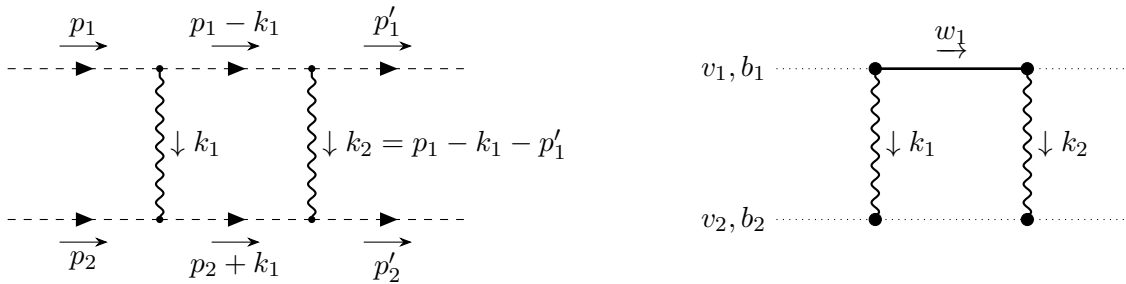
As exemplarily illustrated in figure \ref{fig:prop_glued_together} for the two photon exchange, one of the photon momenta is fixed by the external momenta in the QFT amplitude, whereas in the WQFT we integrate over all photon momenta. As a consequence, an additional integral over the total exchanged photon momentum needs to be added on the left-hand side of \eqref{eq:S-matrix_via_momentum_dressed_prop} when replacing the dressed scalar propagators by the WQFT analogue. Integrating over the total momentum exchange and using equation \eqref{link WQFT and scalar QED propagator} to replace the dressed scalar propagators by the WQFT analogue we obtain:
\begin{equation}\label{eq:s-matrix_dressed_prop_replaced_by_WQFT}
\begin{array}{l}\displaystyle
   \int \frac{\mathrm{d}^D q}{(2\pi)^{D-2}}\delta (q\cdot v_1) \delta (q\cdot v_2) e^{iq\cdot b} \left\langle\phi_{1} \phi_{2}|S| \phi_{1} \phi_{2}\right\rangle \\ \displaystyle \hspace{3cm} ={\mathcal{Z}_0}_\text{WQFT}^{-1}\int\mathcal{D}\left[A_{\mu}\right]G_A^\mathrm{WQFT}(b_1,v_1)G_A^\mathrm{WQFT}(b_2,v_2) e^{i\left(S_{\mathrm{A}}+S_{\mathrm{gf}}\right)} ,
\end{array}
\end{equation}
where $b=b_1-b_2$ and $q=\sum_{i=1}^N k_i=p_1-p_1^\prime=p_2^\prime-p_2$ is the total momentum transfer from particle 1 to particle 2. The constant ${\mathcal{Z}_0}_\text{WQFT}^{-1}=\mathcal{Z}^{-1}\left(G^\mathrm{WQFT}_0 \right)^{-2}$ is the normalization constant ensuring that $\mathcal{Z}_\text{WQFT}=1$ in the non-interacting case. The right-hand side of \eqref{eq:s-matrix_dressed_prop_replaced_by_WQFT} is nothing else than the WQFT partition function \eqref{eq:WQFT_partition_function}. Consequently, we find
\begin{equation}\label{eq:link_WQFTpartion_function_s-matrix}
    \mathcal{Z}_\text{WQFT} = \int \frac{\mathrm{d}^D q}{(2\pi)^{D-2}}\delta (q\cdot v_1) \delta (q\cdot v_2) e^{iq\cdot b} \left\langle\phi_{1} \phi_{2}|S| \phi_{1} \phi_{2}\right\rangle
\end{equation}
in the classical limit. Thus, the WQFT partition function is given by the Fourier transform of the S-matrix with respect to the momentum transfer $q$ into impact parameter space transverse to the $(D-2)$-dimensional scattering plane. This is exactly the definition of the exponentiated eikonal phase $e^{i\chi}$, which has proven to be a useful tool to extract classical observables in gauge theory and gravity directly from scattering amplitudes \cite{eikonal_phase_1, eikonal_phase_2, eikonal_phase_3}. As a result, we find 
\begin{equation}
     \mathcal{Z}_\text{WQFT} = e^{i\chi},
\end{equation}
which has already been observed for the WQFT coupled to gravity in \cite{Jan_2021} and was used to obtain the deflection $\Delta p_i^\mu$ from classical black hole scattering events.

Equation \eqref{eq:link_WQFTpartion_function_s-matrix} can be further generalized for general expectation values in the WQFT. For instance, we can consider a final state photon. In this case \eqref{eq:link_WQFTpartion_function_s-matrix} generalizes to
\begin{equation}\label{eq:link_WQFT_s-matrix_final_state_photon}
    \bigl\langle A_\mu \bigr\rangle_\text{WQFT} = \int \frac{\mathrm{d}^D q}{(2\pi)^{D-2}}\delta (q\cdot v_1) \delta (q\cdot v_2) e^{iq\cdot b} \left\langle\phi_{1} \phi_{2}A_\mu|S| \phi_{1} \phi_{2}\right\rangle .
\end{equation}

\chapter{Adding Spin}\label{Adding Spin}
Realistic physical objects, no matter if they are extended objects like black holes or neutron stars in the context of gravitational wave physics or simply point particles as in particle physics, carry internal spin. For extended objects, it stems from the internal rotational degree of freedom, i.e. the rotation of the object around an inner axis. 
In particle physics, we consider quantized spin having integer or half-integer spin values, which cannot be thought of as spatial rotations. As shown in \cite{Jakobsen_2021_Susy} up to quadratic order in spin, a worldline theory of spin-1/2 or spin-1 particles directly corresponds to classical spatial rotations to linear or quadratic order in spin, respectively. In this chapter, we will consider objects carrying spin-1/2, but the approach used here can also be extended to higher spin.

\section{Adding spin to the worldline}\label{sec:Adding spin to the worldline}

If we seek to describe a particle carrying spin, the approach used in chapter 1 is not sufficient anymore. Instead, we need a worldline theory of a relativistic spinning particle. It has been shown that upon quantization, a description of spin-$N/2$ particles can be obtained by augmenting the worldline theory of a non-spinning relativistic particle by adding $N$ real anti-commuting vector fields  $\Psi^\mu ( \tau )$ to the worldline degrees of freedom \cite{Howe_1988, BRINK1976435, Bastianelli_I_2005, Bastianelli_II_2005}. Spin is then represented by these anti-commuting vector fields, as we will dwell on in a bit. This approach has already been suggested for the WQFT for spinning particles coupled to gravity in \cite{jakobsen_2021gravitational_bremstrahlung_spinning} and \cite{Jakobsen_2021_Susy}. 

For spin-1/2 objects we only need one anti-commuting vector field $\Psi^\mu ( \tau )$. Because of the Grassmann nature of the vector field $\Psi^\mu ( \tau )$, the kinetic term has to be a first-order term:
\begin{equation}
    S_{\mathrm{spin}}=-\frac{1}{2}i\int\mathrm{d}\tau \Psi_\mu\dot{\Psi}^\mu.
\end{equation}
Furthermore, we can couple it to any anti-symmetric tensor. If we want to couple the spin field to the electromagnetic field $A_\mu(x)$, it is obvious to use the electromagnetic tensor $F_{\mu\nu} = \partial_\mu A_\mu-\partial_\nu A_\nu$. Consequently, the interaction term is given by
\begin{equation}
    S_{\mathrm{intSpin}}=ie\int\mathrm{d}\tau F_{\mu\nu}\Psi^\mu\Psi^\nu,
\end{equation}
where the agreement with quantum electrodynamics determines the overall constant.
But how does the Grassmann field $\Psi^\mu ( \tau )$ describe spin? The traditional description of spin is usually given in terms of the spin tensor $S^{\mu\nu}$ representing the internal part of the total angular momentum tensor $J^{\mu\nu} = -J^{\nu\mu}$. The total angular momentum tensor is the conserved charge due to Lorentz invariance and is defined by
\begin{equation}
    J^{\mu\nu} = \underbrace{z^\mu p^\nu-p^\mu z^\nu}_{M^{\mu\nu}} + S^{\mu\nu},
\end{equation}
where $M^{\mu\nu}$ is the orbital part of the angular momentum. Thus, the spin tensor is given by the difference of the total angular momentum and its orbital part. However, there is an ambiguity in this definition. Since a different choice for the yet arbitrary coordinates, $z^\mu$ will result in a different spin $S^{\mu\nu}$ ($J^{\mu\nu}$ is conserved and therefore being unchanged).\footnote{The here established WQFT exhibits a $\mathcal{N}=1$ supersymmetry (SUSY). The ambiguity in the definition of the spin tensor is closely connected to the SUSY of our WQFT since a SUSY transformation simultaneously changes both the coordinates and the spin tensor. A SUSY transformation shifting the position $x^\mu$ by $b^\mu$ also causes a shift in the spin tensor $\delta S^{\mu\nu} \sim b$.} 
The Spin tensor has to obey the SO(1,3) Lorentz algebra
\begin{equation}\label{eq:lorentz_algebra}
    \{S^{ab},S^{cd}\}_{_{\mathrm{P.B.}}} = \eta^{ac}S^{bd} + \eta^{bd}S^{ac} -\eta^{bc}S^{ad} - \eta^{ad}S^{bc}
\end{equation}
under Poisson brackets. Before discussing how the $\psi$-field can be identified with the spin tensor, 
let us briefly address how Poisson brackets behave when acting on Grassmann odd or even quantities, respectively. For this, the results of a more detailed discussion in \cite{Woit13quantummechanics} are repeated here. If both arguments anti-commute the Poisson brackets act like anti-commutators and like a commutator if one or none of the arguments is anti-commuting. For two variables $F_1$ and $F_2$ this means that
\begin{equation}\label{eq:poisson_bracket_grassmann}
    \{ F_1 , F_2 \}_{_{\mathrm{P.B.}}} = -(-1)^{|F_1||F_2|}\{ F_2 , F_1 \}_{_{\mathrm{P.B.}}},
\end{equation}
where $|F_1|$ and $|F_2|$ are the Grassmann degrees of $F_1$ and $F_2$. As a consequence also the Leibniz rule needs to be adjusted depending on the order of the involved anti-commuting variables:
\begin{equation}\label{eq:leipniz_rule_grassman}
  \{F_1, F_2F_3 \}_{_{\mathrm{P.B.}}} = \{F_1,F_2\}_{_{\mathrm{P.B.}}} F_3 + (-1)^{|F_1||F_2|} F_2\{F_1,F_3\}_{_{\mathrm{P.B.}}}.
\end{equation}
Imposing 
\begin{equation}
    \{\Psi^\mu, \Psi^\nu \}_{_{\mathrm{P.B.}}} = -i\eta^{\mu\nu} 
\end{equation}
and using equation \eqref{eq:poisson_bracket_grassmann} and \eqref{eq:leipniz_rule_grassman} it is not hard to show that
\begin{equation}\label{eq:definition_spin_tensor}
    \mathcal{S}^{\mu\nu} = -i \Psi^\mu \Psi^\nu
\end{equation}
obeys the Lorentz algebra \eqref{eq:lorentz_algebra}. In light of this, it is well motivated to identify \eqref{eq:definition_spin_tensor} with the spin tensor. Thus, equation \eqref{eq:definition_spin_tensor} translates the description of spin in terms of the anti-commuting Grassmann fields $\Psi^\mu$ to the traditional classical description of spin in terms of the spin tensor $\mathcal{S}^{\mu\nu}$. But note that this representation of the spin tensor only holds to linear order in spin since the components of the Grassmann vectors $\Psi^\mu$ square to zero.

Resulting from the preceding discussion, the action of the full worldline theory for spin-1/2 particles is given by
\begin{equation}\label{eq:action_WQFT_spin_not writtenout}
    S = S_{\text{pm}} + S_\text{A} + S_\text{int} +S_{\text{spin}} + S_{\text{intSpin}} + S_\mathrm{gf}.
\end{equation}
In order to describe a scattering scenario, we expand the worldline fields around a constant straight line background as already done in chapter 1, but now also including $\Psi^\mu(\tau)$:
\begin{equation}
    x^\mu(\tau) = b^\mu + v^\mu \tau + z^\mu (\tau) \hspace{2cm}\Psi^\mu(\tau) = \eta^\mu + \psi^\mu(\tau)
\end{equation}
and identify
\begin{equation}\label{eq:definition_backround_spin_tensor}
    S^{\mu\nu} = -i \eta^\mu \eta^\nu
\end{equation}
with the constant background spin.
With this background expansion of the worldline fields the action \eqref{eq:action_WQFT_spin_not writtenout} yields
\begin{equation}\label{eq:action_WQFT_spin}
    S=-\int \mathrm{d}\tau \left(\frac{\dot{z}(\tau)^2}{4}+e\left(v+\dot{z}(\tau)\right)\cdot A+
    i\frac{1}{2}\psi_\mu \dot{\psi}^\mu -ieF_{\mu\nu}\left(\psi+\eta\right)^\mu\left(\psi+\eta\right)^\nu\right) + S_\mathrm{A}+ S_\mathrm{gf},
\end{equation}
where constant terms and boundary terms were ignored.
Treating the photon field $A^\mu(x)$ and the worldline fields $z^\mu(\tau)$ and $\psi (\tau)$ on equal footing, physical quantities are then computed via 
\begin{equation}
\begin{array}{l}
\displaystyle
\left\langle\mathcal{O}\left(A,\left\{x_{i}\right\}\right)\right\rangle_{\text {WQFT }}=\mathcal{Z}_{\text {WQFT }}^{-1} \int D\left[A_{\mu}\right] \int \prod_{i=1}^{n} D\left[z_{i}, \psi_i\right] \mathcal{O}\left(A,\left\{x_{i}\right\}\right) e^{iS}
\end{array}
\end{equation}
with
\begin{equation}
\begin{array}{r}\displaystyle
\mathcal{Z}_{\mathrm{WQFT}}:=\text { const } \times \int D\left[A_{\mu}\right] \int \prod_{i=1}^{n} D\left[z_{i}, \psi_i\right] e^{iS} .
\end{array}
\end{equation}
\subsection{WQFT Feynman rules for spin-1/2 particles}
For the derivation of Feynman rules for the spin-1/2 WQFT we Fourier-transform the fields $z^\mu(\tau)$, $\psi^\mu(\tau)$ and $A^\mu(x)$:
\begin{equation}\label{eq:fourierfields_with_spin}
    \begin{aligned}
      z^\mu (\tau)=\int_\omega e^{-i\omega \tau}z^\mu(\omega), &\hspace{1cm}& \psi^\mu(\tau)=\int_\omega e^{-i\omega\tau}\psi^\mu(\omega),&\hspace{1cm}& A^\mu(x)=\int_k A^\mu(-k) e^{ik\cdot x}.
      \end{aligned}
\end{equation}
The kinetic term for $z(\tau)$ is the same as in the non-spinning case. Thus, the $z$-propagator is again given by \eqref{eq:propagatorz}. For the spin field $\psi(\tau)$ we can read off the propagator from the action \eqref{eq:action_WQFT_spin}:
\begin{align}\label{eq:propagatorPsi}
    \begin{tikzpicture}[baseline={(current bounding box.center)}]
    \coordinate (in) at (-2,0);
    \coordinate (out) at (2,0);
    \coordinate (x) at (1,0);
    \coordinate (y) at (-1,0);
    \draw [dotted] (in) -- (y);
    \draw [dotted] (x) -- (out);
    \draw [psiParticle] (x) -- (y) node [midway, below] {$\omega$};
    \draw [fill] (x) circle (.08) node [above]{$\psi^\nu(\tau_2)$};
    \draw [fill] (y) circle (.08) node [above]{$\psi^\mu(\tau_1)$};
    \end{tikzpicture}=-i\eta^{\mu\nu}\int_\omega\frac{e^{-i\omega(\tau_1-\tau_2)}}{\omega\pm i\epsilon}.
\end{align}
To read off the vertices, we look at the interaction part of \eqref{eq:action_WQFT_spin}:
\begin{equation}\label{eq:Sint_with_spin}
    S^\text{int}=-\int \mathrm{d}\tau \left[e\left(v+\dot{z}(\tau)\right)\cdot A-ie2\partial_\mu A_\nu (\psi+\eta)^\mu(\psi+\eta)^\nu\right].
\end{equation}
As in the non-spinning case, $A^\mu$ inherits the $x(\tau)$ dependence given in \eqref{eq:Atau}, which has to be born in mind when deriving Feynman rules.
Using \eqref{eq:Atau} and \eqref{eq:fourierfields_with_spin} the interaction part of the action not depending on the spin field $\psi(\tau)$ can be written as:
\begin{equation}\label{eq:Sint1}
\begin{aligned}
    \left.S^\text{int}\right|_{\psi=0}=&-e\sum_{n=0}^\infty\frac{i^n}{n!}\int_{k,\omega_1,\dots,\omega_n }\dd(k\cdot v+\textstyle\sum_{i=1}^{n}\omega_i)\displaystyle e^{ik\cdot b}\left(\prod_{i=1}^{n} z^{\rho_i}(\omega_i)\right) \\
    &\times A_\mu(k) \left(\left(\prod_{i=1}^{n}k_{\rho_i}\right)\left(v^\mu +2ik_\alpha S^{\alpha\mu}\right)+\sum_{i=1}^n\omega_i\left(\prod_{j\neq i}^nk_{\rho_j}\right)\delta^\mu_{\rho_i}\right),
\end{aligned}
\end{equation}
which is the same as the interaction part in the non-spinning case with $v^\mu$ replaced by $v^\mu \rightarrow v^\mu +2ik_\alpha S^{\alpha\mu}$. The $\psi(\tau)$ dependent part reads

\begin{equation}\label{eq:Sint2}
    \begin{aligned}
      S^{\text{int}}_\text{spin}=&-2e\sum_{n=0}^\infty\frac{i^n}{n!}\int_{k,\omega_1,\dots,\omega_n } e^{ik\cdot b}A_\mu(k)\left(\prod_{i=1}^{n} k\cdot z(\omega_i)\right) \\
    & \qquad\times\left\{ \int_\omega\dd(k\cdot v+\textstyle\sum_{i=1}^{n}\omega_i+\omega)\displaystyle \left(\psi(\omega)\cdot k \ \eta^\mu-\psi^\mu(\omega)k\cdot \eta\right)\right.\\
    &\qquad\qquad\left.+\int_\omega\int_{\omega^\prime} \dd(k\cdot v+\textstyle\sum_{i=1}^{n}\omega_i+\omega+\omega^\prime)\displaystyle \psi(\omega)\cdot k \ \psi^\mu(\omega^\prime)\right\} .
    \end{aligned}
\end{equation}
Depending on the number of involved $\psi(\tau)$'s we obtain three types of vertices from \eqref{eq:Sint1} and \eqref{eq:Sint2}. The first one being a vertex without any $\psi(\tau)$-field:
\begin{figure}[H] \vspace{-1cm}
\begin{align}\label{eq:vertexAZ_allOrder_with_spin}\def\arraystretch{2.4}
\begin{aligned}
  \begin{tikzpicture}[baseline={(current bounding box.center)}]
  \coordinate (in) at (-1,0);
  \coordinate (out1) at (1,0);
  \coordinate (out2) at (1,0.5);
  \coordinate (outn) at (1,1.3);
  \coordinate (x) at (0,0);
  \node (k) at (0,-1.3) {$A_{\mu}(k)$};
  \node (dots) at (0.7,1) {$\vdots$};
  \draw (out1) node [right] {$z^{\rho_1}(\omega_{\rho_1})$};
  \draw (out2) node [right] {$z^{\rho_2}(\omega_{\rho_2})$};
  \draw (outn) node [right] {$z^{\rho_n}(\omega_{\rho_n})$};
  \draw [dotted] (in) -- (x);
  \draw [zUndirected] (x) -- (out1);
  \draw [zUndirected] (x) to [out=45, in=180] (out2);
  \draw [zUndirected] (x) to [out=90, in=180] (outn);
  \draw [photon] (x) -- (k);
  \draw [fill] (x) circle (.08);
  \end{tikzpicture}\begin{array}{ll} & \\ = &\displaystyle e\ i^{n-1} e^{ik\cdot b} \ \dd\left(k\cdot v+\textstyle\sum^n_{i=1}\omega_i\right) \left(\prod^n_{i=1} k_{\rho_i} \left( v^\mu +2ik_\alpha S^{\alpha\mu}\right)\right. \\ & \displaystyle
  \hspace{4cm}\left. + \sum_{i=1}^n\omega_i \left(\prod_{i\neq j}^n k_{\rho_j}\right)\delta^\mu_{\ \rho_i}\right)\end{array}
\end{aligned}
\end{align}\end{figure}
\noindent which is similar to the vertex from the non-spinning case \eqref{eq:vertexAZ_allOrder} with $v^\mu\rightarrow v^\mu +2ik_\alpha S^{\alpha\mu}$. In \eqref{eq:vertexAZ_allOrder_with_spin} and for the two following vertices all legs are taken to be out-going.
The other two vertices arise from \eqref{eq:Sint2} and are given by
\begin{align}\label{eq:vertexAZPsi_allOrder}
\begin{aligned}
  \begin{tikzpicture}[baseline={(current bounding box.center)}]
  \coordinate (in) at (-1,0);
  \coordinate (out1) at (1,0);
  \coordinate (out2) at (1,0.5);
  \coordinate (outn) at (1,1.3);
  \coordinate (x) at (0,0);
  \node (k) at (0,-1.3) {$A_{\mu}(k)$};
  \node (dots) at (0.7,1) {$\vdots$};
  \draw (out1) node [right] {$\psi^{\rho_0}(\omega_{\rho_0})$};
  \draw (out2) node [right] {$z^{\rho_1}(\omega_{\rho_1})$};
  \draw (outn) node [right] {$z^{\rho_n}(\omega_{\rho_n})$};
  \draw [dotted] (in) -- (x);
  \draw [psiParticleDirected] (x) -- (out1);
  \draw [zUndirected] (x) to [out=45, in=180] (out2);
  \draw [zUndirected] (x) to [out=90, in=180] (outn);
  \draw [photon] (x) -- (k);
  \draw [fill] (x) circle (.08);
  \end{tikzpicture}  = &\displaystyle -2e\ i^{n-1} e^{ik\cdot b} \ \dd\left(k\cdot v+\textstyle\sum^n_{i=0}\omega_i\right) \prod^n_{i=1} k_{\rho_i} \Bigl(k_{\rho_0}\eta^\mu - k\cdot\eta \ \delta^\mu_{\ \rho_0}\Bigr)
\end{aligned}
\end{align}
and
\begin{align}\label{eq:vertexAZ2Psi_allOrder}
\begin{aligned}
  \begin{tikzpicture}[baseline={(current bounding box.center)}]
  \coordinate (in) at (-1,0);
  \coordinate (out-1) at (1,0);
  \coordinate (out0) at (1,0.5);
  \coordinate (out1) at (1,1);
  \coordinate (outn) at (1,1.7);
  \coordinate (x) at (0,0);
  \node (k) at (0,-1.3) {$A_{\mu}(k)$};
  \node (dots) at (0.7,1.4) {$\vdots$};
  \draw (out-1) node [right] {$\psi^{\rho_{-1}}(\omega_{\rho_{-1}})$};
  \draw (out0) node [right] {$\psi^{\rho_0}(\omega_{\rho_0})$};
  \draw (out1) node [right] {$z^{\rho_1}(\omega_{\rho_1})$};
  \draw (outn) node [right] {$z^{\rho_n}(\omega_{\rho_n})$};
  \draw [dotted] (in) -- (x);
  \draw [psiParticleDirected] (x) -- (out-1);
  \draw [psiParticleDirected] (x) to [out=35, in=180] (out0);
  \draw [zUndirected] (x) to [out=70, in=180] (out1);
  \draw [zUndirected] (x) to [out=90, in=180] (outn);
  \draw [photon] (x) -- (k);
  \draw [fill] (x) circle (.08);
  \end{tikzpicture}  = &\displaystyle \ 2e\ i^{n-1} e^{ik\cdot b} \ \dd\left(k\cdot v+\textstyle\sum^n_{i=-1}\omega_i\right) \prod^n_{i=1} k_{\rho_i} \Bigl(k_{\rho_0}\delta^\mu_{ \ \rho_{-1}} - k_{\rho_{-1}} \ \delta^\mu_{\ \rho_0}\Bigr) .
\end{aligned}
\end{align}
The $\psi$-lines are equipped with an arrow because the order of the anti-commuting field $\psi^\mu$ and the Grassmann vector $\eta^\mu$ matters when vertices with $\psi$-legs are connected and Wick's theorem is applied. To take into account the correct order of the anti-commuting variables we only connect $\psi$-legs with out-pointing arrows with $\psi$-legs with in-pointing arrows. The vertices with the opposite order of the anti-commuting variables are given by:
\begin{equation}
   \begin{tikzpicture}[baseline]
  \coordinate (in) at (-1,0);
  \coordinate (out1) at (1,0);
  \coordinate (out2) at (1,0.5);
  \coordinate (outn) at (1,1.3);
  \coordinate (x) at (0,0);
  \node (k) at (0,-1.3) {$A_{\mu}(k)$};
  \node (dots) at (0.7,1) {$\vdots$};
  \draw (out1) node [right] {$\psi^{\rho_0}$};
  \draw (out2) node [right] {$z^{\rho_1}$};
  \draw (outn) node [right] {$z^{\rho_n}$};
  \draw [dotted] (in) -- (x);
  \draw [psiParticleDirected] (x) -- (out1);
  \draw [zUndirected] (x) to [out=45, in=180] (out2);
  \draw [zUndirected] (x) to [out=90, in=180] (outn);
  \draw [photon] (x) -- (k);
  \draw [fill] (x) circle (.08);
  \end{tikzpicture}  = \ - \ \ \begin{tikzpicture}[baseline]
  \coordinate (in) at (-1,0);
  \coordinate (out1) at (1,0);
  \coordinate (out2) at (1,0.5);
  \coordinate (outn) at (1,1.3);
  \coordinate (x) at (0,0);
  \node (k) at (0,-1.3) {$A_{\mu}(k)$};
  \node (dots) at (0.7,1) {$\vdots$};
  \draw (out1) node [right] {$\psi^{\rho_0}$};
  \draw (out2) node [right] {$z^{\rho_1}$};
  \draw (outn) node [right] {$z^{\rho_n}$};
  \draw [dotted] (in) -- (x);
  \draw [psiParticleDirected] (out1) -- (x);
  \draw [zUndirected] (x) to [out=45, in=180] (out2);
  \draw [zUndirected] (x) to [out=90, in=180] (outn);
  \draw [photon] (x) -- (k);
  \draw [fill] (x) circle (.08);
  \end{tikzpicture} \hspace{1.35cm}
   \begin{tikzpicture}[baseline]
  \coordinate (in) at (-1,0);
  \coordinate (out-1) at (1,0);
  \coordinate (out0) at (1,0.5);
  \coordinate (out1) at (1,1);
  \coordinate (outn) at (1,1.7);
  \coordinate (x) at (0,0);
  \node (k) at (0,-1.3) {$A_{\mu}(k)$};
  \node (dots) at (0.7,1.4) {$\vdots$};
  \draw (out-1) node [right] {$\psi^{\rho_{-1}}$};
  \draw (out0) node [right] {$\psi^{\rho_0}$};
  \draw (out1) node [right] {$z^{\rho_1}$};
  \draw (outn) node [right] {$z^{\rho_n}$};
  \draw [dotted] (in) -- (x);
  \draw [psiParticleDirected] (x) -- (out-1);
  \draw [psiParticleDirected] (x) to [out=35, in=180] (out0);
  \draw [zUndirected] (x) to [out=70, in=180] (out1);
  \draw [zUndirected] (x) to [out=90, in=180] (outn);
  \draw [photon] (x) -- (k);
  \draw [fill] (x) circle (.08);
  \end{tikzpicture} = \ - \ \  
  \begin{tikzpicture}[baseline]
  \coordinate (in) at (-1,0);
  \coordinate (out-1) at (1,0);
  \coordinate (out0) at (1,0.5);
  \coordinate (out1) at (1,1);
  \coordinate (outn) at (1,1.7);
  \coordinate (x) at (0,0);
  \node (k) at (0,-1.3) {$A_{\mu}(k)$};
  \node (dots) at (0.7,1.4) {$\vdots$};
  \draw (out-1) node [right] {$\psi^{\rho_{-1}}$};
  \draw (out0) node [right] {$\psi^{\rho_0}$};
  \draw (out1) node [right] {$z^{\rho_1}$};
  \draw (outn) node [right] {$z^{\rho_n}$};
  \draw [dotted] (in) -- (x);
  \draw [psiParticleDirected] (out-1) -- (x);
  \draw [psiParticleDirected] (out0) to [out=180, in=35] (x);
  \draw [zUndirected] (x) to [out=70, in=180] (out1);
  \draw [zUndirected] (x) to [out=90, in=180] (outn);
  \draw [photon] (x) -- (k);
  \draw [fill] (x) circle (.08);
  \end{tikzpicture}  .
\end{equation}

\subsection{One and two photon radiation from a single worldline}
With the set of Feynman rules established in the previous subsection, we can now build diagrams in the spin-1/2 WQFT. As in chapter 1 we will compare the here introduced spin-1/2 WQFT to QED by comparing dressed propagators in both theories later on in this chapter. For this, it will be enlightening to work out the one and two photon radiation diagrams in the WQFT (the first two diagrams of the dressed propagator analogue). In the course of that, we will again use the time-symmetric version of the worldline propagators given by the average of the retarded and advanced worldline  propagators:

\begin{align}\label{eq:propagatorzpsi_averaged}
    \begin{tikzpicture}[baseline]
    \coordinate (in) at (-1,0);
    \coordinate (out) at (1,0);
    \coordinate (x) at (-0.5,0);
    \coordinate (y) at (0.5,0);
    \draw [dotted] (in) -- (x);
    \draw [dotted] (y) -- (out);
    \draw [zUndirected] (x) -- (y) node [midway, below] {$\omega$};
    \draw [fill] (x) circle (.08) node [above]{$\mu$};
    \draw [fill] (y) circle (.08) node [above]{$\nu$};
    \end{tikzpicture}=-i\eta^{\mu\nu}\left(\frac{1}{(\omega+i\epsilon)^2}+\frac{1}{(\omega-i\epsilon)^2}\right), \hspace{0.2cm}
      \begin{tikzpicture}[baseline]
    \coordinate (in) at (-1,0);
    \coordinate (out) at (1,0);
    \coordinate (x) at (-0.5,0);
    \coordinate (y) at (0.5,0);
    \draw [dotted] (in) -- (x);
    \draw [dotted] (y) -- (out);
    \draw [psiParticle] (x) -- (y) node [midway, below] {$\omega$};
    \draw [fill] (x) circle (.08) node [above]{$\nu$};
    \draw [fill] (y) circle (.08) node [above]{$\mu$};
    \end{tikzpicture}=-i\frac{\eta^{\mu\nu}}{2}\left(\frac{1}{\omega+i\epsilon}+\frac{1}{\omega-i\epsilon}\right).
\end{align}

\subsubsection{One photon radiation}
The one photon radiation from a single worldline is just given by the zeroth order of \eqref{eq:vertexAZ_allOrder_with_spin}:
\begin{align}\label{eq:WQFT1photon_with_spin}
	  \begin{tikzpicture}[baseline={(current bounding box.center)}]
	  \coordinate (in) at (-1,0);
	  \coordinate (out) at (1,0);
	  \coordinate (x) at (0,0);
	  \node (k) at (0,-1.3) {$\mu$};
	  \draw [dotted] (in) -- (x);
	  \draw [dotted] (x) -- (out);
	  \draw [photon] (x) -- (k) node [midway, right] {$k$};
	  \draw [fill] (x) circle (.08);
	  \end{tikzpicture}=-ie \ e^{ik\cdot b}\dd(k\cdot v)\left(v^\mu +2i k_\alpha \ S^{\alpha\mu} \right),
\end{align}
where \eqref{eq:definition_backround_spin_tensor} was used to identify the $\eta^\mu$'s with the spin tensor.
\subsubsection{Two photon radiation} 
There are three diagrams contributing to the two photon radiation from a single worldline, one coming from each vertex type. The one without any propagating $\psi$-field is the same as in the non-spinning case with $v^\mu\rightarrow v^\mu +2ik_\alpha S^{\alpha\mu}$, that is why a detailed discussion is disregarded here. However, the result is given by
\begin{figure}[H]
\vspace{-1.6cm}
\begin{equation}\def\arraystretch{1.5}
\begin{aligned}\label{WQFT_4pt_diagramm1_with_spin}
  \begin{tikzpicture}[baseline]
  \coordinate (in) at (-2,0);
  \coordinate (out) at (2,0);
  \coordinate (x) at (-1,0);
  \coordinate (y) at (1,0);
  \coordinate (top) at (0,1.5);
  \node (second) at (0,0.5) {$\underrightarrow{\omega_2}$};
  \node (topw) at (0,1.8) {$\underrightarrow{\omega_n}$};
  \node (k1) at (-1,-1.5) {$\mu_1$};
  \node (k2) at (1,-1.5) {$\mu_2$};
  \node (dots) at (0,1.2) {$\vdots$};
  \draw (out) node [right] {};
  \draw [dotted] (in) -- (x);
  \draw [zUndirected] (x) -- (y) node [midway, below] {$\overrightarrow{\omega_1}$};
  \draw [zUndirected] (x) to [out=40, in=200] (second) to [out=-20, in=140] (y);
  \draw [zUndirected] (x) to [out=90, in=180] (top)  to [out=0, in=90]  (y);
  \draw [dotted] (y) -- (out);
  \draw [photon] (x) -- (k1) node [midway, right]{$k_1$};
  \draw [photon] (y) -- (k2) node [midway, right]{$k_2$};
  \draw [fill] (x) circle (.08);
  \draw [fill] (y) circle (.08);
  \end{tikzpicture}& \begin{array}{ll} & \\ & \\ & \\= & (-ie)^2 i \ e^{i(k_1+k_2)\cdot b} \ \dd((k_1+k_2\cdot v) \biggl[ \ -  \ 2 \eta_{\mu_1\mu_2} \ \delta_{1n}  \biggr.\\ 
& \ + \ (k_1\cdot k_2)^n \ \boldsymbol{P^{n+1}_+} \ \left(v_{\mu_2}+2i(k_2\cdot S)_{\mu_2}\right)\left(v_{\mu_1}+2i(k_1\cdot S)_{\mu_1}\right)  \\ 
& \ + \ \biggl. (k_1\cdot k_2)^{n-1} \ \boldsymbol{P^{n}_-} \ \bigl(k_{2\mu_1}(v_{\mu_2} + 2i (k_2\cdot S)_{\mu_2}))- (1 \leftrightarrow 2)\bigr) \ \delta_{n \geq 1} \\ & \ - \ (k_1\cdot k_2)^{n-2} \ \boldsymbol{P^{n-1}_+} \ k_{1\mu_2}k_{2\mu_1}\ \delta_{n\geq2}\biggr] ,
  \end{array}
\end{aligned}
\end{equation}
\end{figure}
\noindent where $(k\cdot S)^\mu = -(S\cdot k)^\mu :=k_\nu S^{\nu\mu}$ and
\begin{equation}
    \boldsymbol{P^n_\pm}(\omega) :=\left(\frac{1}{(\omega + i \epsilon )^n} \pm \frac{(-1)^n}{(\omega - i \epsilon )^n}\right) .
\end{equation}
 From vertex \eqref{eq:vertexAZPsi_allOrder} we can construct diagrams with one internal $\psi$-line:
\begin{figure}[H] \vspace{-1.6cm}
\begin{equation}\def\arraystretch{2}
\begin{aligned}\label{WQFT_4pt_diagramm2_with_spin}
  \begin{tikzpicture}[baseline]
  \coordinate (in) at (-2,0);
  \coordinate (out) at (2,0);
  \coordinate (x) at (-1,0);
  \coordinate (y) at (1,0);
  \coordinate (top) at (0,1.5);
  \node (second) at (0,0.5) {$\underrightarrow{\omega_1}$};
  \node (topw) at (0,1.8) {$\underrightarrow{\omega_n}$};
  \node (k1) at (-1,-1.5) {$\mu_1$};
  \node (k2) at (1,-1.5) {$\mu_2$};
  \node (dots) at (0,1.2) {$\vdots$};
  \draw (out) node [right] {};
  \draw [dotted] (in) -- (x);
  \draw [psiParticleDirected] (x) -- (y) node [midway, below] {$\overrightarrow{\omega_0}$};
  \draw [zUndirected] (x) to [out=40, in=200] (second) to [out=-20, in=140] (y);
  \draw [zUndirected] (x) to [out=90, in=180] (top)  to [out=0, in=90]  (y);
  \draw [dotted] (y) -- (out);
  \draw [photon] (x) -- (k1) node [midway, right]{$k_1$};
  \draw [photon] (y) -- (k2) node [midway, right]{$k_2$};
  \draw [fill] (x) circle (.08);
  \draw [fill] (y) circle (.08);
  \end{tikzpicture}& \begin{array}{ll} & \\ &  \\=& -\displaystyle\frac{2}{n!} e^2 i^{n-1} e^{i(k_1+k_2)\cdot b} \textstyle \ \dd\left((k_1+k_2)\cdot v\right)\\
  &\times\displaystyle\bigintsss_{\omega_0, \dots, \omega_n} \hspace{-0.4cm}\textstyle\dd\left(k_2\cdot v -\sum_{i=0}^n\omega_i\right)\displaystyle\prod_{i=1}^n D^2(\omega_i)D^1(\omega_0) \ (k_1\cdot k_2)^n \\
  &\times\bigl(k_{2{\rho_0}} \eta_{\mu_2}-k_2\cdot\eta \ \eta_{\mu_2\rho_0}\bigr)\bigl(k_1^{\rho_0} \eta_{\mu_1}-k_1\cdot\eta \  \delta^{\rho_0}_{\ \mu_1}\bigr) .
  \end{array}
\end{aligned}
\end{equation}
\end{figure}
\noindent and from vertex \eqref{eq:vertexAZ2Psi_allOrder} diagrams with two internal $\psi$-lines:
\begin{figure}[H] \vspace{-.6cm}
\begin{equation}\def\arraystretch{2}
\begin{aligned}\label{WQFT_4pt_diagramm3_with_spin}
  \begin{tikzpicture}[baseline]
  \coordinate (in) at (-2,0);
  \coordinate (out) at (2,0);
  \coordinate (x) at (-1,0);
  \coordinate (y) at (1,0);
  \coordinate (top) at (0,2.5);
  \coordinate (second) at (0,0.5);
  \node (third) at (0,1.5) {$\underrightarrow{\omega_1}$};
  \node (topw) at (0,2.8) {$\underrightarrow{\omega_n}$};
  \node (k1) at (-1,-1.5) {$\mu_1$};
  \node (k2) at (1,-1.5) {$\mu_2$};
  \node (dots) at (0,2.2) {$\vdots$};
  \draw (out) node [right] {};
  \draw [dotted] (in) -- (x);
  \draw [psiParticleDirected] (x) -- (y) node [midway, below] {$\overrightarrow{\omega_{-1}}$};
  \draw [psiParticleDirected] (x) to [out=40, in=180] (second) node [above] {$\underrightarrow{\omega_0}$} to [out=0, in=140] (y);
  \draw [zUndirected] (x) to [out=80, in=200] (third) to [out=-20, in=100] (y);
  \draw [zUndirected] (x) to [out=90, in=180] (top)  to [out=0, in=90]  (y);
  \draw [dotted] (y) -- (out);
  \draw [photon] (x) -- (k1) node [midway, right]{$k_1$};
  \draw [photon] (y) -- (k2) node [midway, right]{$k_2$};
  \draw [fill] (x) circle (.08);
  \draw [fill] (y) circle (.08);
  \end{tikzpicture}& \begin{array}{ll} & \\ &  \\=& -\displaystyle\frac{2}{n!} \  e^2 \ i^{n} \ e^{i(k_1+k_2)\cdot b} \textstyle \ \dd\left((k_1+k_2)\cdot v\right)\\
  &\times\displaystyle\bigintsss_{\omega_{-1}, \omega_0, \dots, \omega_n} \hspace{-0.9cm}\textstyle\dd\left(k_2\cdot v -\sum_{i=-1}^n\omega_i\right)\displaystyle\prod_{i=1}^n D^2(\omega_i) \ \frac{D^1(\omega_0)}{2} \ \frac{D^1(\omega_{-1})}{2}  \\
  &\times(k_1\cdot k_2)^n\bigl(k_{2{\rho_0}} \eta_{\mu_2\rho_{-1}} - k_{2\rho_{-1}} \eta_{\mu_2\rho_0}\bigr) \bigl( k_1^{\rho_0} \delta^{\rho_{-1}}_{\ \mu_1} - k_1^{\rho_{-1}} \delta^{\rho_0}_{\ \mu_1}\bigr) .
  \end{array}
\end{aligned}
\end{equation}
\end{figure}
\noindent The integrals in these two diagrams are the same as in the non-spinning case and were already solved in appendix \ref{appendix:worldline_integrals}. With the master formulas from appendix \ref{appendix:worldline_integrals} the solutions yield
\begin{equation}
    \begin{aligned}
    \eqref{WQFT_4pt_diagramm2_with_spin} = \ &  -(-ie)^2 \ 2 \ e^{i (k_1+k_2)\cdot b} \ \dd ((k_1+k_2)\cdot v) \ (k_1\cdot k_2)^n \ \boldsymbol{P^{n+1}_-} \\ & \qquad\times \biggl[k_1\cdot k_2 S_{\mu_2\mu_1} + k_2\cdot S\cdot k_1 \eta_{\mu_1\mu_2} - k_{2\mu_1}(S\cdot k_1)_{\mu_2} + k_{1\mu_2}(S\cdot k_2)_{\mu_1} \biggr]
    \end{aligned}
\end{equation}
and 
\begin{equation}
    \begin{aligned}
    \eqref{WQFT_4pt_diagramm3_with_spin} = \ & - (-ie)^2 e^{i (k_1+k_2)\cdot b} \ \dd ((k_1+k_2)\cdot v) \ (k_1\cdot k_2)^n \ \boldsymbol{P^{n+1}_+} \ \bigl( k_1\cdot k_2 \eta_{\mu_1\mu_2} - k_{2\mu_1} k_{1\mu_2} \bigr) .
    \end{aligned}
\end{equation}
As in the first chapter, the 3 diagrams can now be summed up to all orders:
\begin{figure}[H] \vspace{-1.8cm}
\begin{equation} \label{eq:WQFT_4pt_diagram1_with_spin_summed_up} \def\arraystretch{2}
    \begin{aligned}\hspace{-.8cm}
   \mathlarger{\mathlarger{\mathlarger{\sum}}}_{n=0}^\infty\ \ 
    \begin{tikzpicture}[baseline]
  \coordinate (in) at (-1.2,0);
  \coordinate (out) at (1.2,0);
  \coordinate (x) at (-0.6,0);
  \coordinate (y) at (0.6,0);
  \coordinate (third) at (0,0.5);
  \coordinate (fourth) at (0,0.8);
  \coordinate (fifth) at (0,1.4);
  \node (n) at (0,1.6) {$n$};
  \node (dots) at (0,1.2) {$\vdots$};
  \node (k1) at (-0.6,-1) {};
  \node (k2) at (0.6,-1) {};
  \draw [dotted] (in) -- (out);
  \draw [photon] (x) -- (k1);
  \draw [photon] (y) -- (k2);
  \draw [fill] (x) circle (.06);
  \draw [fill] (y) circle (.06);
  \draw [zUndirected] (x) to [out=50, in=130] (y);
  \draw [zUndirected] (x) to [out=70, in=180] (third) to [out=0, in=110] (y);
  \draw [zUndirected] (x) to [out=80, in=180] (fourth) to [out=0, in=100] (y);
  \draw [zUndirected] (x) to [out=90, in=180] (fifth) to [out=0, in=90] (y);
  \end{tikzpicture} \begin{array}{ll} & \\ & \\ =& (-ie)^2 \ i \ e^{i(k_1+k_2)\cdot b} \ \dd((k_1+k_2)\cdot v) \ \biggl[ \ - 2 \eta_{\mu_1\mu_2}  \biggr.\\
  & \hspace{-.8cm}\displaystyle + \sum_{n=0}^\infty  (k_1\cdot k_2)^n\boldsymbol{P^{n+1}_+} \ \biggl(\left(v_{\mu_2}+2i(k_2\cdot S)_{\mu_2}\right)\left(v_{\mu_1}+2i(k_1\cdot S)_{\mu_1}\right)-k_{1\mu_2}k_{2\mu_1}\biggr) \\
  & \hspace{-.8cm}\displaystyle  - \sum_{n=0}^n (k_1\cdot k_2)^n\boldsymbol{P^{n+1}_-}\biggl( k_{1\mu_2}\left(v_{\mu_1}+2i(k_1\cdot S)_{\mu_1}\right)-k_{2\mu_1}\left(v_{\mu_2}+2i(k_2\cdot S)_{\mu_2}\right)\biggr) \biggl. \biggr] , \end{array} 
    \end{aligned}
\end{equation}\end{figure}

\begin{figure}[H] \vspace{-1.2cm}
\begin{equation} \label{eq:WQFT_4pt_diagram2_with_spin_summed_up} \def\arraystretch{2}
    \begin{aligned}\hspace{-.8cm}
   \mathlarger{\mathlarger{\mathlarger{\sum}}}_{n=0}^\infty\ \ 
    \begin{tikzpicture}[baseline]
  \coordinate (in) at (-1.2,0);
  \coordinate (out) at (1.2,0);
  \coordinate (x) at (-0.6,0);
  \coordinate (y) at (0.6,0);
  \coordinate (third) at (0,0.5);
  \coordinate (fourth) at (0,0.8);
  \coordinate (fifth) at (0,1.4);
  \node (n) at (0,1.6) {$n$};
  \node (dots) at (0,1.2) {$\vdots$};
  \node (k1) at (-0.6,-1) {};
  \node (k2) at (0.6,-1) {};
  \draw [dotted] (in) -- (x);
  \draw [dotted] (y) -- (out);
  \draw [photon] (x) -- (k1);
  \draw [photon] (y) -- (k2);
  \draw [fill] (x) circle (.06);
  \draw [fill] (y) circle (.06);
  \draw [psiParticleDirected] (x) -- (y);
  \draw [zUndirected] (x) to [out=50, in=130] (y);
  \draw [zUndirected] (x) to [out=70, in=180] (third) to [out=0, in=110] (y);
  \draw [zUndirected] (x) to [out=80, in=180] (fourth) to [out=0, in=100] (y);
  \draw [zUndirected] (x) to [out=90, in=180] (fifth) to [out=0, in=90] (y);
  \end{tikzpicture} \begin{array}{ll} & \\ =& -(-ie)^2 \ 2 \ e^{i(k_1+k_2)\cdot b} \ \dd((k_1+k_2)\cdot v) \ \displaystyle \sum_{n=0}^\infty \ (k_1\cdot k_2)^n \ \boldsymbol{P^{n+1}_-} \\
  &  \times \biggl[ k_1\cdot k_2 S_{\mu_2\mu_1} + k_2\cdot S \cdot k_1 \eta_{\mu_1\mu_2} - k_{2\mu_1} (S\cdot k_1)_{\mu_2} + k_{1\mu_2} (S\cdot k_2)_{\mu_1} \biggr] , \end{array} 
    \end{aligned}
\end{equation}\end{figure}
\begin{figure}[H] \vspace{-0.8cm}
\begin{equation} \label{eq:WQFT_4pt_diagram3_with_spin_summed_up} \def\arraystretch{2}
    \begin{aligned}\hspace{-1.2cm}
   \mathlarger{\mathlarger{\mathlarger{\sum}}}_{n=0}^\infty\ \ 
    \begin{tikzpicture}[baseline]
  \coordinate (in) at (-1.2,0);
  \coordinate (out) at (1.2,0);
  \coordinate (x) at (-0.6,0);
  \coordinate (y) at (0.6,0);
  \coordinate (third) at (0,0.5);
  \coordinate (fourth) at (0,0.8);
  \coordinate (fifth) at (0,1.4);
  \node (n) at (0,1.6) {$n$};
  \node (dots) at (0,1.2) {$\vdots$};
  \node (k1) at (-0.6,-1) {};
  \node (k2) at (0.6,-1) {};
  \draw [dotted] (in) -- (x);
  \draw [dotted] (y) -- (out);
  \draw [photon] (x) -- (k1);
  \draw [photon] (y) -- (k2);
  \draw [fill] (x) circle (.06);
  \draw [fill] (y) circle (.06);
  \draw [psiParticleDirected] (x) -- (y);
  \draw [psiParticleDirected] (x) to [out=50, in=130] (y);
  \draw [zUndirected] (x) to [out=70, in=180] (third) to [out=0, in=110] (y);
  \draw [zUndirected] (x) to [out=80, in=180] (fourth) to [out=0, in=100] (y);
  \draw [zUndirected] (x) to [out=90, in=180] (fifth) to [out=0, in=90] (y);
  \end{tikzpicture} \begin{array}{ll} & \\ =& -i (-ie)^2 \ 2 \ e^{i(k_1+k_2)\cdot b} \ \dd((k_1+k_2)\cdot v) \ \displaystyle \sum_{n=0}^\infty \ (k_1\cdot k_2)^n \ \boldsymbol{P^{n+1}_+} \\
  &  \qquad\times \biggl[ k_1\cdot k_2 \eta_{\mu_2\mu_1}  - k_{2\mu_1} k_{1\mu_2} \biggr] . \end{array} 
    \end{aligned}
\end{equation}\end{figure}
\noindent Finally, adding up the three diagram types results in:
\begin{equation}\label{eq:WQFT_4pt_with_spin_All_summed_up}
    \begin{aligned}
   & \eqref{eq:WQFT_4pt_diagram1_with_spin_summed_up} + \eqref{eq:WQFT_4pt_diagram2_with_spin_summed_up} + \eqref{eq:WQFT_4pt_diagram3_with_spin_summed_up} = (-ie)^2 \ i \ e^{i(k_1+k_2)\cdot b} \ \dd((k_1+k_2)\cdot v) \Biggl[ -2 \eta_{\mu_1\mu_2} \Biggr. \\ & \qquad\qquad
    + \sum_{n=0}^\infty \ (k_1\cdot k_2)^n \boldsymbol{P^{n+1}_+} \biggl(\left(v_{\mu_2}+2i(k_2\cdot S)_{\mu_2}\right)\left(v_{\mu_1}+2i(k_1\cdot S)_{\mu_1}\right)-k_1\cdot k_2 \eta_{\mu_2\mu_1}\biggr) \\ & \qquad\qquad
     - \sum_{n=0}^n \ (k_1\cdot k_2)^n \ \boldsymbol{P^{n+1}_-}\biggl( k_{1\mu_2}\left(v_{\mu_1}+2i(k_1\cdot S)_{\mu_1}\right)-k_{2\mu_1}\left(v_{\mu_2}+2i(k_2\cdot S)_{\mu_2}\right) \biggr. \\ & \qquad\qquad\qquad\qquad
     \biggl. -2i\Bigl( k_1\cdot k_2 S_{\mu_2\mu_1} + k_2\cdot S \cdot k_1 \eta_{\mu_1\mu_2} - k_{2\mu_1} (S\cdot k_1)_{\mu_2} + k_{1\mu_2} (S\cdot k_2)_{\mu_1} \Bigr)\biggr)\Biggl. \Biggr] .
    \end{aligned}
\end{equation}
Note that the bracket multiplied with $\boldsymbol{P^{n+1}_+}$ is symmetric in $(1 \leftrightarrow 2)$ and the bracket multiplied with $\boldsymbol{P^{n+1}_-}$ is anti-symmetric under the exchange of photon one and two. This separation into symmetric and anti-symmetric terms occurred already in the non-spinning case, and also sorting the scalar QED expression into symmetric and anti-symmetric parts gave us a matching expression when expanded into a Taylor series for small photon momenta. Interestingly, the same Taylor expansion appears in \eqref{eq:WQFT_4pt_with_spin_All_summed_up}. Recalling the Taylor expansion \eqref{eq:taylor_expansion_internal_propagator1} and \eqref{eq:taylor_expansion_internal_propagator2} of the internal scalar propagators of a scalar QED 4pt function we can write:
\begin{equation}
    \sum_{n=0}^n \ (k_1\cdot k_2)^n \ \boldsymbol{P^{n+1} _\pm} =   \frac{1}{(p^\prime+k_1)^2-m^2+i\epsilon} \pm\frac{1}{(p^\prime+k_2)^2-m^2+i\epsilon}
\end{equation}
if we impose $v^\mu = (p+ p^\prime )^\mu$. Thus, \eqref{eq:WQFT_4pt_with_spin_All_summed_up} can be written as
\begin{equation}\label{eq:WQFT_4pt_in_terms_of_scalar_propagators}
    \begin{aligned}
   & \eqref{eq:WQFT_4pt_with_spin_All_summed_up} = (-ie)^2 \ e^{i(k_1+k_2)\cdot b} \ \dd((k_1+k_2)\cdot v)  \\ &  \qquad\times \Biggl\{ -2i \eta_{\mu_1\mu_2} \Biggr.
    + \Biggl[ \Biggr.\frac{1}{(p^\prime+k_1)^2-m^2+i\epsilon} \Biggl(\left(v_{\mu_2}+2i(k_2\cdot S)_{\mu_2}\right)\left(v_{\mu_1}+2i(k_1\cdot S)_{\mu_1}\right)\Biggr. \\ & \qquad\qquad
     -k_1\cdot k_2 \eta_{\mu_2\mu_1} + k_{1\mu_2}\left(v_{\mu_1}+2i(k_1\cdot S)_{\mu_1}\right)-k_{2\mu_1}\left(v_{\mu_2}+2i(k_2\cdot S)_{\mu_2}\right) \\ & \qquad\qquad
      -2i\Bigl( k_1\cdot k_2 S_{\mu_2\mu_1} + k_2\cdot S \cdot k_1 \eta_{\mu_1\mu_2} - k_{2\mu_1} (S\cdot k_1)_{\mu_2} + k_{1\mu_2} (S\cdot k_2)_{\mu_1} \Bigr)\Biggl. \Biggr) \Biggl. +  (1 \leftrightarrow 2) \Biggr]\Biggl. \Biggr\} .
    \end{aligned}
\end{equation}
where the (anti-)symmetry of the respective terms in \eqref{eq:WQFT_4pt_with_spin_All_summed_up} was exploited. With \eqref{eq:WQFT_4pt_in_terms_of_scalar_propagators} we found an expression that looks much more like a modified scalar QED than actual QED. Most notable is the appearance of scalar propagators instead of fermionic propagators, as one would may have expected for a theory aiming to describe spin-1/2 particles.

After the detailed study of the here established WQFT for spin-1/2 particles in this chapter so far, the next section will be dedicated to the study of dressed propagators in actual QED.

\section{Electron propagator in an abelian background field}\label{chapter:elec_prop}
In this section, we want to derive a worldline representation (Feynman-Schwinger representation)  of the fermionic propagator in an abelian background field, as we did for the scalar case in chapter 1. Again we are aiming for a path integral representation which is akin to the WQFT expression and which we will use to find a link between the WQFT describing a spin-1/2 particle and QED similar to the one in equation \eqref{link WQFT and scalar QED propagator}. The procedure will be very alike to the scalar case, hence some steps in the derivation will be shortened a bit. The approach in this section is based on a similar derivation in \cite{Schubert_2020} and reproduces their results.

In order to find the appropriate Green's function corresponding to the dressed fermionic propagator, we can again look at the action describing a fermion coupled to an electromagnetic field. If we treat the photon field as a constant background, the kinematic part of the QED-lagrangian \eqref{actionQED} given in appendix \ref{apendix:FeynmanRules} is $\overline{\psi}\left(i\slashed D-m\right)\psi$. The dressed propagator is the Green's function to the kinematic operator, i.e. it has to satisfy
\begin{equation}
    \left(i\slashed D -m\right) S_A (x,x^\prime ) = -i \delta^{(D)}(x-x^\prime ),
\end{equation}
where $D_\mu=\partial_\mu+ieA_\mu$. Thus, the position space expression for the dressed fermionic propagator is
\begin{equation}
   S^{x^\prime x}[A] =\left\langle x^\prime \left|-i \left[i\slashed D-m\right]^{-1}\right|x\right\rangle .
\end{equation}
The term in the square brackets can be rewritten as
\begin{equation}
\begin{aligned}
\left[i\slashed{D}-m\right]^{-1} &=[i \slashed D+m][i \slashed D+m]^{-1}[i \slashed D-m]^{-1} \\
&=-\left[i \slashed D+m\right]\left[m^{2}+D_{\mu} D^{\mu}+\frac{i}{2} e \gamma^{\mu} \gamma^{\nu} F_{\mu \nu}\right]^{-1} ,
\end{aligned}
\label{inv dirac rewritten}
\end{equation}
where the Gordon identity
\begin{equation}
\slashed D^{2}=D_{\mu} D^{\mu}+\frac{i}{2} e \gamma^{\mu} \gamma^{\nu} F_{\mu \nu}
\end{equation}
was used in the last line. Note that we just as well could have inserted the one factor on the right side, which would just switch the positions of the two brackets in the second line of \eqref{inv dirac rewritten}. From now on, we will focus on finding a representation of the second bracket in \eqref{inv dirac rewritten}, which we refer to as the \textit{kernel} of the dressed electron propagator and keep in mind that we have to multiply the kernel by a factor of $-\left[i\slashed{D}+m\right]$ in order to get the propagator. Consequently, the object we are looking for, namely the kernel of the photon dressed electron propagator, is
\begin{equation}
    K^{x^\prime x}[A]=\left\langle x^\prime \left|-i \left[m^{2}+D_{\mu} D^{\mu}+\frac{i}{2} e \gamma^{\mu} \gamma^{\nu} F_{\mu \nu}\right]^{-1}\right|x\right\rangle .
\end{equation}
The kernel looks exactly like the scalar propagator apart from having an additional term $\frac{i}{2} e \gamma^{\mu} \gamma^{\nu} F_{\mu \nu}$.
Using operator notation $\partial_\mu\rightarrow -i\hat{p}_\mu$ and introducing a Schwinger-proper-time parameter exactly as in the scalar case, the kernel can be written as
\begin{equation}
\label{kernel_schwinger}
     K^{x^\prime x}[A]=\int_0^\infty \mathrm{d}T e^{-iTm^2} \langle x^\prime| e^{-iT\hat{H}}|x\rangle
\end{equation}
but this time with 
\begin{equation}
    \hat{H}=-(\hat{p}-eA(\hat{x})^2+\frac{i}{2}e\gamma^\mu\gamma^\nu F_{\mu\nu}(\hat{x}).
\end{equation}
Once again, we can write the transition amplitude in \eqref{kernel_schwinger} as a quantum mechanical path integral
\begin{equation}
    \langle x^\prime|e^{-i\hat{H}T}|x\rangle = \int_{x(0)=x}^{x(T)=x^\prime} Dx \ \mathcal{P} \exp\left(i\int\mathrm{d}\tau\mathcal{L}(x,\dot{x})\right) ,
\end{equation}
but since the gamma matrices do not commute, including a path ordering operator $\mathcal{P}$. Legendre transforming $\hat{H}$ yields
\begin{equation}
\label{lagranian_lagrange_transf_spin}
   \mathcal{L} = \hat{p}^\mu\frac{\partial\hat{H}}{p^\mu}-\hat{H} = -(\hat{p}-eA(\hat{x}))^2-2eA(\hat{x})\cdot(\hat{p}-eA(\hat{x})-\frac{i}{2}e\gamma^\mu\gamma^\nu F_{\mu\nu}(\hat{x})
\end{equation} 
and the Heisenberg equations of motion for translations in $\tau$ allow us to write:
\begin{equation}
    \dot{\hat{x}}^\mu=\frac{\mathrm{d}\hat{x}}{d\tau}=i\left[\hat{H},\hat{x}^\mu\right]=i\left[-(\hat{p}-eA(\hat{x}))^2,\hat{x}^\mu\right]=2\left(\hat{p}-eA(\hat{x})\right)^\mu,
\end{equation}
where the additional term does not change anything since $\left[\hat{x}^\mu,\hat{x}^\nu\right]=0$. Hence, we can write the lagrangian \eqref{lagranian_lagrange_transf_spin} as
\begin{equation}
  \mathcal{L} =-\left(\frac{\dot{x}}{2}\right)^2-e\dot{x}\cdot A-\frac{i}{2}e\gamma^\mu\gamma^\nu F_{\mu\nu}. 
\end{equation}
Taken all together, we have found the following path integral expression for the kernel of the dressed propagator:
\begin{equation}
     K^{x^\prime x}[A]=\int_0^\infty \mathrm{d}T e^{-iTm^2} \int_{x(0)=x}^{x(T)=x^\prime} Dx \ \mathcal{P} e^{-i\int_0^T\mathrm{d}\tau\frac{1}{4}\dot{x}^2+e\dot{x}\cdot A+\frac{i}{2}e\gamma^\mu\gamma^\nu F_{\mu\nu}}.
\end{equation}
As shown in \cite{Fredkin_1991} and the appendix of \cite{Schubert_2020} the path ordering and the gamma matrices can be removed from the expression by using Wick's theorem and introducing Grassmann valued functions $\psi^\mu(\tau)$ and Grassman vectors $\eta^\mu$ together with the so-called \textit{symbol} map:
\begin{equation}
\begin{aligned}
    K^{x^{\prime} x}\left[A\right]=&2^{-\frac{D}{2}}\int_0^\infty \mathrm{d}T e^{-iTm^2}\int_{x(0)=x}^{x(T)=x^\prime} Dx \ \exp{\left[-i\int_{0}^{T}\mathrm{d}\tau\left( \frac{1}{4}\dot{x}^2+e\dot{x}\cdot A\right)\right]}\\
    &\times\mathrm{symb}^{-1}\int_{\psi(0)+\psi(T)=0} D\psi \ \exp{\left[\int_{0}^{T}\mathrm{d}\tau\left(\frac{1}{2}\psi_\mu \dot{\psi}^\mu -eF_{\mu\nu}\left(\psi+\eta\right)^\mu\left(\psi+\eta\right)^\nu\right)\right]}.
\end{aligned}
\label{kernel start}
\end{equation} 
(In \cite{Schubert_2020} different conventions are used. The authors of \cite{Schubert_2020} work in Euclidean spacetime. Hence, one has to perform a wick rotation $\tau\rightarrow i\tau$ and insert a factor of minus one for every spacetime index contraction to compare with their expression). This expression looks very similar to the WQFT expression suggested at the beginning of this chapter. Of course, the variable names $\psi(\tau)^\mu$ and $\eta^\mu$ were not chosen randomly as in the WQFT in the previous section. Although at this point merely introduced as a mathematical trick, we will see that they will play the same role as the Grassmann variables in the WQFT later on. 

The symbol map is given by
\begin{equation}
\operatorname{symb}\left(\gamma^{\alpha_{1} \alpha_{2} \cdots \alpha_{n}}\right) \equiv(i \sqrt{2})^{n} \eta^{\alpha_{1}} \eta^{\alpha_{2}} \ldots \eta^{\alpha_{n}}
\end{equation}
with
\begin{equation}
\gamma^{\alpha_{1} \alpha_{2} \cdots \alpha_{n}} \equiv \frac{1}{n !} \sum_{\pi \in S_{n}} \operatorname{sign}(\pi) \gamma^{\alpha_{\pi(1)}} \gamma^{\alpha_{\pi(2)}} \cdots \gamma^{\alpha_{\pi(n)}}.
\end{equation}
Thus, symb$^{-1}$ simply replaces products of $\eta$'s by anti-symmetric products of $-i\frac{\gamma^\mu}{\sqrt{2}}$. The huge advantage of the symbol map is that we do not have to be concerned with any gamma matrices in our calculation until the very end when we finally apply the inverse symbol map.
\subsection{N-photon dressed kernel}
As for the $N$-photon dressed scalar propagator, we write the gauge field $A^\mu$ as a sum of $N$ plane waves $A^\mu(x)=\sum_{i=1}^N\varepsilon_i^\mu e^{ik_i\cdot x}$ and only keep terms containing each polarisation vector linearly to get the $N$-photon dressed kernel. Additionally, expanding the path integration variable $x$ around a straight-line trajectory 
\begin{equation}
x(\tau)=x+(x^\prime -x)\frac{\tau}{T}+q(\tau) ,
\end{equation}
which implies Dirichlet boundary conditions $q(0)=q(T)=0$, we get
\begin{equation}
\begin{aligned}
K_{N}^{x^{\prime} x}\left(k_{1}, \varepsilon_{1} ; \ldots ; k_{N}, \varepsilon_{N}\right)=&(-i e)^{N} 2^{-\frac{D}{2}} \int_{0}^{\infty} \mathrm{d} T \ e^{-im^{2} T} e^{-\frac{1}{4} i \frac{\left(x-x^{\prime}\right)^{2}}{T}} \int_{q(0)=0}^{q(T)=0} D q \ e^{-i \int_{0}^{T} \mathrm{d} \tau \frac{\dot{q}^{2}}{4}} \\
& \times \mathrm{symb}^{-1} \int_{\psi(0)+\psi(T)=0} D\psi \ e^{\int_{0}^{T} \mathrm{d} \tau \frac{1}{2} \psi \dot{\psi}} \ V_{\eta}^{x^{\prime} x}\left[k_{1}, \varepsilon_{1}\right] \cdots V_{\eta}^{x^{\prime}}\left[k_{N}, \varepsilon_{N}\right]
\end{aligned}
\end{equation}
with a modified photon vertex operator
\begin{equation}\label{eq:vertex_operator_with_spin}
\begin{aligned}
V_{\eta}^{x^{\prime} x}[k, \varepsilon]=&\int_{0}^{T} \mathrm{d} \tau\left[\varepsilon \cdot\left(\frac{x^{\prime}-x}{T}+\dot{q}\right)-2 \varepsilon \cdot(\psi+\eta) k \cdot(\psi+\eta)\right] e^{i k \cdot\left(x+\left(x^{\prime}-x\right) \frac{\tau}{T}+q(\tau)\right)} \\
=&\left.\int_{0}^{T} \mathrm{d} \tau\left[\varepsilon \cdot\partial_\alpha-2 \varepsilon \cdot\partial_\theta \ k\cdot\partial_\theta\right]e^{\alpha\cdot\left(\frac{x^{\prime}-x}{T}+\dot{q}\right)+\theta\cdot\left(\psi(\tau)+\eta\right)} e^{i k \cdot\left(x+\left(x^{\prime}-x\right) \frac{\tau}{T}+q(\tau)\right)}\right|_{\alpha=\theta=0}. 
\end{aligned}
\end{equation}
In the last line, we introduced the auxiliary polarization vectors $\alpha^\mu$ and $\theta^\mu$, where the latter of which is anti-commuting. These auxiliary vectors lift all $q(\tau)$ and $\psi(\tau)$-dependence to the exponent and reduce the $\psi$-integral to a simple Gaussian form, which will simplify the evaluation of the integral later on.

The $q$-integral is the same as in the scalar case and can be solved with the Green's function $\Delta\left(\tau, \tau^{\prime}\right)$ defined in \eqref{eq:greens_function_for_scalar_integral}.
For the evaluation of the fermionic path integral over the Grassmann field $\psi$ we need a Green's function that satisfies anti-symmetric boundary conditions and $\frac{1}{2}\frac{\partial}{\partial\tau}G_F(\tau,\tau^\prime)=\delta(\tau-\tau^\prime)$. It is given by:
\begin{equation}
\label{electron kernel result}
\begin{aligned}
\left\langle \psi^\mu(\tau)\psi^\nu(\tau^\prime)\right\rangle=&-\frac{1}{2}G_F(\tau,\tau^\prime)\eta^{\mu\nu}\\
G_F(\tau,\tau^\prime)=&\ \mathrm{sign}(\tau-\tau^\prime).       
\end{aligned}
\end{equation}
Performing the $\psi$-integral, as described in appendix \ref{appendix:gaussian_grassmann}, yields
\begin{equation}
\int_{\psi(0)=\psi(T)=0} D\psi \ e^{\int_0^T\mathrm{d}\tau \frac{1}{2}\psi\cdot\dot{\psi}-\sum\limits_{i=1}^N\psi(\tau_i)\cdot\theta_i}=Z_0[\psi] \ e^{\frac{1}{4}\sum\limits_{i,j=1}^N\theta_i\cdot\theta_jG_F(\tau_i,\tau_j)}.
\end{equation}
with
\begin{equation}
    Z_0[\psi]=\int_{\psi(0)=\psi(T)=0} D\psi \ e^{\int_0^T\mathrm{d}\tau \frac{1}{2}\psi\cdot\dot{\psi}}=2^{\frac{D}{2}}.
\end{equation}
In the next step, we transform to momentum space 
\begin{equation}
K_{N}^{p^{\prime} p}\left(k_{1}, \varepsilon_{1} ; \ldots ; k_{N}, \varepsilon_{N}\right)=\int d^{D} x \int d^{D} x^{\prime} \mathrm{e}^{-i p \cdot x+i p^{\prime} \cdot x^{\prime}} K_{N}^{x^{\prime} x}\left(k_{1}, \varepsilon_{1} ; \ldots ; k_{N}, \varepsilon_{N}\right)
\end{equation}
and change the spacetime variables to
\begin{align}
    x_+=\frac{1}{2}\left(x+x^\prime\right), && x_-=x^\prime-x.
\end{align}
The $x_-$-integral is again simply a Gauss integral and the $x_+$-integral just produces a delta-function imposing momentum conservation. The above-described steps lead finally to the following result:

\begin{equation}
\begin{aligned}
K_{N}^{p^{\prime} p}\left(k_{1}, \varepsilon_{1} ; \ldots ; k_{N}, \varepsilon_{N}\right)=&(-i e)^N (2\pi)^D\delta^D\left(p-p^\prime-\sum\limits_{i=1}^{N}k_i\right)\mathrm{symb}^{-1}\int_{0}^{\infty} \mathrm{d} T \ e^{iT\left({p^\prime}^2-m^{2}\right)}\\&\left. \left(\prod\limits_{i=1}^{N}\int_0^T\mathrm{d}\tau_i \ \varepsilon_i\cdot\left[\partial_{\alpha_i}-2\partial_{\theta_i} \ k_i\cdot\partial_{\theta_i}\right]\right) 
 e^\mathrm{Exp}\right|_{\alpha_i=\theta_i=0}
\end{aligned}
\end{equation}
with
\begin{equation}
\begin{aligned}
    \mathrm{Exp}=&\sum\limits_{i=1}^N\left(ik_i\tau_i+\alpha_i\right)\cdot\left(p^\prime+p\right)-i\frac{1}{2}\sum\limits_{i,j=1}^N k_i\cdot k_j |\tau_i-\tau_j|
    -i\sum\limits_{i,j=1}^N \alpha_i\cdot\alpha_j\delta(\tau_i-\tau_j)\\&+\sum\limits_{i,j=1}^N k_i\cdot\alpha_j\mathrm{sign}(\tau_i-\tau_j)+\frac{1}{4}\sum\limits_{i,j=1}^N \theta_i\cdot\theta_j\mathrm{sign}(\tau_i-\tau_j)+\sum\limits_{i=1}^N\theta_i\cdot\eta .
\end{aligned}
\end{equation}

\subsection{Cutting of external scalar legs}
Even though the external legs of the actual propagator are fermion legs, the kernel has free scalar propagators as external legs. If we are only interested in the on-shell expression, these external legs can be stripped of as explained in the previous chapter in section \ref{sec:Cutting of external scalar legs} by extending the $\tau$-integrals to $\mathbb{R}$, dropping the $T$-integral and inserting a delta function factor $N\delta\left(\sum\limits_{i=1}^N\tau_i\right)$. The amputated photon dressed electron kernel then yields:
\begin{equation}
\begin{aligned}
 \widehat{K}_{N}^{p^{\prime} p}\left(k_{1}, \varepsilon_{1} ; \ldots ; k_{N}, \varepsilon_{N}\right)&=({p^\prime}^2-m^2) K_{N}^{p^{\prime} p} (p^2-m^2)\\
 &=(-i e)^N (2\pi)^D\delta^D\left(-p+p^\prime+\sum\limits_{i=1}^{N}\right)\mathrm{symb}^{-1}\\& \left(\prod\limits_{i=1}^{N}\int_{-\infty}^\infty\mathrm{d}\tau_i \ \varepsilon_i\cdot\left[\partial_{\alpha_i}-2\partial_{\theta_i} \ k_i\cdot\partial_{\theta_i}\right]\right) N\delta\left(\sum\limits_{i=1}^N\tau_i\right)
 e^\mathrm{Exp}  \Bigg|_{\small\begin{aligned}\alpha_i=&\theta_i=0 \\ p^2=&m^2={p^\prime}^2\end{aligned}}    
\end{aligned}
\label{form factor kernel}
\end{equation}

\subsection{From the kernel to the propagator}
As already pointed out at the beginning of this chapter, the Dirac propagator can be obtained from the kernel by
\begin{equation}
    S^{x^\prime x}[A]=-\left[i\slashed{D}^\prime+m\right]K^{x^\prime x}[A].
\end{equation}
If we Fourier transform to momentum space again and expand the gauge field $A^\mu$ as a sum of plane waves, this leads to
\begin{equation}
\begin{aligned}
S_{N}^{p^{\prime} p}\left[\varepsilon_{1}, k_{1} ; \ldots ; \varepsilon_{N}, k_{N}\right]=&-\left(\slashed p^{\prime}+m\right) K_{N}^{p^{\prime} p}\left[\varepsilon_{1}, k_{1} ; \ldots ; \varepsilon_{N}, k_{N}\right] \\
&+e \sum_{i=1}^{N} \slashed{\varepsilon}_{i} K_{(N-1)}^{p^{\prime}+k_{i}, p}\left[\varepsilon_{1}, k_{1} ; \ldots ; \varepsilon_{i-1}, k_{i-1};\varepsilon_{i+1}, k_{i+1} ; \ldots \varepsilon_{N}, k_{N}\right].
\end{aligned}
\label{relation S K}
\end{equation}
The second term on the right-hand side appears because the covariant derivative $D$ also contains the gauge field $A^\mu$ and therefore, the polarization vectors $\varepsilon^\mu_i$. As already mentioned before, there is an ambiguity in the relation between $S^{x^\prime x}$ and $K^{x^\prime x}$ due to the fact that we had the freedom to either enter the one factor in \eqref{inv dirac rewritten} on the left or the right-hand side. If one prefers to insert the one factor on the right-hand side, \eqref{relation S K} becomes 
\begin{equation}
\begin{aligned}
S_{N}^{p^{\prime} p}\left[\varepsilon_{1}, k_{1} ; \ldots ; \varepsilon_{N}, k_{N}\right]=& -K_{N}^{p^{\prime} p}\left[\varepsilon_{1}, k_{1} ; \ldots ; \varepsilon_{N}, k_{N}\right] \left(\slashed p+m\right)\\
&+e \sum_{i=1}^{N} K_{(N-1)}^{p^{\prime},p+k_{i}}\left[\varepsilon_{1}, k_{1} ; \ldots ; \varepsilon_{i-1}, k_{i-1};\varepsilon_{i+1}, k_{i+1} ; \ldots \varepsilon_{N}, k_{N}\right]\slashed{\varepsilon}_{i}.
\end{aligned}
\end{equation}
We will stick to the first convention, i.e. \eqref{relation S K}.\\
From the untruncated propagator $S^{p^\prime p}_N$ we can easily move on to the amputated propagator by
\begin{equation}
\begin{aligned}
  \widehat{S}^{p^\prime p}_N=&(-i)^2(\slashed p^\prime-m)S^{p^\prime p}_N (\slashed p-m)\\&
  =\left({p^{\prime}}^2-m^2\right) K_{N}^{p^{\prime} p}\left[\varepsilon_{1}, k_{1} ; \ldots ; \varepsilon_{N}, k_{N}\right] (\slashed p-m)\\
&-(\slashed p^\prime-m) \ e \sum_{i=1}^{N} \slashed{\varepsilon}_{i} K_{(N-1)}^{p^{\prime}+k_{i}, p}\left[\varepsilon_{1}, k_{1} ; \ldots ; \varepsilon_{i-1}, k_{i-1};\varepsilon_{i+1}, k_{i+1} ; \ldots \varepsilon_{N}, k_{N}\right](\slashed p-m).
\end{aligned}
\end{equation}
If we are only interested in an on-shell expression we can simplify this expression. Due to the fact that the $K^{p^\prime+k_i,p}$ terms do not have a pole at $\slashed p^\prime=m$, the last line in the equation above is set to zero if we impose $\slashed p^\prime=m$\footnote{to be more accurate; going on-shell of course means setting ${p^\prime}^2=m^2$, but $(\slashed p^\prime-m)=\frac{{p^\prime}^2-m^2}{\slashed p^\prime+m}=0$. Moreover, the dressed propagator will be sandwiched by polarization vectors which obey the Dirac equation $(\slashed p^\prime -m) u(p^\prime)=0$ when applying the LSZ reduction.}, and we can drop it on-shell. Hence, we find the on-shell relation:
\begin{equation}\label{StoKamputated}
  \left.\widehat{S}^{p^\prime p}_N\right|_\mathrm{on-shell}
  =\left.\left({p^{\prime}}^2-m^2\right) K_{N}^{p^{\prime} p}\left[\varepsilon_{1}, k_{1} ; \ldots ; \varepsilon_{N}, k_{N}\right] (\slashed p-m)\right|_\mathrm{on-shell}
\end{equation}
or alternatively in terms of $\widehat{K}^{p^\prime p}_N$:
\begin{equation}\label{StoOmega amputated}
  \left.\widehat{S}^{p^\prime p}_N\right|_\mathrm{on-shell}
  =\left. \widehat{K}_{N}^{p^{\prime} p}\left[\varepsilon_{1}, k_{1} ; \ldots ; \varepsilon_{N}, k_{N}\right] (\slashed p+m)^{-1}\right|_\mathrm{on-shell}.
\end{equation}
Note that at this point the phrasing ``on-shell" only concerns the $p$- and $p^\prime$-legs and not the photon legs.
\subsection{Symbol map in 4 dimensions}
Before we look at the explicit one and two photon examples, we have to find out how the inverse symbol map translates the $\eta$'s into gamma matrices. But first of all, we can observe that we can have at most four factors of $\eta$'s in 4 dimensions due to their Grassmann property. Hence, there is only a limited number of mappings:
\begin{equation}\label{eq:symbol_map_in_4d}
\begin{aligned}
\operatorname{symb}^{-1}(1) &=\mathbb{1} \\
\operatorname{symb}^{-1}\left(\eta^{\alpha_{1}} \eta^{\alpha_{2}}\right) &=-\frac{1}{4}\left[\gamma^{\alpha_{1}}, \gamma^{\alpha_{2}}\right] \\
\operatorname{symb}^{-1}\left(\eta^{\alpha_{1}} \eta^{\alpha_{2}} \eta^{\alpha_{3}} \eta^{\alpha_{4}}\right) &=\frac{1}{96} \sum_{\pi \in S_{4}} \operatorname{sign}(\pi) \gamma^{\alpha_{\pi(1)}} \gamma^{\alpha_{\pi(2)}} \gamma^{\alpha_{\pi(3)}} \gamma^{\alpha_{\pi(4)}}=-\frac{i}{4} \varepsilon^{\alpha_{1} \alpha_{2} \alpha_{3} \alpha_{4}} \gamma_{5}
\end{aligned}
\end{equation}
with $\varepsilon^{0123}=+1$.
\subsection{One and two photon example}
To test the beforehand derived formulation of the dressed QED propagator, let us look at the explicit one and two photon examples.
\subsubsection{3pt function}
The expression for the amputated one photon kernel is:
\begin{equation}\label{eq:3pt kernel}
\begin{aligned}
\widehat{K}_{1}^{p^{\prime} p}=&(-ie)(2\pi)^D\delta(p-p^\prime-k) \operatorname{symb}^{-1} \left.  \int_{-\infty}^{\infty} \mathrm{d} \tau \
\varepsilon\cdot \left[\partial_\alpha -2\partial_\theta k\cdot \partial_\theta\right] \delta(\tau) \ e^{(i\tau k+\alpha)\cdot (p^\prime+p)+\theta\cdot\eta}\right|_{\alpha=\theta=0} \\
=&(-ie)(2\pi)^D\delta(p-p^\prime-k) \left[\varepsilon\cdot (p^\prime+p)\mathbb{1}+\frac{1}{2}[\slashed \varepsilon,\slashed k]\right]
=(-ie)(2\pi)^D\delta(p+p^\prime-k) \left[\slashed\varepsilon \slashed p+\slashed p^{\prime} \slashed\varepsilon\right]\\=&(-ie)(2\pi)^D\delta(p-p^\prime-k) \left[\slashed\varepsilon(\slashed p+m)+\left(\slashed p^{\prime}-m\right) \slashed\varepsilon\right] .
\end{aligned}
\end{equation}
Proceeding to the amputated propagator, we simply multiply by $(\slashed p+m)^{-1}$ and set the external momenta on shell:
\begin{equation}
  \widehat{S}^{p^\prime p}_1= -ie (2\pi)^D\delta(p-p^\prime-k) \ \slashed \varepsilon.
\end{equation}
Evidently, the result above perfectly matches the known QED vertex given in appendix \ref{QED-feynman_rules} contracted with the polarization vector of the photon: 
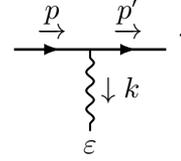
\hspace{.7cm} \begin{tikzpicture}[baseline=5pt]
	  \coordinate (in) at (-1,0);
	  \coordinate (out) at (1,0);
	  \coordinate (x) at (0,0);
	  \node (k) at (0,-1.3) {$\varepsilon$};
	  \draw [zParticle] (in) -- (x) node [midway, above] {$\underrightarrow{p}$};
	  \draw [zParticle] (x) -- (out) node [midway, above] {$\underrightarrow{p^\prime}$};
	  \draw [photon2] (x) -- (k) node [midway, right] {$\downarrow k$};
	  \end{tikzpicture} .
\subsubsection{4pt function} The expression for the amputated  two photon kernel is:
\begin{equation}
    \begin{aligned}
        \widehat{K}^{p^\prime p}_2 &= &(-ie)^2 (2\pi)^D \delta(p-p^\prime -k_1-k_2) \ \mathrm{symb}^{-1}\int_{-\infty}^\infty \mathrm{d}\tau_1\int_{-\infty}^\infty \mathrm{d}\tau_2\varepsilon_1\cdot \left[ \partial_{\alpha_1} - 2\partial_{\theta_1} k_1\cdot\partial_{\theta_1}\right] \\
        &&\times \ \varepsilon_2\cdot \left[ \partial_{\alpha_2} - 2\partial_{\theta_2} k_2\cdot\partial_{\theta_2}\right] 2\delta(\tau_1+\tau_2)e^\mathrm{Exp}\Biggl.\Biggr|_{\alpha=\theta=0} 
    \end{aligned}
\end{equation}
with
\begin{equation}
\begin{aligned}
    \mathrm{Exp}=&\sum\limits_{i=1}^2\left(ik_i\tau_i+\alpha_i\right)\cdot\left(p^\prime+p\right)-i k_1\cdot k_2 |\tau_1-\tau_2|
    -2i \alpha_1\cdot\alpha_2\ \delta(\tau_1-\tau_2)\\&+ (k_1\cdot\alpha_2-k_2\cdot\alpha_1)\ \mathrm{sign}(\tau_1-\tau_2)+\frac{1}{2} \theta_1\cdot\theta_2\ \mathrm{sign}(\tau_1-\tau_2)+\sum\limits_{i=1}^2\theta_i\cdot\eta.
\end{aligned}
\end{equation}
Performing the partial derivatives and applying the inverse symbol map yields
\begin{equation}
    \begin{aligned}
        \widehat{K}^{p^\prime p}_2 = &(-ie)^2 (2\pi)^D \delta(p-p^\prime -k_1-k_2) \ \mathrm{symb}^{-1}\int_{-\infty}^\infty \mathrm{d}\tau_1\int_{-\infty}^\infty \mathrm{d}\tau_2 \ 2\ \delta(\tau_1+\tau_2) \\
        &\times \ \biggl[ -i 2\varepsilon_1\cdot\varepsilon_2\delta(\tau_1-\tau_2)-\varepsilon_1\cdot \varepsilon_2 k_1\cdot k_2 +\varepsilon_1\cdot (p^\prime +p)\epsilon_2\cdot (p^\prime+p) \biggr.\\
        & \hspace{0,7cm}-\frac{1}{2}\varepsilon\cdot (p^\prime+p)\left[ \slashed\varepsilon_2,\slashed k_2\right] -\frac{1}{2}\varepsilon_2\cdot(p^\prime+p)\left[ \slashed\varepsilon_1,\slashed k_1\right] \\
        & \hspace{.7cm}+\biggl( \biggr. \varepsilon_1\cdot(p^\prime+p)\varepsilon_2\cdot k_1-\varepsilon_1\cdot k_2\varepsilon_2\cdot(p^\prime+p) +\frac{1}{2}\varepsilon_1\cdot k_2\left[\slashed \varepsilon_2, \slashed k_2\right] \\
        & \hspace{0,7cm}-\frac{1}{2}\varepsilon_2\cdot k_1\left[\slashed \varepsilon_1,\slashed k_1\right]+\frac{1}{2}\varepsilon_2\cdot k_1\left[ \slashed \varepsilon_1, \slashed k_2\right]+\frac{1}{2}\varepsilon_1\cdot \varepsilon_2\left[ \slashed k_2, \slashed k_1\right] - \frac{1}{2}k_1\cdot k_2 \left[ \slashed \varepsilon_1, \slashed \varepsilon_2\right] \\
        & \hspace{0,7cm}  - \frac{1}{2}\varepsilon_1\cdot k_2\left[ \slashed\varepsilon_2 , \slashed k_1\right]\biggl. \biggr) \ \mathrm{sign}(\tau_1-\tau_2) - i \varepsilon(\varepsilon_1, k_1, \varepsilon_2, k_2)\gamma_5\biggl. \biggr]
        e^\mathrm{Exp(\alpha=\theta=0)},
    \end{aligned}
\end{equation}
where we used the notation $\varepsilon(\varepsilon_1, k_1, \varepsilon_2, k_2) = \varepsilon^{\alpha\beta\gamma\delta}{\varepsilon_1}_\alpha, {k_1}_\beta, {\varepsilon_2}_\gamma, {k_2}_\delta$. The integrals in this expression are exactly the same three integral types as in the scalar case in chapter 1. Therefore, we can use the already derived solutions given in equation \eqref{eq:integral1}, \eqref{eq:integral2} and \eqref{eq:integral3}. This way, we obtain
\begin{equation} \label{eq: 4pt QED kernel}
\begin{aligned}
\widehat{K}_{2}^{p^{\prime} p}= &(-ie)^2 (2\pi)^D \delta(p-p^\prime -k_1-k_2)\\&\times\left\{-2 i\varepsilon_{1} \cdot \varepsilon_{2}+\left[\frac { i } { ( p ^ { \prime } + k _ { 1 } ) ^ { 2 }-m^2 } \biggl(\varepsilon_{1} \cdot\left(p^{\prime}+p-k_{2}\right) \varepsilon_{2} \cdot\left(p^{\prime}+p+k_{1}\right)\right.\right.\biggr.\\ 
& \hspace{6mm} +\varepsilon_{1} \cdot k_{2} \varepsilon_{2} \cdot k_{1}-k_{1} \cdot k_{2} \varepsilon_{1} \cdot \varepsilon_{2}-\frac{1}{2} \varepsilon_{1} \cdot \varepsilon_{2}\left[\slashed k_{1}, \slashed k_{2}\right]-\frac{1}{2} k_{1} \cdot k_{2}\left[\slashed\varepsilon_{1}, \slashed\varepsilon_{2}\right] \\ 
& \hspace{6mm}-\frac{1}{2} \varepsilon_{1} \cdot k_{2}\left[\slashed\varepsilon_{2},\slashed k_{1}\right]+\frac{1}{2}\left[\slashed \varepsilon_{1}, \slashed k_{2}\right] \varepsilon_{2} \cdot k_{1}+\frac{1}{2} \varepsilon_{1} \cdot\left(p^{\prime}+p-k_{2}\right)\left[\slashed\varepsilon_{2}, \slashed k_{2}\right]\\
&\hspace{6mm}+\frac{1}{2}\left[\slashed\varepsilon_{1},\slashed k_{1}\right] \varepsilon_{2} \cdot\left(p^{\prime}+p+k_{1}\right) 
\biggl.\left.+i \gamma_{5} \varepsilon\left(\varepsilon_{1}, \varepsilon_{2}, k_{1}, k_{2}\right)\biggr)+(1 \leftrightarrow 2)\bigg]\right\}.
\end{aligned}
\end{equation}
Using the relations given in appendix \ref{appendix:gamma matrix} this expression can be brought to a form looking more like a QED-4pt-function:
\begin{equation}
\begin{aligned}
\widehat{K}_{2}^{p^{\prime} p}=(-ie)^2(2\pi)^D \delta(p-p^\prime -k_1-k_2)&\biggl\{\biggr.\frac { i } {( p ^ { \prime } + k _ { 1 } ) ^ { 2 } -m^2 } \biggl[ \biggr. \slashed\varepsilon_{1}\left(\slashed p^{\prime}+\slashed k_{1}+m\right) \slashed\varepsilon_{2}(\slashed p+m)\\
&+\left(\slashed p^{\prime}+m\right) \slashed\varepsilon_{1} \slashed\varepsilon_{2}(\slashed p-m)\\
&+\left(\slashed p^{\prime}-m\right) \slashed\varepsilon_{1}\left(\slashed p^{\prime}+\slashed k_{1}-m\right) \slashed\varepsilon_{2}\biggl. \biggr] +(1 \leftrightarrow 2)\biggl. \biggr\}.
\end{aligned}
\end{equation}
Using \eqref{StoOmega amputated} we can deduce from here the amputated two photon dressed QED propagator:
\begin{equation}
    \widehat{S}^{p^\prime p}_2=(-ie)^2(2\pi)^D \delta(p-p^\prime -k_1-k_2)\left(\slashed \varepsilon_1\frac{i(\slashed p^\prime+\slashed k_1 +m)}{(p^\prime+k_1)^2-m^2}\slashed\varepsilon_2+\slashed \varepsilon_2\frac{i(\slashed p^\prime+\slashed k_2 +m)}{(p^\prime+k_2)^2-m^2}\slashed\varepsilon_1\right) .
\end{equation}
Again, we reproduced the result one would obtain using the QED Feynman rules given in appendix \ref{QED-feynman_rules} of the two contributing diagrams:
\begin{figure}[H]
\centering
  \begin{tikzpicture}[baseline={(current bounding box.center)}]
  \coordinate (in) at (-1,0);
  \coordinate (out) at (2,0);
  \coordinate (x) at (0,0);
  \coordinate (y) at (1,0);
  \node (k2) at (0,-1.3) {$\varepsilon_2$};
  \node (k1) at (1,-1.3) {$\varepsilon_1$};
  \draw (out) node [right] {};
  \draw [zParticle] (in) -- (x) node [near start, above] {$\underrightarrow{p}$};
  \draw [zParticle] (x) -- (y) node [midway, above] {$p^\prime\underrightarrow{+} k_1$};
  \draw [zParticle] (y) -- (out) node [near end, above] {$\underrightarrow{p^\prime}$};
  \draw [photon2] (x) -- (k2) node [midway, left] {$k_2\downarrow$};
  \draw [photon2] (y) -- (k1) node [midway, right] {$\downarrow k_1$};
  \end{tikzpicture}+
    \begin{tikzpicture}[baseline={(current bounding box.center)}]
  \coordinate (in) at (-1,0);
  \coordinate (out) at (2,0);
  \coordinate (x) at (0,0);
  \coordinate (y) at (1,0);
  \node (k2) at (0,-1.3) {$\varepsilon_2$};
  \node (k1) at (1,-1.3) {$\varepsilon_1$};
  \draw (out) node [right] {};
  \draw [zParticle] (in) -- (x) node [near start, above] {$\underrightarrow{p}$};
  \draw [zParticle] (x) -- (y) node [midway, above] {$p^\prime\underrightarrow{+} k_2$};
  \draw [zParticle] (y) -- (out) node [near end, above] {$\underrightarrow{p^\prime}$};
  \draw [photon2] (y) -- (k2) node [near end,left] {$k_2\swarrow$};
  \draw [photon2] (x) -- (k1) node [near end, right] {$\searrow k_1$};
  \end{tikzpicture}
  \caption{QED Feynman diagrams contributing to the 4pt function}
\end{figure}
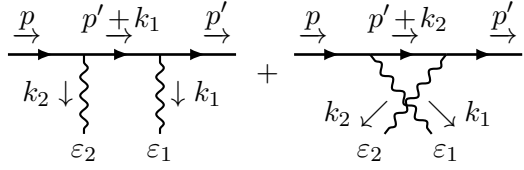
\noindent Consequently, we explicitly confirmed that the afore-derived Feynman-Schwinger representation reproduces the dressed QED-propagator. Therefore, we initially started with computing the kernel of the dressed propagator and were then able to deduce the dressed QED-propagator from the kernel using \eqref{StoKamputated} or \eqref{StoOmega amputated}.

\section{Second Order QED}\label{chapter:2nd_order_form}
An alternative way to understand the prior described formulation of the QED-dressed propagator in terms of the kernel is the \textit{second order formalism} of QED \cite{Morgan_1995_second_order_fermions}. The second order QED provides a formalism that is very similar to that of scalars. It can be thought of as a reordering of the QED Feynman rules, which gives rise to new Feynman rules, or as an independent, effective field theory.
In the next section, our intention will be to rearrange the QED-vertices and propagators in a way, such that the propagators become scalar propagators and all gamma matrix dependency is stored in the vertices.

\subsection{Second-order rules from first-order rules}
In order to store all gamma matrix dependency in the vertices, we combine the QED-vertex given in appendix \ref{QED-feynman_rules} with the numerator of the fermionic propagator:
\begin{equation}\label{combination_vert_prop}
    A^\mu_{p^\prime ,k} = -ie \ \gamma^\mu\left(\slashed p^\prime + \slashed k + m\right).
\end{equation}
The remaining propagator is now just a scalar propagator:
\begin{equation}
\begin{tikzpicture}[]
  \coordinate (in) at (-1,0);
  \coordinate (out) at (1,0);
  \draw [zParticle] (in) -- (out) node [midway, above] {$\underrightarrow{p}$};
  \end{tikzpicture} = \frac{i}{p^2-m^2}.
\end{equation}
Together with a scalar propagator, $A^\mu_{p^\prime ,k}$ will give the same expression as a QED-vertex followed by a fermionic propagator.
Using momentum conservation and that $\gamma^\alpha\gamma^\beta=\eta^{\alpha\beta}+\frac{1}{2}\left[\gamma^\alpha,\gamma^\beta\right]$, equation \eqref{combination_vert_prop} can be rewritten as
\begin{equation}
    A^\mu_{p^\prime ,k} =  -ie\left(({p^\prime} + p)^\mu +\frac{1}{2}\left[ \gamma^\mu, \slashed k \right] -( \slashed p^\prime - m) \gamma^\mu\right) = B^\mu_{p^\prime , k} + C^\mu_{p^\prime , k}
\end{equation}
where we defined $B^\mu_{p^\prime , k} := -ie\left(({p^\prime} + p)^\mu +\left[ \gamma^\mu, \slashed k \right]\right)$ and $C^\mu_{p^\prime , k} := ie( \slashed p^\prime - m) \gamma^\mu$.
Since $\overline{u}(p^\prime)(\slashed p^\prime - m)=0$, the second term $C^\mu_{p^\prime , k}$ does not contribute if the $p^\prime$ leg is set on-shell and will only contribute to higher order diagrams. For that reason, we define our new trivalent vertex as
\begin{equation} \label{second_order_3vertex}
    \begin{tikzpicture}[baseline={(current bounding box.center)}]
	  \coordinate (in) at (-1,0);
	  \coordinate (out) at (1,0);
	  \coordinate (x) at (0,0);
	  \node (k) at (0,-1.3) {$\mu$};
	  \draw [zParticle] (in) -- (x) node [midway, above] {$\underrightarrow{p}$};
	  \draw [zParticle] (x) -- (out) node [midway, above] {$\underrightarrow{p^\prime}$};
	  \draw [photon2] (x) -- (k) node [midway, right] {$\downarrow k$};
	  \end{tikzpicture} := B^\mu_{p^\prime , k} = -ie\left(({p^\prime} + p)^\mu +\frac{1}{2}\left[ \gamma^\mu, \slashed k \right]\right)
\end{equation}
but keep in mind that we have to take into account the $C^\mu_{p^\prime , k}$-term for higher order diagrams. Note that the trivalent vertex \eqref{second_order_3vertex} looks very similar to a vertex in scalar QED and shows great similarity with the zeroth-order WQFT vertex in \eqref{eq:WQFT1photon_with_spin} with $v\widehat{=}p+p^\prime$ and $S^{\mu\nu}\widehat{=}\frac{i}{4}\left[ \gamma^\mu,\gamma^\nu\right]$. To see what happens when we look at higher order diagrams, let us now consider the following example:
\begin{align}\label{4pt_for_second_order}
\begin{aligned}
 \begin{tikzpicture}[baseline={(current bounding box.center)}]
  \coordinate (in) at (-1,0);
  \coordinate (out) at (2,0);
  \coordinate (x) at (0,0);
  \coordinate (y) at (1,0);
  \node (k1) at (0,-1.3) {$\mu$};
  \node (k2) at (1,-1.3) {$\nu$};
  \draw (out) node [right] {};
  \draw [zParticle] (in) -- (x) node [near start, above] {$\underrightarrow{p}$};
  \draw [zParticle] (x) -- (y) node [midway, above] {$p^\prime\underrightarrow{+} k_2$};
  \draw [zParticle] (y) -- (out) node [near end, above] {$\underrightarrow{p^\prime}$};
  \draw [photon2] (x) -- (k1) node [midway, left] {$k_1\downarrow$};
  \draw [photon2] (y) -- (k2) node [midway, right] {$\downarrow k_2$};
 \end{tikzpicture}+&
 \begin{tikzpicture}[baseline={(current bounding box.center)}]
  \coordinate (in) at (-1,0);
  \coordinate (out) at (2,0);
  \coordinate (x) at (0,0);
  \coordinate (y) at (1,0);
  \node (k1) at (0,-1.3) {$\mu$};
  \node (k2) at (1,-1.3) {$\nu$};
  \draw (out) node [right] {};
  \draw [zParticle] (in) -- (x) node [near start, above] {$\underrightarrow{p}$};
  \draw [zParticle] (x) -- (y) node [midway, above] {$p^\prime\underrightarrow{+} k_1$};
  \draw [zParticle] (y) -- (out) node [near end, above] {$\underrightarrow{p^\prime}$};
  \draw [photon2] (y) -- (k1) node [near end,left] {$k_1\swarrow$};
  \draw [photon2] (x) -- (k2) node [near end, right] {$\searrow k_2$};
  \end{tikzpicture} 
  \\ &=A^\nu_{p^\prime,k_2}\frac{i}{(p^\prime+k_2)^2-m^2} \ A^\mu_{p^\prime+k_2 , k_1}
  + A^\mu_{p^\prime,k_1}\frac{i}{(p^\prime+k_1)^2-m^2} \ A^\nu_{p^\prime+k_1 , k_2}.
\end{aligned}
\end{align}
Once again, we can neglect the $C^\mu_{p^\prime , k}$-contribution in the $A^\mu_{p^\prime , k}$-terms on the left-hand side if we set the external legs on shell:
\begin{equation}
    \eqref{4pt_for_second_order} = B^\nu_{p^\prime,k_2}\frac{i}{(p^\prime+k_2)^2-m^2}  \left(B^\mu_{p^\prime+k_2 , k_1} +C^\mu_{p^\prime+k_2 , k_1}\right) \ + \ \bigl((k_1, \mu) \leftrightarrow (k_2, \nu)\bigr) .
\end{equation}
Consequently, we see that the only contributions, which are not considered yet by the new defined trivalent vertex \eqref{second_order_3vertex}, are:
\begin{align}
    A^\nu_{p^\prime,k_2}\frac{i}{(p^\prime+k_2)^2-m^2} \ C^\mu_{p^\prime+k_2 , k_1} &=-i e \gamma^\nu (\slashed p^\prime + \slashed k_2 +m) \frac{i}{(p^\prime +k_2)^2-m^2}\left( ie(\slashed p^\prime + \slashed k_2 -m)\gamma^\mu\right) \nonumber \\
    &= ie^2 \gamma^\nu \gamma^\mu \label{2nd_1contr} \\ \ & \ \nonumber \\
     A^\nu_{p^\prime,k_1}\frac{1}{(p^\prime+k_1)^2-m^2} \ C^\mu_{p^\prime+k_1 , k_2} &= ie^2 \gamma^\mu \gamma^\nu \label{2nd_2contr}
\end{align}
Adding both contributions above yields
\begin{equation}
    \eqref{2nd_1contr} + \eqref{2nd_2contr} = ie^2 \{ \gamma^\nu, \gamma^\mu \} = 2ie^2\eta^{\mu\nu},
\end{equation}
which leads us to the definition of a new quartic vertex:
\begin{align}\label{second_order_QED_4vertex}
    \begin{aligned}
    \begin{tikzpicture}[baseline={(current bounding box.center)}]
  \coordinate (in) at (-1.3,0);
  \coordinate (out) at (1.3,0);
  \coordinate (x) at (0,0);
  \node (k1) at (-1,-1.3) {$\mu$};
  \node (k2) at (1,-1.3) {$\nu$};
  \draw (out) node [right] {};
  \draw [zParticle] (in) -- (x) node [midway, above] {$\underrightarrow{p}$};
  \draw [zParticle] (x) -- (out) node [midway, above] {$\underrightarrow{p^\prime}$};
  \draw [photon2] (x) -- (k1) node [midway, left] {$k_1\swarrow$};
  \draw [photon2] (x) -- (k2) node [midway, right] {$\searrow k_2$};
 \end{tikzpicture} 
    \end{aligned} := 2i e^2 \eta^{\mu\nu},
\end{align}
which is the same as the quartic vertex in scalar QED given in \eqref{eq:vertices_scalar_QED}.
The above derived vertices \eqref{second_order_3vertex} and \eqref{second_order_QED_4vertex}, together with a scalar propagator form the Feynman rules for the second order formalism. By multiplying with $(\slashed p +m)^{-1}$ from the right for every ingoing external fermion, we can construct QED-diagrams with the just derived second order Feynman rules.

\subsection{Second order theory as an independent field theory}
In contrast to the above-described procedure, the above derived Feynman rules can also be deduced from an independent field theory with some set of fields $\varPsi$ and the following lagrangian:
\begin{equation}
    \mathcal{L} = \varPsi^\dagger \left(D^2+m^2-\frac{e}{2}\sigma^{\mu\nu} F_{\mu\nu}\right)\varPsi,
\end{equation}
where $\sigma^{\mu\nu} = \frac{i}{2}\left[\gamma^\mu,\gamma^\nu\right]$. Evidently, this theory gives rise to a trivalent and a quartic vertex. In fact, they are the same as the ones in the section above. Since the lagrangian only contains second-order derivatives, the propagator is the same as for scalars. From this point of view, it is immediately clear where the term ``second order formalism" comes from.
\subsection{Dressed propagator from second-order Feynman rules}
Let us now explicitly consider some easy examples in the second-order formalism.
\paragraph{Three-point function:} The 3pt-function is simply given by the trivalent vertex of the second order theory \eqref{second_order_3vertex}:
\begin{equation}
    \begin{tikzpicture}[baseline={(current bounding box.center)}]
	  \coordinate (in) at (-1,0);
	  \coordinate (out) at (1,0);
	  \coordinate (x) at (0,0);
	  \node (k) at (0,-1.3) {$\mu$};
	  \draw [zParticle] (in) -- (x) node [midway, above] {$\underrightarrow{p}$};
	  \draw [zParticle] (x) -- (out) node [midway, above] {$\underrightarrow{p^\prime}$};
	  \draw [photon2] (x) -- (k) node [midway, right] {$\downarrow k$};
	  \end{tikzpicture} = e\left(({p^\prime} + p)^\mu +\frac{1}{2}\left[ \gamma^\mu, \slashed k \right]\right),
\end{equation}
which agrees with the $N=1$  kernel \eqref{eq:3pt kernel}.

\paragraph{Four-point function:} The 4pt-function is given by the sum of 3 diagrams:

\begin{equation*}\def\arraystretch{1.7}
    \begin{aligned}
 \begin{tikzpicture}[baseline={(current bounding box.center)}]
  \coordinate (in) at (-1,0);
  \coordinate (out) at (2,0);
  \coordinate (x) at (0,0);
  \coordinate (y) at (1,0);
  \node (k2) at (0,-1.3) {$\mu_2$};
  \node (k1) at (1,-1.3) {$\mu_1$};
  \draw (out) node [right] {};
  \draw [zParticle] (in) -- (x) node [near start, above] {$\underrightarrow{p}$};
  \draw [zParticle] (x) -- (y) node [midway, above] {$p^\prime\underrightarrow{+} k_1$};
  \draw [zParticle] (y) -- (out) node [near end, above] {$\underrightarrow{p^\prime}$};
  \draw [photon2] (x) -- (k2) node [midway, left] {$k_2\downarrow$};
  \draw [photon2] (y) -- (k1) node [midway, right] {$\downarrow k_1$};
 \end{tikzpicture}&\begin{array}{l} = \displaystyle(-ie)^2 \left(\left(p^\prime+p-k_2\right)^{\mu_1}-i \sigma^{\mu_1\alpha}{k_1}_\alpha\right) \\ \hspace{4cm}\displaystyle\times \frac{i}{(p^\prime+k_1)^2-m^2}\left(\left(p^\prime+p+k_1\right)^{\mu_2}-i \sigma^{\mu_2\beta}{k_2}_\beta\right)\end{array} \\
  &+ \\
 \begin{tikzpicture}[baseline={(current bounding box.center)}]
  \coordinate (in) at (-1,0);
  \coordinate (out) at (2,0);
  \coordinate (x) at (0,0);
  \coordinate (y) at (1,0);
  \node (k2) at (0,-1.3) {$\mu_2$};
  \node (k1) at (1,-1.3) {$\mu_1$};
  \draw (out) node [right] {};
  \draw [zParticle] (in) -- (x) node [near start, above] {$\underrightarrow{p}$};
  \draw [zParticle] (x) -- (y) node [midway, above] {$p^\prime\underrightarrow{+} k_2$};
  \draw [zParticle] (y) -- (out) node [near end, above] {$\underrightarrow{p^\prime}$};
  \draw [photon2] (y) -- (k2) node [near end,left] {$k_2\swarrow$};
  \draw [photon2] (x) -- (k1) node [near end, right] {$\searrow k_1$};
  \end{tikzpicture}& \begin{array}{l}=(-ie)^2 \left(\left(p^\prime+p-k_1\right)^{\mu_2}-i \sigma^{\mu_2\alpha}{k_2}_\alpha\right) \\
  \hspace{4cm}\displaystyle\times\frac{i}{(p^\prime+k_2)^2-m^2}\left(\left(p^\prime+p+k_2\right)^{\mu_1}-i \sigma^{\mu_1\beta}{k_1}_\beta\right) \end{array} \\ &+ \\
   \begin{tikzpicture}[baseline={(current bounding box.center)}]
  \coordinate (in) at (-1.3,0);
  \coordinate (out) at (1.3,0);
  \coordinate (x) at (0,0);
  \node (k2) at (-1,-1.3) {$\mu_2$};
  \node (k1) at (1,-1.3) {$\mu_1$};
  \draw (out) node [right] {};
  \draw [zParticle] (in) -- (x) node [midway, above] {$\underrightarrow{p}$};
  \draw [zParticle] (x) -- (out) node [midway, above] {$\underrightarrow{p^\prime}$};
  \draw [photon2] (x) -- (k2) node [midway, left] {$k_2\swarrow$};
  \draw [photon2] (x) -- (k1) node [midway, right] {$\searrow k_1$};
 \end{tikzpicture} &= 2i e^2 \eta^{\mu_1\mu_2} \\
    \end{aligned}
\end{equation*}

\begin{equation}
\begin{aligned}\label{eq:two_photn_second_order}
    =& (-ie)^2 \biggl[ \biggr.  - 2i\eta^{\mu_1\mu_2}+\biggl( \biggr.\frac{i}{(p^\prime+k_1)^2-m^2}  \Bigl(\left(p^\prime+p-k_2\right)^{\mu_1}-i \sigma^{\mu_1\alpha}{k_1}_\alpha\Bigr) \\
    &\hspace{5.92cm}\times\left(\left(p^\prime+p+k_1\right)^{\mu_2}-i \sigma^{\mu_2\beta}{k_2}_\beta\right) +(1 \leftrightarrow 2) \biggl. \biggr)\biggl. \biggr] . \\ \ 
\end{aligned}
\end{equation}
Using the gamma matrix identities listed in appendix \ref{appendix:gamma matrix}, the result for the second order 4-pt function calculated above can be brought to the same form as the $N=2$  kernel in \eqref{eq: 4pt QED kernel}. 
With the two examples above, we explicitly showed that the second order formalism indeed reproduces the kernel introduced in section \ref{chapter:elec_prop}.

\section{Dressed propagators in QED vs. spin-1/2 WQFT}\label{chapter:WQFT_elec:prop}
Now that we have considered explicit examples in both, QED and the spin-1/2 WQFT, the natural question is how the two theories are related. As we have explicitly seen, the WQFT analogue to the dressed electron propagator looked more like a modified version of scalar QED, and the two photon diagram even exhibited an internal scalar propagator. After that, we derived a worldline representation of the dressed electron propagator in terms of the kernel, which also showed great similarities with scalar QED. The same applies to the second order QED which was introduced as an alternative access to the kernel. 

From this point of view, it seems intuitive to rather compare the QED kernel/second order QED to the spin-1/2 WQFT instead of directly relating it to actual QED. For the one photon example, we already noticed a match between the QED kernel/second order QED and the spin-1/2 WQFT if we identify $v=p+p^\prime$ and \begin{equation}
S^{\mu\nu}=-i \eta^\mu\eta^\nu \ \widehat{=} \ \frac{i}{4}\left[\gamma^\mu,\gamma^\nu\right]. \end{equation}
In fact, this identification is not surprising since $S^{\mu\nu}=\frac{i}{4}\left[\gamma^\mu,\gamma^\nu\right]$ is just an alternative representation of the Lorentz algebra \eqref{eq:lorentz_algebra}. It is not a hard task to explicitly show that also the 2 photon example in the second order QED \eqref{eq:two_photn_second_order}, the two photon kernel \eqref{eq: 4pt QED kernel} and the two photon diagram in the spin-1/2 WQFT \eqref{eq:WQFT_4pt_in_terms_of_scalar_propagators} match up under the identification of the different spin representations (apart from a factor of $\delta((k_1+k_2)\cdot v) e^{i(k_1+k_2)\cdot b}$). But instead of further elaborating on the explicit examples, let us now move on to a more general comparison for electron propagators dressed with an arbitrary number of photons.
For this, we integrate out the worldline fields in the spin-1/2 WQFT:
\begin{equation}
\begin{aligned}
    K_A^{\mathrm{WQFT}}\left(b,v\right)=&\int Dx \ \exp{\left[-i\int_{-\infty}^{\infty}\mathrm{d}\tau\left( \frac{1}{4}\dot{x}^2+e\dot{x}\cdot A\right)\right]}\\
    &\times\int D\psi \ \exp{\left[\int_{-\infty}^{\infty}\mathrm{d}\tau\left(\frac{1}{2}\psi_\mu \dot{\psi}^\mu -eF_{\mu\nu}\Psi^\mu\Psi^\nu \right)\right]}
\end{aligned}
\end{equation}
and proceed exactly like in the corresponding section in the non-spinning case in the first chapter.
Again writing the gauge field $A^\mu$ as a collection of plane waves $A^\mu(x)=\sum_{i=1}^N\varepsilon_i^\mu e^{ik_i x}$ and considering the background field expansion 
\begin{equation}
    x^\mu(\tau) = b^\mu + v^\mu \tau + z^\mu (\tau), \hspace{2cm}\Psi^\mu(\tau) = \eta^\mu + \psi^\mu(\tau)
\end{equation}
the $N$-photon dressed WQFT-Kernel yields:
\begin{equation}
\begin{aligned}
  K^{\mathrm{WQFT}}_N\left(b,v,\{k_1,\varepsilon_1,...,k_N,\varepsilon_N\}\right)=&\left(-ie\right)^N\int Dz\int D\psi \ e^{-\int_{-\infty}^{\infty}\mathrm{d}\tau\left[\frac{i}{4}(v+\dot{z})^2-\frac{1}{2}\psi_\mu\dot{\psi}^\mu\right]} \\&
  V_\eta^{b,v}\left[k_1,\varepsilon_1\right]...V_\eta^{b,v}\left[k_N,\varepsilon_N\right],
\end{aligned}
\end{equation}
where the vertex operator $V_\eta^{b,v}\left[k,\varepsilon\right]$ is the vertex operator $V_\eta^{x,x^\prime}\left[k,\varepsilon\right]$ given in \eqref{eq:vertex_operator_with_spin} with $\frac{x^\prime-x}{T}$ replaced by $v$, $q$ replaced by $z$ and $x$ replaced by $b$. Ignoring boundary terms and imposing generic translation invariant Green's functions:
\begin{equation}
\begin{aligned}
    \left\langle z^\mu(\tau)z^\nu(\tau^\prime) \right\rangle=& \ 2i\Delta(\tau -\tau^\prime)\eta^{\mu\nu}&\mathrm{with}&& \frac{\partial^2}{\partial\tau^2}\Delta(\tau-\tau^\prime)=&\delta(\tau-\tau^\prime)\\
    \left\langle\psi^\mu(\tau)\psi^\nu(\tau^\prime)\right\rangle=&-\frac{1}{2}G_F(\tau-\tau^\prime)\eta^{\mu\nu}&\mathrm{with}&& \frac{1}{2}\frac{\partial^2}{\partial\tau^2}G_F(\tau-\tau^\prime)=&\delta(\tau-\tau^\prime)
    \end{aligned}
\end{equation}
the WQFT-kernel becomes
\begin{equation}
\begin{aligned}
K^{\mathrm{WQFT}}_N\left(b,v,\{k_1,\varepsilon_1,...,k_N,\varepsilon_N\}\right)=&K^\mathrm{WQFT}_0[z,\psi] \  e^{
i\sum_{i=1}^N k_i \cdot b} (-ie)^N \\&
\left. \left(\prod\limits_{i=1}^{N}\int_{-\infty}^\infty\mathrm{d}\tau_i \ \varepsilon_i\cdot\left[\partial_{\alpha_i}-2\partial_{\theta_i} \ k_i\cdot\partial_{\theta_i}\right]\right) 
 e^\mathrm{Exp}\right|_{\alpha_i=\theta_i=0}
\end{aligned}
\label{wqft-kernel}
\end{equation}
with
\begin{equation}
\begin{aligned}
    \mathrm{Exp}=&\sum\limits_{i=1}^N\left(ik_i\tau_i+\alpha_i\right)\cdot v-i\sum\limits_{i,j=1}^N k_i\cdot k_j \Delta(\tau_i-\tau_j)
    +i\sum\limits_{i,j=1}^N \alpha_i\cdot\alpha_j{}^\bullet\Delta^\bullet(\tau_i-\tau_j)\\&-2\sum\limits_{i,j=1}^N k_i\cdot\alpha_j{}^\bullet\Delta(\tau_i-\tau_j)+\frac{1}{4}\sum\limits_{i,j=1}^N \theta_i\cdot\theta_j G_F(\tau_i-\tau_j)+\sum\limits_{i=1}^N\theta_i\cdot\eta .
\end{aligned}
\end{equation}
and
\begin{equation}
    K^\mathrm{WQFT}_0[z,\psi]:=\lim\limits_{T\rightarrow\infty} e^{-i\frac{v^2}{2}T}Z^0[z]Z^0[\psi].
\end{equation}
$Z^0[z]$ and $Z^0[\psi]$ are the free path integral normalization factors of the two path integrals.\\
As in the corresponding section in chapter 1 we introduce ``center of mass" proper time coordinates:
\begin{equation}\begin{aligned}
    \Tilde{\tau_i}:=\tau_i -\tau_+&&\mathrm{with}&& \tau_+=\frac{1}{N}\sum_{i=1}^N\tau_i 
\end{aligned}\end{equation}
and rewrite the $\tau_i$-integrations
\begin{equation}
    \prod\limits_{i=1}^N\int_{-\infty}^\infty\mathrm{d}\tau_i=\prod\limits_{i=1}^N\int_{-\infty}^\infty\mathrm{d}\Tilde{\tau}_i\int_{-\infty}^\infty\mathrm{d}\tau_+ \delta\left(\frac{1}{N}\sum\limits_{i=1}^N\Tilde{\tau}_i\right) .
\end{equation}
The $\tau_+$-integral just produces a $\delta\left(\sum_{i=1}^Nk_i\cdot v\right)$ and we obtain
\begin{equation}
\begin{aligned}
&K^{\mathrm{WQFT}}_N\left(b,v,\{k_1,\varepsilon_1,...,k_N,\varepsilon_N\}\right)=K^\mathrm{WQFT}_0[z,\psi] \ \delta\left(\sum\limits_{i=1}^N k_i\cdot v\right) e^{
i\sum_{i=1}^N k_i \cdot b} (-ie)^N  \\& \hspace{3cm}
\left. \left(\prod\limits_{i=1}^{N}\int_{-\infty}^\infty\mathrm{d}\tau_i \ \varepsilon_i\cdot\left[\partial_{\alpha_i}-2\partial_{\theta_i} \ k_i\cdot\partial_{\theta_i}\right]\right) N\delta\left(\sum\limits_{i=1}^N \tau_i\right)
 e^\mathrm{Exp}\right|_{\alpha_i=\theta_i=0}.
\end{aligned}
\label{wqft-kernel final}
\end{equation}
By comparing \eqref{wqft-kernel final} and \eqref{form factor kernel} and ignoring the delta function imposing overall momentum conservation, we find (under the premise that we also use the inverse symbol map on the WQFT expression, i.e. identify the different representations of the spin tensor with each other):
\begin{equation}
\boxed{
   \frac{K^\mathrm{WQFT}_N(b,v;\{\varepsilon_i,k_i\})}{K^\mathrm{WQFT}_0}=\delta\left(\sum\limits_{i=1}^N k_i\cdot v\right) e^{i\sum\limits_{i=1}^N k_i\cdot b} \widehat{K}^{p^\prime p}_N(k_i,\varepsilon_i;\dots;k_N,\varepsilon_N)}
   \label{link WQFT and QED Kernel}
\end{equation}
if we identify $v=(p^\prime +p)$ and choose the following Green's functions:
\begin{equation}
\begin{aligned}
\Delta(\tau)=\frac{|\tau|}{2},&\hspace{1.5cm}& G_F(\tau)=\operatorname{sign}(\tau) .
\end{aligned}
\end{equation}
Equation \eqref{link WQFT and QED Kernel} tells us that there is a direct link between the kernel of the QED propagator (not the QED propagator itself) and the corresponding expression in the WQFT.
Using \eqref{StoOmega amputated} the link can be reformulated in terms of the amputated on-shell dressed propagator:
\begin{equation}
\boxed{
   \left.\frac{K^\mathrm{WQFT}_N(b,v;\{\varepsilon_i,k_i\})}{K^\mathrm{WQFT}_0}(\slashed p+m)^{-1}\right|_{p \ \mathrm{and}\ p^\prime \mathrm{on-shell}} \hspace{-8mm}= \delta\left(\sum\limits_{i=1}^N k_i\cdot v\right) e^{i\sum\limits_{i=1}^N k_i\cdot b} \widehat{S}^{p^\prime p}_N(k_i,\varepsilon_i;\dots;k_N,\varepsilon_N)}.
   \label{link WQFT and QED dressed S}
\end{equation}

\section{S-matrices in QED vs. expectation values in the WQFT}
To generalize the comparison of QED and the spin-1/2 WQFT, let us first of all revisit the discussion about S-matrices in the classical limit in section \ref{sec:S-matrices}. There we argued that S-matrices can be expressed via dressed propagators in the classical limit on the basis of scalar QED, but the same reasoning also applies to QED. Ignoring diagrams that include fermion loops, we can integrate out the fermions treating the gauge field as a constant background and express correlators via dressed fermionic propagators:
\begin{equation}
\begin{aligned} \label{5pt_correlator_via_dressed_propagator_QED}
\langle\Omega| T &\ \{\left(A_{\mu}(x)\right) \psi_{1}\left(x_{1}\right) \overline{\psi}_{1}\left(x_{1}^{\prime}\right) \psi_{2}\left(x_{2}\right) \overline{\psi}_{2}\left(x_{2}^{\prime}\right)\}|\Omega\rangle \\ &
=\tilde{\mathcal{Z}}^{-1} \int \mathcal{D}\left[A_{\mu}, \overline{\psi}_1,\psi_{1},\overline{\psi}_2, \psi_{2}\right] \left(A_{\mu}(x)\right) \psi_{1}\left(x_{1}\right) \overline{\psi}_{1}\left(x_{1}^{\prime}\right) \psi_{2}\left(x_{2}\right) \overline{\psi}_{2}\left(x_{2}^{\prime}\right) e^{i S_\mathrm{QED}}\\
&=\mathcal{Z}^{-1} \int \mathcal{D}\left[A_{\mu}\right] \left(A_{\mu}(x)\right) S_{A,1}\left(x_{1}, x_{1}^{\prime}\right) S_{A,2}\left(x_{2}, x_{2}^{\prime}\right) e^{i\left(S_{\mathrm{A}}+S_{\mathrm{gf}}\right)},
\end{aligned}
\end{equation}
where we again write an optional final state photon in parentheses.
The dressed fermionic propagators are a functional of the gauge field $A^\mu(x)$ and are defined as
\begin{equation}
S_{A,i}\left(x, x^{\prime}\right)=\langle A|T\{\psi_i(x)\overline{\psi}_i(x^\prime)\}|A\rangle=\mathcal{Z}_{i}^{-1} \int \mathcal{D}\left[\overline{\psi}_i,\psi_{i}\right] \psi_{i}(x) \overline{\psi}_{i}\left(x^{\prime}\right) e^{i S_{i}}.
\end{equation}
From \eqref{5pt_correlator_via_dressed_propagator_QED} we can move on to the S-matrix via LSZ reduction, i.e. set the external legs on-shell and replace the external propagators by polarization vectors and Fourier transform to momentum space:
\begin{equation}
\begin{aligned}
\label{Smatrix_via_dressed_propagator_QED}
\left\langle\psi_{1} \psi_{2}(A)|S| \psi_{1} \psi_{2}\right\rangle=&\mathcal{Z}^{-1} \left(\int \mathrm{d}^{D}\left[x\right] e^{-i k \cdot x} \right)\int \mathcal{D}\left[A_{\mu}\right]\left(\varepsilon^{\mu}(k) A_{\mu}(x)\right)\\ & 
\overline{u}\left(p^\prime_1\right)\widehat{S}_{A,1}\left(p_{1}, p_{1}^{\prime}\right)u\left(p_1\right) \overline{u}\left(p^\prime_2\right)\widehat{S}_{A,2}\left(p_{2}, p_{2}^{\prime}\right)u\left(p_2\right) e^{i\left(S_{\mathrm{A}}+S_{\mathrm{gf}}\right)}.
\end{aligned}
\end{equation}
The $\widehat{S}_{A,i}\left(p_{1}, p_{1}^{\prime}\right)$ are the amputated momentum space dressed propagators with the external fermionic legs on-shell. Using equation \eqref{link WQFT and QED dressed S} the amputated dressed propagators can be replaced by the WQFT analogues:
\begin{equation}\label{eq:s-matrix_dressed_prop_replaced_by_WQFT_with_spin}\def\arraystretch{2.3}
\begin{array}{l}\displaystyle
   \int \frac{\mathrm{d}^D q}{(2\pi)^{D-2}}\delta (q\cdot v_1) \delta (q\cdot v_2) e^{iq\cdot b} \left\langle\psi_{1} \psi_{2}(A)|S| \psi_{1} \psi_{2}\right\rangle \\ \displaystyle \hspace{.5cm} ={\mathcal{Z}_0}_\text{WQFT}^{-1}\int\mathcal{D}\left[A_{\mu}\right]\overline{u}\left(p^\prime_1\right)\frac{K_A^\mathrm{WQFT}(b_1,v_1)}{(\slashed p_1+m_1)}u\left(p_1\right) \overline{u}\left(p^\prime_2\right)\frac{K_A^\mathrm{WQFT}(b_2,v_2)}{(\slashed p_2 +m_2)}u\left(p_2\right) e^{i\left(S_{\mathrm{A}}+S_{\mathrm{gf}}\right)} \\
   \hspace{.5cm} =\displaystyle\frac{{\mathcal{Z}_0}_\text{WQFT}^{-1}}{4m_1m_2}\int\mathcal{D}\left[A_{\mu}\right]\overline{u}\left(p^\prime_1\right)K_A^\mathrm{WQFT}(b_1,v_1)u\left(p_1\right) \overline{u}\left(p^\prime_2\right)K_A^\mathrm{WQFT}(b_2,v_2)u\left(p_2\right) e^{i\left(S_{\mathrm{A}}+S_{\mathrm{gf}}\right)} \\
   \hspace{.5cm} =\displaystyle\frac{1}{4m_1m_2}\overline{u}\left(p^\prime_1\right)\overline{u}\left(p^\prime_2\right) \ \bigl\langle A_\mu \bigr\rangle_\text{WQFT} \ 
   u\left(p_1\right)u\left(p_2\right)
\end{array}
\end{equation}
where $b=b_1-b_2$ and $q=\sum_{i=1}^N k_i=p_1-p_1^\prime=p_2^\prime-p_2$ and where we used that the polarization vectors obey the Dirac equation:
\begin{equation}
    (\slashed p+m)^{-1} u(p)= \frac{1}{2m} u(p) .
\end{equation}
As already explained in chapter \ref{sec:S-matrices in scalar QED vs. expectation values in the WQFT} an additional integral over the total momentum transfer $q$ had to be introduced when replacing the dressed propagators with the WQFT analogues. 

At first glance, it may seem weird that the fermionic polarization vectors, implicitly carrying spinor indices, act on a WQFT expectation value in equation \eqref{eq:s-matrix_dressed_prop_replaced_by_WQFT_with_spin}. But remember that all the relations given in this subsection only work under the premise that we use the inverse symbol map on WQFT expressions to identify the representation of spin in terms of Grassmann vectors with antisymmetric products of gamma matrices. Thus, although not explicitly written, the spinor indices of the fermionic polarization vectors have to be contracted with the spinor indices of the gamma matrices.


\chapter{Coupling to Gravity}
When gravity is introduced, the usual strategy is to replace the Minkowski metric with a general metric $ \eta_{\mu\nu}\rightarrow g_{\mu\nu}$, which incorporates the possibility of curved spacetime. Moreover, derivatives are replaced with covariant derivatives obeying $\nabla_\rho \ g_{\mu\nu}=0$ to make the theory independent of the choice of coordinates.
However, when spin is under consideration, this creates a new problem: We need a representation of spin in curved space. But this problem can be elegantly circumvented with the so-called \textit{vielbein} or \textit{tetrad} formalism. Briefly summarized, this formalism maps the usual coordinate description in general relativity into a local inertial frame allowing us to simply use the flat space spin description given in the previous chapter. In the next section, we will introduce the vielbein/tetrad formalism building on references \cite{Weinberg:1972kfs} and \cite{carroll_spacetime_2016} and explain how it can be used to formulate a worldline theory of spin-1/2 particles coupled to gravity.


\section{Spin in curved Space}
Although one can always find a coordinate system so that at a given point $x^\mu$ the metric $g_{\mu\nu}(x)$ locally reduces to the Minkowski metric, in general, this can not be achieved globally when using a single coordinate basis as conventionally done in general relativity. However, we can introduce a set of basis vectors $\boldsymbol{e}_{(a)}$ with local flat space coordinates $\xi^a(x)$ at each point on the manifold and demand that they are orthonormal in the sense that their inner product fulfills
\begin{equation}
    g(\boldsymbol{e}_{(a)},\boldsymbol{e}_{(b)})=\eta_{ab},
\end{equation}
where $g(\ , \ )$ denotes the usual metric tensor written in a coordinate independent way.
The metric in any ordinary coordinate basis is then related to the locally inertial system by
\begin{equation}\label{eq:g_vs_eta_flat_space_coordiantes}
    g_{\mu\nu}(x) = \frac{\partial \xi^a(x)}{\partial x^\mu}\frac{\partial\xi^b(x)}{\partial x^\nu} \eta_{ab},
\end{equation}
where the $\xi^a(x)$ are the locally flat space coordinates. This can be thought of as defining a different locally flat coordinate system at each point on the manifold and then defining a coordinate transformation separately for each point on the manifold that map the flat space coordinate systems to a genuine globally defined coordinate system.
For each point in spacetime, we can define a matrix 
\begin{equation}
    e_\mu^{\ a}(x)=\left(\frac{\partial \xi^a(x^\prime)}{\partial x^{\prime\mu}}\right)_{x^\prime=x}
\end{equation} to relate it to the coordinate basis vectors
\begin{equation}
    \boldsymbol{e}_{(\mu)} = e_\mu^{\ a}(x) \boldsymbol{e}_{(a)},
\end{equation}
which, of course, depends on the point in spacetime. This matrix, which can also be thought of as a set of 4 vectors ($e^{\ 0}_{ \mu}$, $e^{\ 1}_{\mu}$, $e^{\ 2}_{ \mu}$, $e^{\ 3}_{ \mu}$), is know as \textit{vielbein} or \textit{tetrad}. Here and in the following, coordinate (curved) space indices will be denoted with Greek letters and the local inertial frame indices with Roman letters. In line with this, Greek indices will be raised and lowered with $g_{\mu\nu}$
and Roman indices with the Minkowski metric $\eta_{\mu\nu}$:
\begin{equation}\label{eq:vielbein_raise_lower_indices}
    e^\mu_{\ a} = g^{\mu\nu}\eta_{ab} e^{\ b}_{\nu}.
\end{equation}
Using the vielbein we can rewrite equation \eqref{eq:g_vs_eta_flat_space_coordiantes} as
\begin{equation}\label{eq:vielbein_g_eta}
    g_{\mu\nu}=e_\mu^{\ a} e_\nu^{\ b} \eta_{ab}
\end{equation}
or equally as
\begin{equation}\label{eq:vielbein_g_eta_respectively}
    g_{\mu\nu}e^\mu_{\ a}e^\nu_{\ b} = \eta_{ab},
\end{equation}
which is why the vielbein is sometimes referred to as the ``square root" of the metric. 
From equations \eqref{eq:vielbein_raise_lower_indices} and \eqref{eq:vielbein_g_eta} we see that raising and lowering both indices provides the inverse (contravariant) vielbein, yielding 
\begin{equation}\label{eq:orthonormality_vielbein}
    e^\mu_{\ a}e_\nu^{\ a}=\delta^\mu_\nu , \hspace{2cm} e_\mu^{\ a} e^\mu_{\ b} = \delta^a_b.
\end{equation}
In short, the vielbeins are the components of the coordinate basis vectors in terms of the orthonormal basis vectors, and the inverse vielbeins are the components of the orthonormal basis vectors in terms of the coordinate basis. Through the vielbein any vector can be expressed in terms of the other basis as
\begin{equation}
    V^a = e_\mu^{\ a}(x) V^\mu, \hspace{2cm} V^\mu = e^\mu_{\ a}(x)V^a.
\end{equation}
But how do the vielbeins transform if we change coordinates or stated differently, what transformations keep the formalism intact? To answer this, we need the transformations that preserve \eqref{eq:vielbein_g_eta} or \eqref{eq:vielbein_g_eta_respectively}, respectively. The metric $g_{\mu\nu}$ transforms under general coordinate transformations, and therefore the Greek indices of the vielbein have to transform as
\begin{equation}
    e^{a}_{\ \mu}(x^\prime) = \frac{\partial x^\nu}{\partial x^{\prime \mu}} \ e^a_{\ \nu} (x).
\end{equation}
On the other hand, flat Minkowski space is invariant under Lorentz transformations, which is why the Roman indices of the vielbein transform under Lorentz transformations, but with different Lorentz transformations at each point in spacetime:
\begin{equation}
e^{\prime a}_{\ \ \mu}(x)= \Lambda^a_{\ b}(x) e^b_{\ \mu}(x).
\end{equation}
These local position dependent Lorentz transformations fulfill at each point 
\begin{equation}
    \Lambda^a_{\ a^\prime}\Lambda^b_{\ b^\prime}\eta_{ab} = \eta_{a^\prime b^\prime}.
\end{equation}
Since derivatives are intrinsically non-local, the next question that arises is: How can we define a covariant derivative that is compatible with these transformations? The answer to this reminds very much of the Christoffel symbol $\Gamma^\nu_{\mu\sigma}$ constituting the correction term that upgrades a partial derivative to a covariant derivative in curved spacetime:
\begin{equation}
    \nabla_\mu V^\nu_{\ \ \lambda} = \partial_\mu V^\nu_{\ \ \lambda} + \Gamma^\nu_{\mu\sigma} V^\sigma_{\ \ \lambda}-\Gamma^\sigma_{\mu\lambda}V^\nu_{\ \ \sigma}.
\end{equation}
The Christoffel symbol translates vectors from one tangent space into the other, and similarly, one can define a connection that translates from one local inertial frame into the other. As shown in detail in \cite{Weinberg:1972kfs}, the so-called \textit{spin connection} allows us to define the covariant derivative in the local inertial coordinates:
\begin{equation}
    \nabla_\mu V^a_{\ \ b} = \partial_\mu V^a_{\ \ b} + \omega_{\mu \ c}^{\ a} V^c_{\ \ b} -\omega_{\mu \ b}^{\ c} V^a_{\ \ c}
\end{equation}
and is connected to the Christoffel symbol by
\begin{equation}\label{eq:spin_connection_christoffel}
    \omega_{\mu \ b}^a = e_\nu^{\ a} e^\lambda_{\ b}\Gamma^\nu_{\mu\nu} - e^\lambda_{\ b}\partial_\mu e_\lambda^{\ a}.
\end{equation}
If the covariant derivative acts on objects carrying both Roman and Greek indices, it entails the spin connection as well as the Christoffel symbol. Eventually, the spin connection defined in this way obeys
\begin{equation}\label{eq:covariant derivative_on_vielbein}
    \nabla_\mu e_\nu^{\ a}=0
\end{equation}
which allows us to pull the covariant derivative past a vielbein. 

The advantage of the vielbein formalism is that we can now write the Grassmann vectors that we used to represent spin in the second chapter as
\begin{equation}
    \Psi^\mu = e^\mu_{\ a}\Psi^a, \hspace{2cm} \Psi_\mu = e_\mu^{\ a}\Psi_a ,
\end{equation}
where the flat space Grassmann vectors fulfill the same properties as before, most importantly the Poisson bracket \eqref{eq:poisson_bracket_grassmann} and the Lorentz algebra \eqref{eq:lorentz_algebra}. The spin tensor in curved space is then defined as
\begin{equation}
    \mathcal{S}^{\mu\nu} = -i e^\mu_{\ a }e^\nu_{\ b}\Psi^a\Psi^b .
\end{equation}

\section{Spin-1/2 WQFT coupled to gravity}
Let us now use the vielbein formalism to couple the spin-1/2 WQFT to gravity. For this purpose, we simply take the action from chapter two and replace the Minkowski metric with a general metric
\begin{equation}
    \eta_{\mu\nu}\longrightarrow g_{\mu\nu}(x)=e_\mu^{\ a}(x) e_\nu^{\ b}(x) \eta_{ab}
\end{equation}
and replace derivatives with covariant derivatives
\begin{equation}
    \frac{\mathrm{d}}{\mathrm{d}\tau}\longrightarrow\frac{D}{D\tau}:= \dot{x}^\mu\nabla_\mu .
\end{equation}
Doing so, the action yields
\begin{equation}
    S=-\int\mathrm{d}\tau \left(\frac{1}{4}\dot{x}^2+\frac{i}{2}\Psi_\mu\dot{\Psi}^\mu\right) \longrightarrow -\int\mathrm{d}\tau\left(\frac{1}{4} g_{\mu\nu}\dot{x}^\mu\dot{x}^\nu + \frac{i}{2} \Psi_a\frac{D}{D\tau}\Psi^a \right),
\end{equation}
where we used that the covariant derivative can be pulled past the vielbein 
\begin{equation}
    \frac{D}{D\tau}\Psi^\mu= e^\mu_{\ a}\frac{D}{D\tau}\Psi^a 
\end{equation}
and took advantage of the orthonormality of the vielbeins \eqref{eq:orthonormality_vielbein}.
Moreover, we add the Einstein-Hilbert action 
\begin{equation}\label{eq:Einstein-Hilber_action}
    S_\mathrm{EH} = -\frac{2}{\kappa^2}\int \mathrm{d}^Dx\sqrt{-g}R ,
\end{equation} 
where $\kappa^2 = 32\pi G$ is the gravitational coupling, $R$ is the Ricci scalar and $g$ the determinant of the metric $g_{\mu\nu}$,
as well as a possible gauge fixing term $S_{\mathrm{gf}}$. Furthermore, writing out the covariant derivative in terms of the spin connection \begin{equation}
    \frac{D}{D\tau}\Psi^a = \dot{x}^\mu\nabla_\mu\Psi^a = \dot{x}^\mu\left(\partial_\mu\Psi^a + \omega_{\mu \ b}^{\ a}\Psi^b\right) = \dot{\Psi}^a+\dot{x}^\mu\omega_{\mu \ b}^{\ a}\Psi^b
\end{equation}
the full worldline action is given by\footnote{As pointed out in \cite{Jan_2021} the path integral measure becomes metric dependent in curved space which can be controlled by introducing so-called "Lee-Yang" ghosts \cite{Bastianelli_1993}. Albeit these ghosts do not contribute at tree-level, one should be aware that the action given here needs to be modified when loops are considered.}
\begin{equation}
    S = -\frac{1}{2}\int\mathrm{d}\tau\left(\frac{1}{2} g_{\mu\nu}\dot{x}^\mu\dot{x}^\nu + i \Psi_a\dot{\Psi}^a+\dot{x}^\mu\omega_{\mu \ b}^{\ a}\Psi_a\Psi^b\right)+ S_\mathrm{EH}  + S_{\mathrm{gf}} .
\end{equation}
If we now want to use the weak field expansion
\begin{equation}
    g_{\mu\nu} = \eta_{\mu\nu} + \kappa h_{\mu\nu}, \hspace{2cm }  g^{\mu\nu} = \eta^{\mu\nu} - \kappa h^{\mu\nu},
\end{equation}
we also have to expand the spin connection in orders of $\kappa$. For this, we need the weak field expansion for the vielbein, which can be deduced from
\begin{equation}\label{eq:vielbein_expansion_start}
g_{\mu\nu}(x)=\eta_{\mu\nu}+\kappa h_{\mu\nu}(x) \overset{!}{=} e_\mu^{\ a}(x) e_\nu^{\ b}(x) \eta_{ab} .
\end{equation}
First of all, we can write
\begin{equation}
    e_\nu^{\ a} = \eta^{ca}e^\mu_{\ c}g_{\mu\nu}\approx \eta^{ca}e^\mu_{\ c}(\eta_{\mu\nu}+\kappa h_{\mu\nu} ) + \mathcal{O}(\kappa^2) .
\end{equation}
Note that after the weak field expansion, the difference between Greek and Roman indices loses its meaning since all indices can now be raised and lowered with the Minkowski metric in line with the approximation.
By making a general ansatz
\begin{equation}
    e^{\ a}_{ \nu} = \eta^{\mu a}(\alpha_{\nu\mu} +\kappa\beta_{\nu\mu})
\end{equation}
and plugging it into equation \eqref{eq:vielbein_expansion_start} one can easily find the weak field expansion for the einbein
\begin{equation}
    e^{\ a}_\mu = \delta_{\mu}^{\ a}+\frac{\kappa}{2}h_{\mu}^{\ a} + \mathcal{O}(\kappa^2).
\end{equation}
Using equation \eqref{eq:spin_connection_christoffel} together with the weak field expansion of the Christoffel symbol
\begin{equation}
    \Gamma^\rho_{\mu\nu} \approx \frac{\kappa}{2}\eta^{\rho\sigma}\left(\partial_\mu h_{\sigma\nu} + \partial_\nu h_{\mu\sigma} - \partial_\sigma h_{\mu\nu}\right) + \mathcal{O}(\kappa^2)
\end{equation}
we finally find
\begin{equation}
    \omega_{\mu \  b}^{\ a} = \frac{\kappa}{2}\left(\partial_b h_{\mu}^{\ a} - \partial^a h_{\mu b}\right) + \mathcal{O}(\kappa^2) .
\end{equation}
Hence, the action in terms of the graviton field $h_{\mu\nu}$ can be written to linear order in $\kappa$ as
\begin{equation}\label{eq:expanded_gravity_worldline_action}
    S = -\frac{1}{2}\int\mathrm{d}\tau\frac{1}{2}\dot{x}^2 + i\Psi_a\dot{\Psi}^a + \frac{\kappa}{2}\dot{x}^\mu \left(h_{\mu\nu}\dot{x}^\nu +i\left(\partial_b h_{\mu a}-\partial_a h_{\mu b}\right)\Psi^a\Psi^b \right) + \mathcal{O}(\kappa^2) +S_\mathrm{EH}+S_\mathrm{gf} .
\end{equation}
As in the previous chapters, we can further apply the straight line expansion
\begin{equation}
    x^\mu(\tau) = b^\mu + v^\mu \tau + z^\mu (\tau), \hspace{2cm}\Psi^\mu(\tau) = \eta^\mu + \psi^\mu(\tau)
\end{equation}
and ignore all constants and boundary terms in the action, which results in
\begin{equation}\begin{aligned}\label{eq:expanded_WQFT_action_gravity}
     S =& -\frac{1}{2}\int\mathrm{d}\tau\frac{1}{2}\dot{z}^2 + i\psi_a\dot{\psi}^a + \frac{\kappa}{2} \biggl( \biggr. \left(v^\mu v^\nu + \dot{z}^\mu \dot{z}^\nu + v^\mu \dot{z}^\nu+v^\nu \dot{z}^\mu \right) h_{\mu\nu} \\ &  \qquad
     +i(v^\mu+\dot{z}^\mu ) \left(\partial_b h_{\mu a}-\partial_a h_{\mu b}\right)( \eta^a +\psi^a)(\eta^b+ \psi^b ) \biggl.\biggr) + \mathcal{O}(\kappa^2) +S_\mathrm{EH}+S_\mathrm{gf} . \end{aligned}
\end{equation}
To read off the Feynman rules, we transform to Fourier space. The Fourier transform of the worldline fields was already given in equation \eqref{eq:fourierfields_with_spin} and Fourier transforming the graviton field results in
\begin{equation}\label{eq:fourier_h}
    h^{\mu\nu} = \int_k h^{\mu\nu}(-k)e^{ik\cdot b} = \sum_{n=0}^\infty \frac{i^n}{n!}\int_{k, \omega_1,\dots , \omega_n} e^{ik\cdot } e^{i(k\cdot v + \sum_{i=1}^n \omega_i )\tau} \left(\prod_{i=1}^n k\cdot z(-\omega_i)\right) h^{\mu\nu}(-k).
\end{equation}
The kinetic terms in the action \eqref{eq:expanded_WQFT_action_gravity} are the same as in the flat space case and therefore the propagators are again given by \eqref{eq:propagatorz} and \eqref{eq:propagatorPsi}.
Using equation \eqref{eq:fourier_h}, the lowest order vertex in $\kappa$ and in $\hbar$ can be read off the action \eqref{eq:expanded_WQFT_action_gravity}:
\begin{align}\label{eq:vertexh}
	  \begin{tikzpicture}[baseline={(current bounding box.center)}]
	  \coordinate (in) at (-1,0);
	  \coordinate (out) at (1,0);
	  \coordinate (x) at (0,0);
	  \node (k) at (0,-1.3) {$h_{\mu\nu}(k)$};
	  \draw [dotted] (in) -- (x);
	  \draw [dotted] (x) -- (out);
	  \draw [graviton] (x) -- (k);
	  \draw [fill] (x) circle (.08);
	  \end{tikzpicture}=-i\frac{\kappa}{4}\ e^{ik\cdot b}\dd(k\cdot v)\left(v^\mu v^\nu +2k_a\eta^a\eta^{(\nu} v^{\mu )} \right) .
\end{align}
To compare with quantum field theory in the next section, we can translate the description in terms of the Grassmann vectors $\eta^\mu$ to the usual description in terms of gamma matrices. To this end we use the inverse symbol map given in \eqref{eq:symbol_map_in_4d} in 4-dimensions: $\eta^\mu\eta^\nu\rightarrow -\frac{1}{4}\left[\gamma^\mu , \gamma^\nu\right]$
and impose the observation of the previous chapters $v^\mu=(p+p^\prime)^\mu$. We can then write:
\begin{equation}\begin{aligned}
    \eqref{eq:vertexh} &= \dots \left(\frac{1}{2}(p+p^\prime)^\mu \left((p+p^\prime)^\nu - \frac{1}{2} \left[ \slashed k, \gamma^\nu\right] \right) + (\mu \leftrightarrow \nu) \right) \\
    & = \dots \left(\frac{1}{2}(p+p^\prime)^\mu \left(\slashed p^\prime \gamma^\nu+\gamma^\nu \slashed p \right) + (\mu \leftrightarrow \nu) \right) \\
    & = \dots \left(\frac{1}{2}(p+p^\prime)^\mu \left((\slashed p^\prime -m) \gamma^\nu + \gamma^\nu (\slashed p +m)\right) + (\mu \leftrightarrow \nu) \right) \\
     & = \dots \left( (\slashed p^\prime- m)\gamma^{(\mu}(p+p^\prime)^{\nu)} + \gamma^{(\mu}(p+p^\prime)^{\nu )}(\slashed p +m)\right) ,
     \end{aligned}\end{equation}
where momentum conservation and equation \eqref{eq:gamma_matrix_decomp_2} was used from the first to the second line. If we set $p$ and $p^\prime$ on-shell, i.e. use that $\bar{u}(p^\prime)(\slashed p^\prime- m)=0$, the first term is zero, and we can write
\begin{equation}\label{eq:eq:vertexh_on_shell}
     \eqref{eq:vertexh} \longrightarrow-i\frac{\kappa}{4}\ e^{ik\cdot b}\dd(k\cdot v)\gamma^{(\mu}(p+p^\prime)^{\nu )}(\slashed p +m) :=\widehat{K}^{\mu\nu}_1(p,p^\prime )
\end{equation}

\section{Comparison to quantum gravity}
Now let us compare the trivalent vertex in the spin-1/2 WQFT to a fermionic theory of quantum gravity given in appendix \ref{appendix:Feynmanrules_quantum_gravity}. The amputated 3-pt function is simply given by vertex \eqref{eq:3vertex_quantum_gravity}:
\begin{equation}\label{eq:3pt_quantum_gravity}
\begin{aligned}
    \begin{tikzpicture}[baseline={(current bounding box.center)}]
    \coordinate (in) at (-1,0);
	  \coordinate (out) at (1,0);
	  \coordinate (x) at (0,0);
	  \node (k) at (0,-1.3) {$h_{\mu\nu}(k)$};
	  \draw [zParticle] (in) -- (x) node [midway, above] {$\underrightarrow{p}$};
	  \draw [zParticle] (x) -- (out) node [midway, above] {$\underrightarrow{p^\prime}$};
	  \draw [graviton] (x) -- (k);
	  \draw [fill] (x) circle (.04);
	  \end{tikzpicture} &= i \frac{\kappa}{2}\left[ \eta^{\mu\nu}\left(\slashed p - m -\frac{1}{2}\slashed k\right) - \gamma^{(\mu}( p-\frac{1}{2}k)^{\nu )}\right] \\
	  &=i \frac{\kappa}{4}\left[ \eta^{\mu\nu}\left(\slashed p - m + \slashed p^\prime - m \right) - \gamma^{(\mu}( p+ p^\prime)^{\nu )}\right] ,
	  \end{aligned}
\end{equation}
where momentum conservation $p=p^\prime + k$ was used from the first to the second line. On-shell, when sandwiched by polarization vectors, the first term goes to zero since
$(\slashed p-m) u(p)$ and $\bar{u}(p)(\slashed p^\prime -m)$. As a consequence, we can write 
\begin{equation}\label{eq:3pt_quantum_gravity_onshell}
    \eqref{eq:3pt_quantum_gravity} \longrightarrow -i \frac{\kappa}{4}  \gamma^{(\mu}(p+p^\prime)^{\nu )} := \widehat{S}^{\mu\nu}_1(p,p^\prime )
\end{equation}
on-shell.

Comparing \eqref{eq:3pt_quantum_gravity_onshell} and \eqref{eq:eq:vertexh_on_shell}
we find the same match as in the electromagnetic case in the second chapter
\begin{equation}\label{eq:link_1graviton_dressed_propagator}
   \widehat{K}^{\mu\nu}_1(p,p^\prime )(\slashed p +m)^{-1} = e^{ik\cdot b}\dd(k\cdot v) \ \widehat{S}^{\mu\nu}_1(p,p^\prime )
\end{equation}
explicitly confirming that equation \eqref{link WQFT and QED dressed S} also holds for dressed fermionic propagators in a gravitational background field for the one graviton dressed propagator. Verifying this statement for higher orders in the coupling $\kappa$ gets fairly complicated. Since we have an infinite series in $\kappa$ already at the level of the action in the weak field expansion, a general all-order comparison as in QED or scalar QED is not possible. Hence, one is left with comparing order for order. Thus, the natural next step would be to compare with the 4-pt function. But already here, the expressions become very cumbersome, which calls for the need of computer analysis and therefore exceed the scope of this thesis. Another approach could be to derive a second-order theory of quantum gravity which is more similar to the WQFT and therefore easier to compare. To my knowledge, such a theory has not been developed yet and would have to be established. However, this would also be more difficult than in QED since the theory of fermions coupled to gravity has not only one but two different vertices which are each of its own much more involved. Nevertheless, following the discussion in this thesis, we can be confident that \eqref{eq:link_1graviton_dressed_propagator} also holds for higher numbers of graviton interactions.
\unchapter{Conclusions}

In this work, we have successfully derived a spin-1/2 WQFT coupled to an electromagnetic or gravitational field that holds up to linear order in spin and investigated its relation to quantum field theory. We have shown that expectation values in this (spin-1/2) WQFT can be directly related to S-matrices in quantum field theory in the classical limit. For this, we followed a similar strategy as in \cite{Jan_2021}, where a similar link was found for a WQFT for non-spinning particles coupled to gravity. First, we demonstrated that S-matrices in QFT can be expressed via dressed propagators in the classical limit. In the next step, we derived a direct link between dressed propagators in both theories and inferred from it a relation between S-matrices and WQFT expectation values. 

In the first chapter, we looked at a WQFT for non-spinning particles coupled to an electromagnetic field and compared it to scalar QED in the above-described manner. 
In the second chapter, we repeated the procedure in the more advanced case of spin-1/2 particles coupled to an electromagnetic field. In the last chapter, we finally discussed how a spin-1/2 WQFT could be coupled to gravity and implemented in curved space. The there established spin-1/2 WQFT coupled to a gravitational field was then compared with a fermionic theory of quantum gravity.

In all of these instances, we found direct links at the level of on-shell dressed propagators or even for general S-matrices and WQFT expectation values under the condition that we impose $v=p+p^\prime$ (or $v=\frac{p+p^\prime}{2m}$ if interpreted as an actual velocity). Hence, the velocity, one of the natural parameters on the WQFT side, corresponds to the average of the in-going and out-going momenta in a QFT scattering amplitude. This comes from the fact that we had to use time symmetric propagators in the WQFT to find a match with Feynman propagators in QFT. As already anticipated in \cite{Jan_2021} and here again explicitly confirmed, the time symmetric propagators in the WQFT are given by the average of the advanced and retarded worldline propagators. 
In order that the linking relation also holds in the presence of spin, we further had to identify the different representations of spin in the spin-1/2 WQFT (Grassmann vectors) and fermionic QFT (gamma matrices). This identification is given by the symbol map \eqref{eq:symbol_map_in_4d}.

Furthermore, we have seen that the more natural link of the spin-1/2 WQFT to QFT is given with a modified scalar QED, which we refer to as \textit{second order QED} or the \textit{kernel} of the QED dressed propagator. As shown in the second chapter, QED can then be deduced from the second order QED/the kernel.

On the level of dressed propagators, we were able to sum up infinite towers of worldline loops, incorporating all quantum contributions in the sense that all terms of the $\hbar$ expansion are retained. To my knowledge, this has not been done before. For this, we derived a master formula that gives a solution for arbitrary numbers of worldline loops in appendix \ref{appendix:worldline_integrals}. By applying this formula to diagrams describing the radiation of two photons from a single worldline, we found that summing up all internal worldline loops generates an internal scalar propagator even in the spin 1/2 case.
With this, we showed that (at least for the dressed propagator) the WQFT also holds in the non-classical regime if we include all possible worldline loops.

For future projects, it would be interesting to look at WQFTs being able to describe higher spin, e.g., a spin-1 WQFT can be formulated by introducing two instead of one Grassmann field or, alternatively, one complex Grassmann field. In \cite{jakobsen_2021gravitational_bremstrahlung_spinning,Jakobsen_2021_Susy,Jakobsen_22_Conservative_and_Radiative, Shi_2021_Classical_double_copy}, a spin-1 WQFT has already been established, but a general relation to scattering amplitudes in QFT is still missing.

\unchapter{Acknowledgement}

I would like to express my gratitude to everyone who helped and supported me in the process of this thesis. 
\\ 

Foremost, I wish to thank my supervisors Prof. Dr. Jan Plefka and Dr. Jan Steinhoff for giving me the chance to deal with such an interesting topic in the framework of my master thesis, for leading me trough the process and letting me be a part of their research groups at Humboldt-University in Adlershof and the Max Planck Institute for Gravitational Physics in Golm. 
\\ 

Furthermore, I want to thank Dr. Gustav Mogull for taking the time to answer all my questions and giving me advice when I struggled with a problem. I appreciated all the interesting talks we had over Zoom or at the institute when Corona allowed it.
\\ 

I am especially grateful to Lea Rektorschek who supported me particularly in the final phase of this thesis, who was there for me the entire time, tried to make working easier for me and proofread the thesis.
\\

I also want to thank Leo Shaposhnik, Jakob Wintergerst and Vincent Meyer for proofreading the thesis.


\appendix
\chapter{Feynman rules} \label{apendix:FeynmanRules}
\section{Feynman rules of Scalar QED}
The Feynman rules for scalar quantum electrodynamics are given by a scalar Feynman propagator
\begin{equation}
\begin{tikzpicture}
\begin{feynman}
\vertex (x) at (-1,0);
\vertex (y) at (1,0);
\diagram{
(x) -- [charged scalar, momentum={[arrow shorten=0.12mm] \(p\)}] (y)};
\end{feynman}
\end{tikzpicture} = \frac{i}{p^2-m^2+i\epsilon}
\end{equation}
and a trivalent and a quartic vertex:
\begin{equation}\label{eq:vertices_scalar_QED}
\begin{aligned}
    \begin{tikzpicture}[baseline]
    \begin{feynman}
    \coordinate (x) at (0,0);
    \coordinate (in) at (-1.15*0.5,-1.15*0.87);
    \coordinate (out) at (-1.15*0.5,1.15*0.87);
    \coordinate (k) at (1.15,0);
    \diagram{ (in) -- [charged scalar, edge label'=\(p\)] (x) --[charged scalar, edge label'=\(p^\prime\)] (out)};
    \draw [photon2] (x) -- (k) node [right] {$\mu$};
    \draw [fill] (x) circle (.04);
    \end{feynman}
    \end{tikzpicture}  = -ie(p+p^\prime)^\mu
    & \hspace{1cm}& \begin{tikzpicture}[baseline]
    \begin{feynman}
    \node (a) at (-1,1) {$\mu$};
    \node (b) at (1,1) {$\nu$};
    \coordinate (c) at (1,-1);
    \coordinate (d) at (-1,-1);
    \coordinate (x) at (0,0);
    \diagram{(d) -- [charged scalar] (x) -- [charged scalar] (c)};
    \draw [photon2] (a) -- (x);
    \draw [photon2] (b) -- (x);
    \draw [fill] (x) circle (.04);
    \end{feynman}
    \end{tikzpicture} = 2ie^2\eta^{\mu\nu}.
    \end{aligned}
\end{equation}
For scalar QED the conventions in this thesis follow \cite{Peskin:1995ev}.

\section{QED-Feynman rules}
\label{QED-feynman_rules}
The full action in quantum electrodynamics (QED) is given by
\begin{equation}\label{actionQED}
    S_\mathrm{QED}=\sum_{i} \underbrace{\int\mathrm{d}^Dx\overline{\psi}_i\left(i\slashed D-m\right)\psi_i}_{\mathrm{S_i}}\underbrace{-\int\mathrm{d}^Dx\frac{1}{4}F_{\mu\nu}F^{\mu\nu}}_{\mathrm{S_\mathrm{A}}}+S_{\mathrm{gf}}.
\end{equation}
As described in any textbook on quantum field theory (e.g. \cite{SchwartzMatthewD2014Qfta}), the Feynman rules for quantum electrodynamics are given by a fermionic propagator:
\begin{equation}
\begin{tikzpicture}
\begin{feynman}
\vertex (x) at (-1,0);
\vertex (y) at (1,0);
\diagram{
(x) -- [fermion, momentum={[arrow shorten=0.12mm] \(p\)}] (y)};
\end{feynman}
\end{tikzpicture} = \frac{i(\slashed p + m)}{p^2-m^2+i\epsilon}
\end{equation}
and a trivalent vertex:
\begin{equation}\label{eq:vertices_QED}
\begin{aligned}
    \begin{tikzpicture}[baseline]
    \begin{feynman}
    \coordinate (x) at (0,0);
    \coordinate (in) at (-1.15*0.5,-1.15*0.87);
    \coordinate (out) at (-1.15*0.5,1.15*0.87);
    \coordinate (k) at (1.15,0);
    \diagram{ (in) -- [fermion] (x) --[fermion] (out)};
    \draw [photon2] (x) -- (k) node [right] {$\mu$};
    \draw [fill] (x) circle (.04);
    \end{feynman}
    \end{tikzpicture}  = -ie\gamma^\mu.
    \end{aligned}
\end{equation}

\section{Feynman rules for Quantum Gravity}\label{appendix:Feynmanrules_quantum_gravity}
In this appendix, we list the Feynman rules of a fermionic quantum field theory coupled to gravity given in reference \cite{schuster}. The action of this theory is given by
\begin{equation}
    S = \int \mathrm{d}^4 x \sqrt{-g} \bar{\psi} (i\slashed D -m)\psi .
\end{equation}
The propagator arising from this action is a fermionic propagator:
\begin{equation}
\begin{tikzpicture}[]
\begin{feynman}
  \coordinate (in) at (-1,0);
  \coordinate (out) at (1,0);
  \diagram{
(in) -- [fermion, momentum={[arrow shorten=0.12mm] \(p\)}] (out)};
  \end{feynman}\end{tikzpicture} = \frac{1}{\slashed p-m} = \frac{\slashed p + m}{p^2-m^2}.
\end{equation}
As derived in \cite{schuster}, there is one vertex constituting a fermion-graviton interaction of the order $\kappa$:
\begin{equation}\label{eq:3vertex_quantum_gravity}
\begin{aligned}
    \begin{tikzpicture}[baseline]\begin{feynman}
    \coordinate (in) at (-1.2*0.87,-1.2*0.5);
	  \coordinate (out) at (1.2*0.87,-1.2*0.5);
	  \coordinate (x) at (0,0);
	  \node (k) at (0,1.6) {$h_{\mu\nu}(k)$};
	   \diagram{
(in) -- [fermion, momentum={[arrow shorten=0.12mm] \(p\)}] (x)};
 \diagram{
(x) -- [fermion, momentum={[arrow shorten=0.12mm] \(p^\prime\)}] (out)};
	  \draw [graviton] (x) -- (k);
	  \draw [fill] (x) circle (.04);
	  \end{feynman}
	  \end{tikzpicture} &= i \frac{\kappa}{2}\left[ \eta^{\mu\nu}\left(\slashed p - m -\frac{1}{2}\slashed k\right) - \gamma^{(\mu}( p-\frac{1}{2}k)^{\nu )}\right] 
	  \end{aligned}
\end{equation}
and one vertex of $\kappa^2$:
\vspace{-1.5cm}
\begin{equation}\label{eq:4vertex_quantum_gravity}\def\arraystretch{1.7}
\begin{aligned}
    \begin{tikzpicture}[baseline={(current bounding box.center)}]
    \begin{feynman}
    \coordinate (in) at (-1.2,-1.2);
	  \coordinate (out) at (1.2,-1.2);
	  \coordinate (x) at (0,0);
	  \node (k1) at (-1.5,1.5) {$h_{\mu_1\nu_1}(k_1)$};
	  \node (k2) at (1.5,1.5) {$h_{\mu_2\nu_2}(k_2)$};
	  \draw [graviton] (x) -- (k1);
	  \draw [graviton] (x) -- (k2);
	  \draw [fill] (x) circle (.04);
	  \diagram{
(in) -- [fermion, momentum={[arrow shorten=0.12mm] \(p\)}] (x)};
\diagram{
(x) -- [fermion, momentum={[arrow shorten=0.12mm] \(p^\prime\)}] (out)};
	  \end{feynman}
	  \end{tikzpicture} & \begin{array}{ll} & \\ & \\ & \\ = &  i \kappa^2 \biggl[ \biggr. - (\slashed k_1 + \slashed k_2 )\left(\frac{5}{16} \gamma^{(\mu_1} \eta^{\nu_1 )(\mu_2}\gamma^{\nu_2)} +\frac{1}{16}\eta^{\mu_1\mu_1}\eta^{\mu_2\mu_2} \right) \\
	  & \qquad + \gamma^{(\mu_1}\eta^{\nu_1)(\mu_2}(\frac{3}{8}p-\frac{1}{2}k_1+\frac{3}{8}k_2)^{\nu_2)} - \gamma^{(\mu_1} (\frac{1}{4}p - \frac{1}{8}k_2)^{\nu_1)}\eta^{\mu_2\nu_2}\\
	  & \qquad -\frac{1}{4}(\slashed p -m) \mathcal{P}^{\mu_1\nu_1\mu_2\nu_2} + ( 1 \longleftrightarrow 2 ) \biggl. \biggr] \end{array}
	  \end{aligned}
\end{equation}
with $\mathcal{P}^{\mu\nu\alpha\beta}:= \eta^{\mu(\alpha}\eta^{\beta ) \nu} - \frac{1}{2}\eta^{\mu\nu}\eta^{\alpha\beta}$.

\chapter{Gamma matrix identities}\label{appendix:gamma matrix}
Starting from the defining anticommutation relation:
\begin{equation}
    \{\gamma^\mu,\gamma^\nu\}=2\eta^{\mu\nu}
\end{equation}
we can observe that we can decompose expressions involving two gamma matrices in the following manner:
\begin{equation}\label{eq:gamma_matrix_decomp_2}
    \gamma^\alpha\gamma^\beta=\eta^{\alpha\beta}+\frac{1}{2}\left[\gamma^\alpha,\gamma^\beta\right].
\end{equation}
Phrased in more general terms, we see that the appearance of gamma matrices in a given expression is not unique, which makes it hard to compare more involved terms containing gamma matrices. Therefore, it will be convenient to decompose such terms in a unique basis
\begin{equation}
    X = x_A\Gamma^A,
\end{equation}
where the $\Gamma^A$'s are the basis matrices.
\paragraph{Basis of $\mathbb{M}(4,\mathbb{C})$ from gamma matrices} 
\ \\ \ \\ 
Conveniently, a basis of any $4 \times 4$ matrix is given by
\begin{equation}
    \{\Gamma^A\} = \{\mathbb{1}, \gamma^\mu, \sigma^{\mu\nu}, \gamma^\mu\gamma_5, \gamma_5\},
\end{equation}
with $\sigma^{\mu\nu} = \frac{i}{2}\left[\gamma^\mu,\gamma^\nu\right] $ \cite{Peskin:1995ev}. Since these basis matrices are orthogonal under a trace, the coefficients $x_A$ can be found by
\begin{equation} \label{coefficients_gamma_matrix_decomp}
    x_A = \frac{1}{4}\mathrm{tr}\{X\Gamma_A\},
\end{equation}
where $\Gamma_A$ is the inverse of $\Gamma^A$. The inverse basis matrices are given by $\{\Gamma_A\} = \{\mathbb{1}, \gamma_\mu, \sigma_{\mu\nu}, \gamma_5\gamma_\mu, \gamma_5\}$. Using \eqref{coefficients_gamma_matrix_decomp}, one can easily find the decomposition into the basis $\Gamma^A$ of a given matrix $X$. For some useful examples, this decomposition is listed below:

\begin{equation}
\begin{aligned} \label{gamma_matrix_decomp}
    \slashed a \slashed b = & a\cdot b +\frac{1}{2}\left[ \slashed a, \slashed b \right] \\ \ & \ \\
    \slashed a \slashed b \slashed c = & a\cdot b \ \slashed c -a\cdot c \ \slashed b + b\cdot c \ \slashed a +i a_\nu  b_\rho  c_\sigma  \varepsilon^{\mu\nu\rho\sigma} \\ \ & \ \\
    \slashed a \slashed b \slashed c \slashed d = &\left(a\cdot b \ c\cdot d - a\cdot c \ b\cdot d + a\cdot d \ b \cdot c\right)\mathbb{1} - i \varepsilon\left(a,b,c,d\right)\gamma_5 \\
    &+\frac{1}{2}\left(\left[\slashed a,\slashed b\right]c\cdot d -\left[\slashed a,\slashed c\right] b\cdot d + \left[\slashed a, \slashed d\right] b\cdot c + \left[\slashed b ,\slashed c\right] a\cdot d - \left[\slashed b, \slashed d\right] a\cdot c + \left[\slashed c, \slashed d\right] a\cdot b\right)
    \end{aligned}
\end{equation}
with $\varepsilon\left(a,b,c,d\right)=a_\mu b_\nu c_\rho d_\sigma \varepsilon^{\mu\nu\rho\sigma}$ and $\varepsilon^{0123}=1$.
\chapter{Gaussian Integration}\label{appendix:gaussian_integration}
In this appendix, we provide formulas for the Gaussian functional integrals appearing in the prior calculations. 

\section{Gaussian functional integrals}\label{appendix:gaussian_integrals}
First, let us consider Gaussian integrals of the form
\begin{equation}
Z_0 = \int \mathcal{D}\left[ q \right] e^{-\frac{i}{4}\int_0^s \mathrm{d}\tau \int_0^{s^\prime}q(\tau)\mathcal{M}(\tau,\tau^\prime)q(\tau^\prime)},
\label{gauss_Z0}
\end{equation}
where $\mathcal{M}$ is assumed to be a symmetric operator (in our case: $\mathcal{M}(\tau,\tau^\prime) = -\delta(\tau-\tau^\prime)\frac{\partial^2}{{\partial\tau^\prime}^2}$), which also implies that the inverse operator $\mathcal{M}^{-1}$ is symmetric.
Naturally, the inverse operator has to obey:
\begin{equation}
    \int_0^r\mathrm{d}r\mathcal{M}^{-1}(\tau,r)\mathcal{M}(r,\tau^\prime) = \delta(\tau-\tau^\prime).
\end{equation}
By shifting $q(\tau) \rightarrow q(\tau) \pm \int_o^s\mathrm{d}r j(r)\mathcal{M}^{-1}(r,\tau)$ we can introduce a linear term in $q$ in the exponent of \eqref{gauss_Z0} and express the obtained expression via \eqref{gauss_Z0}:
\begin{equation}
\int \mathcal{D}\left[ q \right] e^{-\frac{i}{4}\int_0^s \mathrm{d}\tau \int_0^{s^\prime}q(\tau)\mathcal{M}(\tau,\tau^\prime)q(\tau^\prime) \pm \frac{1}{2}\int_0^s\mathrm{d}\tau q(\tau)j(\tau)}
= Z_0 \ e^{-i\int_0^s\mathrm{d}\tau\int_0^{s^\prime}\mathrm{d}\tau^\prime j(\tau)\mathcal{M}^{-1}(\tau,\tau^\prime)j(\tau^\prime)}.
\end{equation}

\section{Gaussian functional Grassmann integrals}\label{appendix:gaussian_grassmann}
Now, let us look at a similar integral type, but this time involving the Grassmann field $\psi(\tau)$:
\begin{equation}
    Z_0 = \int \mathcal{D}[\psi]e^{\frac{1}{2}\int_0^s\mathrm{d}\tau\int_0^{s^\prime}\mathrm{d}\tau^\prime \psi(\tau)\mathcal{M}(\tau,\tau^\prime)\psi(\tau^\prime)}.
\end{equation}
Here, we assume $\mathcal{M}$ (and also $\mathcal{M}^{-1}$) to be antisymmetric. Again, shifting the Grassman field by $\psi(\tau) \rightarrow q(\tau) \pm \int_o^s\mathrm{d}r j(r)\mathcal{M}^{-1}(r,\tau)$ leads to
\begin{equation}
    \int \mathcal{D}[\psi]e^{\frac{1}{2}\int_0^s\mathrm{d}\tau\int_0^{s^\prime}\mathrm{d}\tau^\prime \psi(\tau)\mathcal{M}(\tau,\tau^\prime)\psi(\tau^\prime) \pm \int_0^s\mathrm{d}\tau\psi(\tau)j(\tau)}
    = Z_0 \ e^{\frac{1}{2}\int_0^s\mathrm{d}\tau\int_0^{s^\prime}\mathrm{d}\tau^\prime j(\tau)\mathcal{M}^{-1}(\tau,\tau^\prime)j(\tau^\prime)}.
\end{equation}

\chapter{Worldline Integrals}\label{appendix:worldline_integrals}
In this appendix, we calculate the worldline integrals emerging in this thesis. All of these integrals are over products of time-symmetric propagators of some order $n$, given by the average of the retarded and the advanced $i\epsilon$-prescription:
\begin{equation}\label{eq:n-th_order_wordline_propagator}
    D^n(\omega):=\left(\frac{1}{(\omega + i \epsilon )^n} + \frac{1}{(\omega - i \epsilon )^n}\right) .
\end{equation}
To be able to also express the results, it will be convenient to define a slightly more general type of equation \eqref{eq:n-th_order_wordline_propagator}:
\begin{equation}\label{eq:definition_of_P}
    \boldsymbol{P^n_\pm}(\omega) :=\left(\frac{1}{(\omega + i \epsilon )^n} \pm \frac{(-1)^n}{(\omega - i \epsilon )^n}\right) .
\end{equation}
This definition contains the $n$-th order propagators and is related to \eqref{eq:n-th_order_wordline_propagator} by
\begin{equation}
    D^n(\omega) = 
    \begin{cases} 
    \boldsymbol{P^n_+} (\omega) & \text{if $n$ is even} \\
    \boldsymbol{P^n_-} (\omega) & \text{if $n$ is odd} 
    \end{cases}
\end{equation}
and has the property $\boldsymbol{P^n_\pm}(-\omega)=\pm \boldsymbol{P^n_\pm}(\omega)$.
The integrals that will be needed are each over a product of two of the $ \boldsymbol{P^n_\pm}$'s:
\begin{equation}\label{eq:considered_worldline_integrals}
    \int_\omega  \boldsymbol{P^n_\pm}(\omega) \boldsymbol{P^m_\pm}(\omega+b) \hspace{1cm}\text{and}\hspace{1cm}  \int_\omega  \boldsymbol{P^n_\pm}(\omega) \boldsymbol{P^m_\mp}(\omega+b) .
\end{equation}
To solve these integrals, let us first consider the individual terms that one obtains when inserting the definition of the $ \boldsymbol{P^n_\pm}$'s \eqref{eq:definition_of_P} into the integrals in \eqref{eq:considered_worldline_integrals}. The first one is
\begin{equation}
    I^{nm}_{\pm\pm}(b) := \int_\omega\left(\frac{1}{\omega\pm i\epsilon)^n}\frac{1}{(\omega+b\pm i\epsilon)^m}\right) .
\end{equation}
Using the Residue theorem (as in any textbook on complex analysis, e.g. in \cite{janich2006funktionentheorie}), this integral is zero for $n>0$ and $m>0$ since one can always find a closed contour representing the integration over the real axis that does not contain any poles. Consequently, we find
\begin{equation}
    I^{nm}_{\pm\pm}(b) = 0
\end{equation}
for  $n>0$ and $m>0$.
\begin{figure}
    \centering
   \begin{tikzpicture}[decoration={
    markings,
    mark=at position 0.5 with {\arrow[scale=1.5,>=stealth]{>}}}]
   \draw[->] (-2.3,0) -- (2.3,0) node (xaxis) [right] {$\operatorname{Re}(\omega)$};
   \draw[->] (0,-1.3) -- (0,2.3) node (yaxis) [above] {$\operatorname{Im}(\omega)$};
   \draw[dashed, very thick, postaction={decorate}] (2,.05) to [out=90, in=0] (0,2) to [out=180, in=90] (-2,.05) to [out=0, in=180] (2,.05);
   \draw [fill] (0,.5) circle (.05) node [right] {$\ \epsilon$};
   \draw [fill] (-1,-.5) circle (.05);
   \node at (0,.5) {$ -$};
    \node at (0,-.5) {$ -$};
    \node at (0,-.5) [right] {\ -$\epsilon$};
    \node at (-1,0) {\rotatebox[origin=c]{90}{$-$}};
    \node at (-1,0.1) [above] {-$b$};
   \end{tikzpicture}
    \caption{Poles and contour of choice for the calculation of $I^{nm}_{-+}(b)$ in equation \eqref{eq:contour_calculation}}
    \label{fig:contour_picture}
\end{figure}
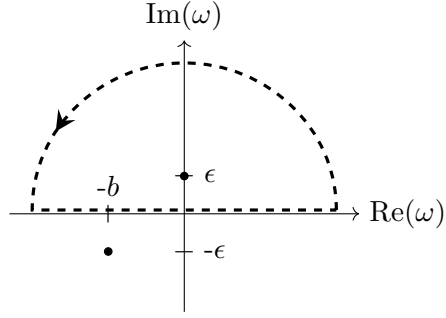
The second integral type present in \eqref{eq:considered_worldline_integrals} is
\begin{equation}
    I^{nm}_{\mp\pm}(b) := \int_\omega\left(\frac{1}{\omega\mp i\epsilon)^n}\frac{1}{(\omega+b\pm i\epsilon)^m}\right) .
\end{equation}
For the upper sign, we can choose the contour pictured in figure \ref{fig:contour_picture} containing one pole. Again using the residue theorem yields:
\begin{equation}\label{eq:contour_calculation}
    \begin{aligned}
     I^{nm}_{\mp\pm}(b) &= \pm i \ \operatorname{Res}_{\pm i\epsilon} \left(\frac{1}{\omega\mp i\epsilon)^n}\frac{1}{(\omega+b\pm i\epsilon)^m}\right)\\ 
     & = \pm i \ \textstyle\frac{1}{(n-1)!}\displaystyle\lim_{\omega\rightarrow\pm i\epsilon} \partial^{(n-1)}_\omega \left(\frac{1}{(\omega+b\pm i\epsilon)^m}\right) \\
     & = \pm i \ (-1)^{n-1} \frac{(m+n-2)!}{(n-1)!(n-1)!} \ \lim_{\omega\rightarrow\pm i\epsilon}\frac{1}{(\omega+b\pm i\epsilon)^{m+n-1}} \\
     & = \pm i \ (-1)^{n-1} \frac{(m+n-2)!}{(n-1)!(n-1)!} \ \frac{1}{(b\pm i\epsilon)^{m+n-1}} . 
    \end{aligned}
\end{equation}
With this preparation, we are now ready to tackle the integrals in \eqref{eq:considered_worldline_integrals}:
\begin{equation} \label{eq:P+P+}
    \begin{aligned}
    \int_\omega  \boldsymbol{P^n_\pm}(\omega) \boldsymbol{P^m_\pm}(\omega+b) & = I^{nm}_{++}(b) \pm (-1)^m I^{nm}_{+-} \pm (-1)^n I^{nm}_{-+} + (-1)^{n+m} I^{nm}_{--} \\
    & = \mp \ i \ \frac{(m+n-2)!}{(n-1)!(n-1)!} \ \boldsymbol{P^{n+m-1}_+}(b)
    \end{aligned}
\end{equation}
and 
\begin{equation}
    \begin{aligned}\label{eq:P+P-}
    \int_\omega  \boldsymbol{P^n_\pm}(\omega) \boldsymbol{P^m_\mp}(\omega+b) & = I^{nm}_{++}(b) \mp (-1)^m I^{nm}_{+-} \pm (-1)^n I^{nm}_{-+} - (-1)^{n+m} I^{nm}_{--} \\
    & = \mp \ i \ \frac{(m+n-2)!}{(n-1)!(n-1)!} \ \boldsymbol{P^{n+m-1}_-}(b) .
    \end{aligned}
\end{equation}
Equipped with equation \eqref{eq:P+P+} and \eqref{eq:P+P-} we can now solve all worldline integrals emerging in this thesis:
\begin{enumerate}
    \item[1.)] The first one is:
    \begin{equation}
    \begin{aligned}
    \bigintss_{\displaystyle\omega_1,\dots,\omega_n } \hspace{-1cm} \dd{\left(\textstyle \sum^n_{i=1}\omega_i-a\right)} \prod_{i=1}^{n}D^2(\omega_i)  = \bigintss_{\displaystyle\omega_1,\dots,\omega_n } \hspace{-1cm} \dd{\left(\textstyle \sum^n_{i=1}\omega_i-a\right)} \prod_{i=1}^{n}\boldsymbol{P^2_+}(\omega_i) \qquad \qquad\qquad  \\
    \qquad = \bigintss_{\displaystyle\omega_1,\dots,\omega_{n-2} } \prod_{i=1}^{n-2}\boldsymbol{P^2_+}(\omega_i)\bigintss_{\displaystyle\omega_{n-1}}\boldsymbol{P^2_+}(\omega_{n-1})\boldsymbol{P^2_+}(\omega_{n-1}+\textstyle \sum^{n-2}_{i=1}\omega_i-a)
    \end{aligned}
\end{equation}
Now, according to equation \eqref{eq:P+P+} each integration raises the index of $\boldsymbol{P^n_+}$ by one, gives a factor of $-i$ and an additional factor coming from the factorial term starting at $2$ and going up to $n$. Consequently, we obtain:
\begin{equation}
    \begin{aligned}
    \bigintss_{\displaystyle\omega_1,\dots,\omega_n } \hspace{-1cm} \dd{\left(\textstyle \sum^n_{i=1}\omega_i-a\right)} \prod_{i=1}^{n}D^2(\omega_i)  = n! \ (-i)^{n-1} \  \boldsymbol{P^{n+1}_+}(a) .
    \end{aligned}
\end{equation}

\item[2.)] The second one is:
\begin{equation}
    \begin{aligned}
    \bigintss_{\displaystyle\omega_1,\dots,\omega_n } \hspace{-1cm} \dd{\left(\textstyle \sum^n_{i=1}\omega_i-a\right)} \prod_{i=1}^{n-1}D^2(\omega_i) \ D^1(\omega_n) = \bigintss_{\displaystyle\omega_1,\dots,\omega_n } \hspace{-1cm} \dd{\left(\textstyle \sum^n_{i=1}\omega_i-a\right)} \prod_{i=1}^{n-1}\boldsymbol{P^2_+}(\omega_i) \ \boldsymbol{P^1_-}(\omega_n) \\
     = - \bigintss_{\displaystyle\omega_1,\dots,\omega_{n-2} } \prod_{i=2}^{n-1}\boldsymbol{P^2_+}(\omega_i)\bigintss_{\displaystyle\omega_{n-1}}\boldsymbol{P^2_+}(\omega_{n-1})\boldsymbol{P^1_-}(\omega_{n-1}+\textstyle \sum^{n-2}_{i=1}\omega_i-a) ,
    \end{aligned}
\end{equation}
where $\boldsymbol{P^n_-}(-\omega)=- \boldsymbol{P^n_-}(\omega)$ was used.
According to equation \eqref{eq:P+P-}, each integration raises again the index of $\boldsymbol{P^n_-}$ by one, gives a factor of $-i$ and an additional factor coming from the factorial term starting at $1$ and going up to $(n-1)$. As a consequence, we get:
\begin{equation}
    \begin{aligned}
    \bigintss_{\displaystyle\omega_1,\dots,\omega_n } \hspace{-1cm} \dd{\left(\textstyle \sum^n_{i=1}\omega_i-a\right)} \prod_{i=1}^{n-1}D^2(\omega_i) \ D^1(\omega_n) = (n-1)! \  (-i)^{n-1} \ \boldsymbol{P^n_-} (a) \ \delta_{n \geq 1},
    \end{aligned}
\end{equation}
where we again used that $\boldsymbol{P^n_-}(\omega)$ is an odd function.

\item[3.)] The third one is:
\begin{equation}
    \begin{aligned}
    \bigintss_{\displaystyle\omega_1,\dots,\omega_n } \hspace{-1cm}& \dd{\left(\textstyle \sum^n_{i=1}\omega_i-a\right)} \prod_{i=1}^{n-2}D^2(\omega_i)\ D^1(\omega_{n-1}) \ D^1(\omega_n)  \\
    &= \bigintss_{\displaystyle\omega_1,\dots,\omega_n } \hspace{-1cm} \dd{\left(\textstyle \sum^n_{i=1}\omega_i-a\right)} \prod_{i=1}^{n-2}\boldsymbol{P^2_+}(\omega_i) \ \boldsymbol{P^1_-}(\omega_{n-1})\ \boldsymbol{P^1_-}(\omega_n) \\
     &= - \bigintss_{\displaystyle\omega_1,\dots,\omega_{n-2} } \prod_{i=2}^{n-2}\boldsymbol{P^2_+}(\omega_i)\bigintss_{\displaystyle\omega_{n-1}}\boldsymbol{P^1_-}(\omega_{n-1})\boldsymbol{P^1_-}(\omega_{n-1}+\textstyle \sum^{n-2}_{i=1}\omega_i-a) \\
     &=-i \bigintss_{\displaystyle\omega_1,\dots,\omega_{n-3} } \prod_{i=1}^{n-3}\boldsymbol{P^2_+}(\omega_i)\bigintss_{\displaystyle\omega_{n-2}}\boldsymbol{P^2_+}(\omega_{n-2})\boldsymbol{P^1_+}(\omega_{n-2}+\textstyle \sum^{n-3}_{i=1}\omega_i-a) ,
    \end{aligned}
\end{equation}
where equation \eqref{eq:P+P+} was used from the third to the last line.
According to equation \eqref{eq:P+P+} each following integration  raises the index of $\boldsymbol{P^n_+}$ by one, gives a factor of $-i$ and an additional factor coming from the factorial term starting at $1$ and going up to $(n-2)$. Resulting from this, we obtain:
\begin{equation}
    \begin{aligned}
    \bigintss_{\displaystyle\omega_1,\dots,\omega_n } \hspace{-1cm}& \dd{\left(\textstyle \sum^n_{i=1}\omega_i-a\right)} \prod_{i=1}^{n-2}D^2(\omega_i)\ D^1(\omega_{n-1}) \ D^1(\omega_n)  = (n-2)! \ (-i)^{n-1} \ \boldsymbol{P^{n-1}_+}(a) \ \delta_{n\geq 2} .
    \end{aligned}
\end{equation}

\item[4.)] The fourth is trivial:
 \begin{equation}
    \begin{aligned}
    \bigintss_{\displaystyle\omega_1,\dots,\omega_n } \hspace{-1cm} \dd{\left(\textstyle \sum^n_{i=1}\omega_i-a\right)} \prod_{i=1}^{n-1}D^2(\omega_i)  = \delta_{1n} .
    \end{aligned}
\end{equation}
\end{enumerate}
The first three integrals can be efficiently summarized into the following master formula:
\begin{equation}\boxed{
    \begin{aligned} \ \\
  \bigintss_{\displaystyle\omega_1,\dots,\omega_n } \hspace{-1cm} \dd{\left(\textstyle \sum^n_{i=1}\omega_i-a\right)} \prod_{i=1}^{n-\alpha-\beta}D^2(\omega_i) \ D^1(\omega_{n-1})^{\displaystyle\alpha}
   \ D^1(\omega_n)^{\displaystyle\beta} \hspace{4cm}\\
  \qquad \qquad = (n-\alpha-\beta)! \ i ^{n-1} \left( \frac{(-1)^{n-1}}{(a + n i \epsilon)^{n+1-\alpha-\beta}} + \frac{1}{(a - n i \epsilon)^{n+1-\alpha-\beta}} \right) \delta_{n\geq \alpha+\beta} ,   \\ \ \end{aligned}}
\end{equation}
with $\alpha , \beta \in [0,1]$.

\newpage
\addcontentsline{toc}{chapter}{\bibname}
\bibliographystyle{unsrtdin} 
\bibliography{references}
\newpage

\end{document}